\documentclass[longauth]{aa}  

\usepackage{graphicx}
\usepackage{tabularx}
\usepackage{txfonts}
\usepackage{longtable} 
\usepackage{hyperref}
\hypersetup{
    colorlinks=true,
    linkcolor=blue,
    citecolor=blue,
    filecolor=blue,      
    urlcolor=blue,
    breaklinks=true,
}
\usepackage{hyperref}
\usepackage{lscape}
\usepackage{xcolor}
\usepackage{color}

\newcommand{\trades}[0]{\textsc{TRADES}}
\newcommand{\pyde}[0]{\textsc{PyDE}}
\newcommand{\emcee}[0]{\textsc{emcee}}

\newcommand{\plb}[0]{\ensuremath{\mathrm{b}}}
\newcommand{\plc}[0]{\ensuremath{\mathrm{c}}}
\newcommand{\pld}[0]{\ensuremath{\mathrm{d}}}

\newcommand{\mps}[0]{\ensuremath{\mathrm{m\ s^{-1}}}}
\newcommand{\unif}[2]{\ensuremath{\mathcal{U}(#1,#2)}}
\newcommand{\norm}[2]{\ensuremath{\mathcal{N}(#1,#2)}}
\newcommand{\bounds}[2]{\ensuremath{\mathcal{B}(#1,#2)}}

\begin{document}

   \title{Transit-timing variations in the AU Mic system observed with CHEOPS\thanks{This article uses data from the CHEOPS programme CH\_PR100010.}}

   \authorrunning{Boldog et al.}


\author{Á.~Boldog\inst{\ref{inst:1},\ref{inst:2},\ref{inst:3}}, 
Gy.~M.~Szabó\inst{\ref{inst:4},\ref{inst:3}}\,$^{\href{https://orcid.org/0000-0002-0606-7930}{\protect\includegraphics[height=0.19cm]{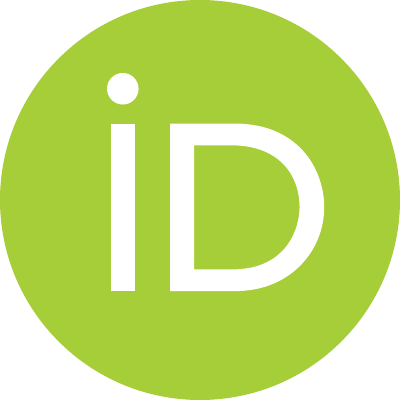}}}$, 
L.~Kriskovics\inst{\ref{inst:1},\ref{inst:2}}, 
L.~Borsato\inst{\ref{inst:5}}\,$^{\href{https://orcid.org/0000-0003-0066-9268}{\protect\includegraphics[height=0.19cm]{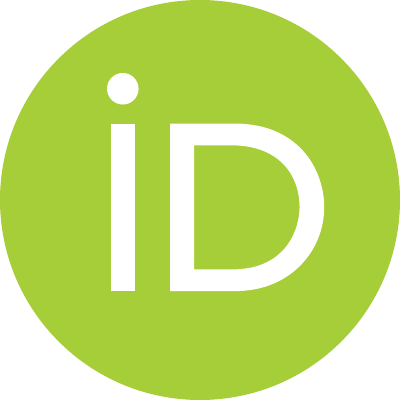}}}$, 
D.~Gandolfi\inst{\ref{inst:6}}\,$^{\href{https://orcid.org/0000-0001-8627-9628}{\protect\includegraphics[height=0.19cm]{figures/orcid.pdf}}}$, 
M.~Lendl\inst{\ref{inst:7}}\,$^{\href{https://orcid.org/0000-0001-9699-1459}{\protect\includegraphics[height=0.19cm]{figures/orcid.pdf}}}$, 
M.~N.~Günther\inst{\ref{inst:8}}\,$^{\href{https://orcid.org/0000-0002-3164-9086}{\protect\includegraphics[height=0.19cm]{figures/orcid.pdf}}}$, 
A.~Heitzmann\inst{\ref{inst:7}}\,$^{\href{https://orcid.org/0000-0002-8091-7526}{\protect\includegraphics[height=0.19cm]{figures/orcid.pdf}}}$, 
T.~G.~Wilson\inst{\ref{inst:9}}\,$^{\href{https://orcid.org/0000-0001-8749-1962}{\protect\includegraphics[height=0.19cm]{figures/orcid.pdf}}}$, 
A.~Brandeker\inst{\ref{inst:10}}\,$^{\href{https://orcid.org/0000-0002-7201-7536}{\protect\includegraphics[height=0.19cm]{figures/orcid.pdf}}}$, 
Z.~Garai\inst{\ref{inst:3},\ref{inst:4},\ref{inst:11}}\,$^{\href{https://orcid.org/0000-0001-9483-2016}{\protect\includegraphics[height=0.19cm]{figures/orcid.pdf}}}$, 
Y.~Alibert\inst{\ref{inst:12},\ref{inst:13}}\,$^{\href{https://orcid.org/0000-0002-4644-8818}{\protect\includegraphics[height=0.19cm]{figures/orcid.pdf}}}$, 
R.~Alonso\inst{\ref{inst:14},\ref{inst:15}}\,$^{\href{https://orcid.org/0000-0001-8462-8126}{\protect\includegraphics[height=0.19cm]{figures/orcid.pdf}}}$, 
T.~Bárczy\inst{\ref{inst:16}}\,$^{\href{https://orcid.org/0000-0002-7822-4413}{\protect\includegraphics[height=0.19cm]{figures/orcid.pdf}}}$, 
D.~Barrado~Navascues\inst{\ref{inst:17}}\,$^{\href{https://orcid.org/0000-0002-5971-9242}{\protect\includegraphics[height=0.19cm]{figures/orcid.pdf}}}$, 
S.~C.~C.~Barros\inst{\ref{inst:18},\ref{inst:19}}\,$^{\href{https://orcid.org/0000-0003-2434-3625}{\protect\includegraphics[height=0.19cm]{figures/orcid.pdf}}}$, 
W.~Baumjohann\inst{\ref{inst:20}}\,$^{\href{https://orcid.org/0000-0001-6271-0110}{\protect\includegraphics[height=0.19cm]{figures/orcid.pdf}}}$, 
W.~Benz\inst{\ref{inst:13},\ref{inst:12}}\,$^{\href{https://orcid.org/0000-0001-7896-6479}{\protect\includegraphics[height=0.19cm]{figures/orcid.pdf}}}$, 
N.~Billot\inst{\ref{inst:7}}\,$^{\href{https://orcid.org/0000-0003-3429-3836}{\protect\includegraphics[height=0.19cm]{figures/orcid.pdf}}}$, 
C.~Broeg\inst{\ref{inst:13},\ref{inst:12}}\,$^{\href{https://orcid.org/0000-0001-5132-2614}{\protect\includegraphics[height=0.19cm]{figures/orcid.pdf}}}$, 
A.~Collier~Cameron\inst{\ref{inst:21}}\,$^{\href{https://orcid.org/0000-0002-8863-7828}{\protect\includegraphics[height=0.19cm]{figures/orcid.pdf}}}$, 
A.~C.~M.~Correia\inst{\ref{inst:22}}\,$^{\href{https://orcid.org/0000-0002-8946-8579}{\protect\includegraphics[height=0.19cm]{figures/orcid.pdf}}}$, 
Sz.~Csizmadia\inst{\ref{inst:23}}\,$^{\href{https://orcid.org/0000-0001-6803-9698}{\protect\includegraphics[height=0.19cm]{figures/orcid.pdf}}}$, 
P.~E.~Cubillos\inst{\ref{inst:20},\ref{inst:24}}, 
M.~B.~Davies\inst{\ref{inst:25}}\,$^{\href{https://orcid.org/0000-0001-6080-1190}{\protect\includegraphics[height=0.19cm]{figures/orcid.pdf}}}$, 
M.~Deleuil\inst{\ref{inst:26}}\,$^{\href{https://orcid.org/0000-0001-6036-0225}{\protect\includegraphics[height=0.19cm]{figures/orcid.pdf}}}$, 
A.~Deline\inst{\ref{inst:7}}, 
O.~D.~S.~Demangeon\inst{\ref{inst:18},\ref{inst:19}}\,$^{\href{https://orcid.org/0000-0001-7918-0355}{\protect\includegraphics[height=0.19cm]{figures/orcid.pdf}}}$, 
B.-O.~Demory\inst{\ref{inst:12},\ref{inst:13}}\,$^{\href{https://orcid.org/0000-0002-9355-5165}{\protect\includegraphics[height=0.19cm]{figures/orcid.pdf}}}$, 
A.~Derekas\inst{\ref{inst:4}}, 
B.~Edwards\inst{\ref{inst:27}}, 
J.~A.~Egger\inst{\ref{inst:13}}\,$^{\href{https://orcid.org/0000-0003-1628-4231}{\protect\includegraphics[height=0.19cm]{figures/orcid.pdf}}}$, 
D.~Ehrenreich\inst{\ref{inst:7},\ref{inst:28}}\,$^{\href{https://orcid.org/0000-0001-9704-5405}{\protect\includegraphics[height=0.19cm]{figures/orcid.pdf}}}$, 
A.~Erikson\inst{\ref{inst:23}}, 
A.~Fortier\inst{\ref{inst:13},\ref{inst:12}}\,$^{\href{https://orcid.org/0000-0001-8450-3374}{\protect\includegraphics[height=0.19cm]{figures/orcid.pdf}}}$, 
L.~Fossati\inst{\ref{inst:20}}\,$^{\href{https://orcid.org/0000-0003-4426-9530}{\protect\includegraphics[height=0.19cm]{figures/orcid.pdf}}}$, 
M.~Fridlund\inst{\ref{inst:29},\ref{inst:30}}\,$^{\href{https://orcid.org/0000-0002-0855-8426}{\protect\includegraphics[height=0.19cm]{figures/orcid.pdf}}}$, 
K.~Gazeas\inst{\ref{inst:31}}\,$^{\href{https://orcid.org/0000-0002-8855-3923}{\protect\includegraphics[height=0.19cm]{figures/orcid.pdf}}}$, 
M.~Gillon\inst{\ref{inst:32}}\,$^{\href{https://orcid.org/0000-0003-1462-7739}{\protect\includegraphics[height=0.19cm]{figures/orcid.pdf}}}$, 
M.~Güdel\inst{\ref{inst:33}}, 
P.~Guterman\inst{\ref{inst:26},\ref{inst:34}}, 
Ch.~Helling\inst{\ref{inst:20},\ref{inst:35}}, 
K.~G.~Isaak\inst{\ref{inst:8}}\,$^{\href{https://orcid.org/0000-0001-8585-1717}{\protect\includegraphics[height=0.19cm]{figures/orcid.pdf}}}$, 
L.~L.~Kiss\inst{\ref{inst:36},\ref{inst:37}},
E.~Kopp\inst{\ref{inst:44}}, 
J.~Korth\inst{\ref{inst:38}}\,$^{\href{https://orcid.org/0000-0002-0076-6239}{\protect\includegraphics[height=0.19cm]{figures/orcid.pdf}}}$, 
K.~W.~F.~Lam\inst{\ref{inst:23}}\,$^{\href{https://orcid.org/0000-0002-9910-6088}{\protect\includegraphics[height=0.19cm]{figures/orcid.pdf}}}$, 
J.~Laskar\inst{\ref{inst:39}}\,$^{\href{https://orcid.org/0000-0003-2634-789X}{\protect\includegraphics[height=0.19cm]{figures/orcid.pdf}}}$, 
A.~Lecavelier~des~Etangs\inst{\ref{inst:40}}\,$^{\href{https://orcid.org/0000-0002-5637-5253}{\protect\includegraphics[height=0.19cm]{figures/orcid.pdf}}}$,
A.~Luntzer\inst{\ref{inst:33}},  
D.~Magrin\inst{\ref{inst:5}}\,$^{\href{https://orcid.org/0000-0003-0312-313X}{\protect\includegraphics[height=0.19cm]{figures/orcid.pdf}}}$, 
G.~Mantovan\inst{\ref{inst:46}},
L.~Marafatto\inst{\ref{inst:5}},
P.~F.~L.~Maxted\inst{\ref{inst:41}}\,$^{\href{https://orcid.org/0000-0003-3794-1317}{\protect\includegraphics[height=0.19cm]{figures/orcid.pdf}}}$, 
B.~Merín\inst{\ref{inst:42}}\,$^{\href{https://orcid.org/0000-0002-8555-3012}{\protect\includegraphics[height=0.19cm]{figures/orcid.pdf}}}$, 
C.~Mordasini\inst{\ref{inst:13},\ref{inst:12}}, 
M.~Munari\inst{\ref{inst:43}}\,$^{\href{https://orcid.org/0000-0003-0990-050X}{\protect\includegraphics[height=0.19cm]{figures/orcid.pdf}}}$, 
V.~Nascimbeni\inst{\ref{inst:5}}\,$^{\href{https://orcid.org/0000-0001-9770-1214}{\protect\includegraphics[height=0.19cm]{figures/orcid.pdf}}}$, 
G.~Olofsson\inst{\ref{inst:10}}\,$^{\href{https://orcid.org/0000-0003-3747-7120}{\protect\includegraphics[height=0.19cm]{figures/orcid.pdf}}}$, 
R.~Ottensamer\inst{\ref{inst:33}}, 
I.~Pagano\inst{\ref{inst:43}}\,$^{\href{https://orcid.org/0000-0001-9573-4928}{\protect\includegraphics[height=0.19cm]{figures/orcid.pdf}}}$, 
E.~Pallé\inst{\ref{inst:14},\ref{inst:15}}\,$^{\href{https://orcid.org/0000-0003-0987-1593}{\protect\includegraphics[height=0.19cm]{figures/orcid.pdf}}}$, 
G.~Peter\inst{\ref{inst:44}}\,$^{\href{https://orcid.org/0000-0001-6101-2513}{\protect\includegraphics[height=0.19cm]{figures/orcid.pdf}}}$, 
D.~Piazza\inst{\ref{inst:45}}, 
G.~Piotto\inst{\ref{inst:5},\ref{inst:46}}\,$^{\href{https://orcid.org/0000-0002-9937-6387}{\protect\includegraphics[height=0.19cm]{figures/orcid.pdf}}}$, 
D.~Pollacco\inst{\ref{inst:9}}, 
K.~Poppenhaeger\inst{\ref{inst:56}}, 
D.~Queloz\inst{\ref{inst:47},\ref{inst:48}}\,$^{\href{https://orcid.org/0000-0002-3012-0316}{\protect\includegraphics[height=0.19cm]{figures/orcid.pdf}}}$, 
R.~Ragazzoni\inst{\ref{inst:5},\ref{inst:46}}\,$^{\href{https://orcid.org/0000-0002-7697-5555}{\protect\includegraphics[height=0.19cm]{figures/orcid.pdf}}}$, 
N.~Rando\inst{\ref{inst:8}}, 
H.~Rauer\inst{\ref{inst:23},\ref{inst:49}}\,$^{\href{https://orcid.org/0000-0002-6510-1828}{\protect\includegraphics[height=0.19cm]{figures/orcid.pdf}}}$, 
I.~Ribas\inst{\ref{inst:50},\ref{inst:51}}\,$^{\href{https://orcid.org/0000-0002-6689-0312}{\protect\includegraphics[height=0.19cm]{figures/orcid.pdf}}}$, 
M.~Rieder\inst{\ref{inst:52},\ref{inst:12}}, 
N.~C.~Santos\inst{\ref{inst:18},\ref{inst:19}}\,$^{\href{https://orcid.org/0000-0003-4422-2919}{\protect\includegraphics[height=0.19cm]{figures/orcid.pdf}}}$, 
G.~Scandariato\inst{\ref{inst:43}}\,$^{\href{https://orcid.org/0000-0003-2029-0626}{\protect\includegraphics[height=0.19cm]{figures/orcid.pdf}}}$, 
D.~Ségransan\inst{\ref{inst:7}}\,$^{\href{https://orcid.org/0000-0003-2355-8034}{\protect\includegraphics[height=0.19cm]{figures/orcid.pdf}}}$, 
A.~E.~Simon\inst{\ref{inst:13},\ref{inst:12}}\,$^{\href{https://orcid.org/0000-0001-9773-2600}{\protect\includegraphics[height=0.19cm]{figures/orcid.pdf}}}$, 
A.~M.~S.~Smith\inst{\ref{inst:23}}\,$^{\href{https://orcid.org/0000-0002-2386-4341}{\protect\includegraphics[height=0.19cm]{figures/orcid.pdf}}}$, 
S.~G.~Sousa\inst{\ref{inst:18}}\,$^{\href{https://orcid.org/0000-0001-9047-2965}{\protect\includegraphics[height=0.19cm]{figures/orcid.pdf}}}$,
R.~Southworth\inst{\ref{inst:53}}, 
M.~Stalport\inst{\ref{inst:54},\ref{inst:32}}, 
S.~Sulis\inst{\ref{inst:26}}\,$^{\href{https://orcid.org/0000-0001-8783-526X}{\protect\includegraphics[height=0.19cm]{figures/orcid.pdf}}}$, 
S.~Udry\inst{\ref{inst:7}}\,$^{\href{https://orcid.org/0000-0001-7576-6236}{\protect\includegraphics[height=0.19cm]{figures/orcid.pdf}}}$, 
S.~Ulmer-Moll\inst{\ref{inst:7},\ref{inst:13},\ref{inst:54}}\,$^{\href{https://orcid.org/0000-0003-2417-7006}{\protect\includegraphics[height=0.19cm]{figures/orcid.pdf}}}$, 
V.~Van~Grootel\inst{\ref{inst:54}}\,$^{\href{https://orcid.org/0000-0003-2144-4316}{\protect\includegraphics[height=0.19cm]{figures/orcid.pdf}}}$, 
J.~Venturini\inst{\ref{inst:7}}\,$^{\href{https://orcid.org/0000-0001-9527-2903}{\protect\includegraphics[height=0.19cm]{figures/orcid.pdf}}}$, 
E.~Villaver\inst{\ref{inst:14},\ref{inst:15}}, 
N.~A.~Walton\inst{\ref{inst:55}}\,$^{\href{https://orcid.org/0000-0003-3983-8778}{\protect\includegraphics[height=0.19cm]{figures/orcid.pdf}}}$, and
T.~Zingales\inst{\ref{inst:46},\ref{inst:5}}\,$^{\href{https://orcid.org/0000-0001-6880-5356}{\protect\includegraphics[height=0.19cm]{figures/orcid.pdf}}}$}

\institute{\label{inst:1} Konkoly Observatory, HUN-REN Research Centre for Astronomy and Earth Sciences, Konkoly Thege út 15-17., H-1121, Budapest, Hungary \and
\label{inst:2} CSFK, MTA Centre of Excellence, Budapest, Konkoly Thege út 15-17., H-1121, Hungary \and
\label{inst:3} HUN-REN-ELTE Exoplanet Research Group, Szent Imre h. u. 112., Szombathely, H-9700, Hungary \and
\label{inst:4} ELTE Gothard Astrophysical Observatory, 9700 Szombathely, Szent Imre h. u. 112, Hungary \and
\label{inst:5} INAF, Osservatorio Astronomico di Padova, Vicolo dell'Osservatorio 5, 35122 Padova, Italy \and
\label{inst:6} Dipartimento di Fisica, Università degli Studi di Torino, via Pietro Giuria 1, I-10125, Torino, Italy \and
\label{inst:7} Observatoire astronomique de l'Université de Genève, Chemin Pegasi 51, 1290 Versoix, Switzerland \and
\label{inst:8} European Space Agency (ESA), European Space Research and Technology Centre (ESTEC), Keplerlaan 1, 2201 AZ Noordwijk, The Netherlands \and
\label{inst:9} Department of Physics, University of Warwick, Gibbet Hill Road, Coventry CV4 7AL, United Kingdom \and
\label{inst:10} Department of Astronomy, Stockholm University, AlbaNova University Center, 10691 Stockholm, Sweden \and
\label{inst:11} Astronomical Institute, Slovak Academy of Sciences, 05960 Tatranská Lomnica, Slovakia \and
\label{inst:12} Center for Space and Habitability, University of Bern, Gesellschaftsstrasse 6, 3012 Bern, Switzerland \and
\label{inst:13} Space Research and Planetary Sciences, Physics Institute, University of Bern, Gesellschaftsstrasse 6, 3012 Bern, Switzerland \and
\label{inst:14} Instituto de Astrofísica de Canarias, Vía Láctea s/n, 38200 La Laguna, Tenerife, Spain \and
\label{inst:15} Departamento de Astrofísica, Universidad de La Laguna, Astrofísico Francisco Sanchez s/n, 38206 La Laguna, Tenerife, Spain \and
\label{inst:16} Admatis, 5. Kandó Kálmán Street, 3534 Miskolc, Hungary \and
\label{inst:17} Depto. de Astrofísica, Centro de Astrobiología (CSIC-INTA), ESAC campus, 28692 Villanueva de la Cañada (Madrid), Spain \and
\label{inst:18} Instituto de Astrofisica e Ciencias do Espaco, Universidade do Porto, CAUP, Rua das Estrelas, 4150-762 Porto, Portugal \and
\label{inst:19} Departamento de Fisica e Astronomia, Faculdade de Ciencias, Universidade do Porto, Rua do Campo Alegre, 4169-007 Porto, Portugal \and
\label{inst:20} Space Research Institute, Austrian Academy of Sciences, Schmiedlstrasse 6, A-8042 Graz, Austria \and
\label{inst:21} Centre for Exoplanet Science, SUPA School of Physics and Astronomy, University of St Andrews, North Haugh, St Andrews KY16 9SS, UK \and
\label{inst:22} CFisUC, Departamento de Física, Universidade de Coimbra, 3004-516 Coimbra, Portugal \and
\label{inst:23} Institute of Planetary Research, German Aerospace Center (DLR), Rutherfordstrasse 2, 12489 Berlin, Germany \and
\label{inst:24} INAF, Osservatorio Astrofisico di Torino, Via Osservatorio, 20, I-10025 Pino Torinese To, Italy \and
\label{inst:25} Centre for Mathematical Sciences, Lund University, Box 118, 221 00 Lund, Sweden \and
\label{inst:26} Aix Marseille Univ, CNRS, CNES, LAM, 38 rue Frédéric Joliot-Curie, 13388 Marseille, France \and
\label{inst:27} SRON Netherlands Institute for Space Research, Niels Bohrweg 4, 2333 CA Leiden, Netherlands \and
\label{inst:28} Centre Vie dans l’Univers, Faculté des sciences, Université de Genève, Quai Ernest-Ansermet 30, 1211 Genève 4, Switzerland \and
\label{inst:29} Leiden Observatory, University of Leiden, PO Box 9513, 2300 RA Leiden, The Netherlands \and
\label{inst:30} Department of Space, Earth and Environment, Chalmers University of Technology, Onsala Space Observatory, 439 92 Onsala, Sweden \and
\label{inst:31} National and Kapodistrian University of Athens, Department of Physics, University Campus, Zografos GR-157 84, Athens, Greece \and
\label{inst:32} Astrobiology Research Unit, Université de Liège, Allée du 6 Août 19C, B-4000 Liège, Belgium \and
\label{inst:33} Department of Astrophysics, University of Vienna, Türkenschanzstrasse 17, 1180 Vienna, Austria \and
\label{inst:34} Division Technique INSU, CS20330, 83507 La Seyne sur Mer cedex, France \and
\label{inst:35} Institute for Theoretical Physics and Computational Physics, Graz University of Technology, Petersgasse 16, 8010 Graz, Austria \and
\label{inst:36} Konkoly Observatory, Research Centre for Astronomy and Earth Sciences, 1121 Budapest, Konkoly Thege Miklós út 15-17, Hungary \and
\label{inst:37} ELTE E\"otv\"os Lor\'and University, Institute of Physics, P\'azm\'any P\'eter s\'et\'any 1/A, 1117 Budapest, Hungary \and
\label{inst:38} Lund Observatory, Division of Astrophysics, Department of Physics, Lund University, Box 118, 22100 Lund, Sweden \and
\label{inst:39} IMCCE, UMR8028 CNRS, Observatoire de Paris, PSL Univ., Sorbonne Univ., 77 av. Denfert-Rochereau, 75014 Paris, France \and
\label{inst:40} Institut d'astrophysique de Paris, UMR7095 CNRS, Université Pierre \& Marie Curie, 98bis blvd. Arago, 75014 Paris, France \and
\label{inst:41} Astrophysics Group, Lennard Jones Building, Keele University, Staffordshire, ST5 5BG, United Kingdom \and
\label{inst:42} European Space Agency, ESA - European Space Astronomy Centre, Camino Bajo del Castillo s/n, 28692 Villanueva de la Cañada, Madrid, Spain \and
\label{inst:43} INAF, Osservatorio Astrofisico di Catania, Via S. Sofia 78, 95123 Catania, Italy \and
\label{inst:44} Institute of Optical Sensor Systems, German Aerospace Center (DLR), Rutherfordstrasse 2, 12489 Berlin, Germany \and
\label{inst:45} Weltraumforschung und Planetologie, Physikalisches Institut, University of Bern, Gesellschaftsstrasse 6, 3012 Bern, Switzerland \and
\label{inst:46} Dipartimento di Fisica e Astronomia "Galileo Galilei", Università degli Studi di Padova, Vicolo dell'Osservatorio 3, 35122 Padova, Italy \and
\label{inst:47} ETH Zurich, Department of Physics, Wolfgang-Pauli-Strasse 2, CH-8093 Zurich, Switzerland \and
\label{inst:48} Cavendish Laboratory, JJ Thomson Avenue, Cambridge CB3 0HE, UK \and
\label{inst:49} Institut fuer Geologische Wissenschaften, Freie Universitaet Berlin, Maltheserstrasse 74-100,12249 Berlin, Germany \and
\label{inst:50} Institut de Ciencies de l'Espai (ICE, CSIC), Campus UAB, Can Magrans s/n, 08193 Bellaterra, Spain \and
\label{inst:51} Institut d'Estudis Espacials de Catalunya (IEEC), 08860 Castelldefels (Barcelona), Spain \and
\label{inst:52} Weltraumforschung und Planetologie, Physikalisches Institut, University of Bern, Sidlerstrasse 5, 3012 Bern, Switzerland \and
\label{inst:53} ESOC, European Space Agency, Robert Bosch Str. 5, 64293 Darmstadt, Germany \and
\label{inst:54} Space sciences, Technologies and Astrophysics Research (STAR) Institute, Université de Liège, Allée du 6 Août 19C, 4000 Liège, Belgium \and
\label{inst:55} Institute of Astronomy, University of Cambridge, Madingley Road, Cambridge, CB3 0HA, United Kingdom \and
\label{inst:56} Leibniz Institute for Astrophysics, An der Sternwarte 16, 14482 Potsdam, Germany }

    \date{Received ; accepted }

  \abstract
   {AU Mic is a very active M dwarf star with an edge-on debris disk and two known transiting sub-Neptunes with a possible third planetary companion. The two transiting planets exhibit significant transit-timing variations (TTVs) that are caused by the gravitational interaction between the bodies in the system. }
   {Using photometrical observations taken with the CHaracterizing ExOPlanet Satellite (CHEOPS), we aim to constrain the planetary radii, the orbital distances, and the periods of AU Mic b and c. Furthermore, our goal is to determine the superperiod of the TTVs for AU Mic b and to update the transit ephemeris for both planets. Additionally, based on the perceived TTVs, we study the possible presence of a third planet in the system. }
   {We conducted ultra-high precision photometric observations with CHEOPS in 2022 and 2023. We used {\tt{Allesfitter}} to fit the planetary transits and to constrain the planetary and orbital parameters. We combined our new measurements with results from previous years to determine the periods and amplitudes of the TTVs. We applied dynamical modelling based on TTV measurements from the 2018-2023 period to reconstruct the perceived variations. }
   {We found that the orbital distances and periods for AU Mic b and c agree with the results from previous works. However, the values for the planetary radii deviate slightly from previous values, which we attribute to the effect of spots on the stellar surface. AU Mic c showed very strong TTVs, with transits that occurred $\sim$80 minutes later in 2023 than in 2021. Through a dynamical analysis of the system, we found that the observed TTVs can be explained by a third planet with an orbital period of $\sim$12.6~days and a mass of 0.203$^{0.022}_{0.024}$~M$_{\oplus}$. We explored the orbital geometry of the system and found that AU Mic c has a misaligned retrograde orbit. The limited number of AU~Mic observations prevented us from determining the exact dynamical configuration and planetary parameters. Further monitoring of the system with CHEOPS might help to improve these results.}
   {}


   \maketitle
%

\section{Introduction}
AU Mic is a young M1-type dwarf star with an age of 22$\pm$3~Myr \citep{MamajekBell2014} that hosts two sub-Neptune-type planets \citep{Plavchan2020, Martioli2021} and a complex debris disk that is viewed edge-on \citep{MathioudakisDoyle1991, Kalas2004}. It is a fairly close system with a distance of 9.71~pc \citep{Gaia2023}, which makes it an excellent target for investigating a planetary system in the early stages of evolution. AU Mic is a very active M dwarf star \citep{Butler1981, TsikoudiKellett2000, Gilbert2022} with a strong magnetic field \citep{Kochukhov2020}. It produces frequent flare events \citep{IlinPoppenhaeger2022} and has large stellar spots \citep{Hebb2007, Plavchan2020}. Given the relative proximity of the planets to the host star, signs of star-planet interactions are expected \citep{Kavanagh2021} and may be reflected in the distribution of stellar flares \citep{IlinPoppenhaeger2022}. On the other hand, the spots on the stellar surface may influence the planetary parameters derived from transit modelling for AU~Mic~b and c \citep{Oshagh2013}. A better understanding of the activity of the host may therefore be beneficial for refined radii of the planets. We study the activity of AU Mic in an accompanying publication (Kriskovics et al., in prep). 

AU Mic hosts an extended debris disk that is viewed edge-on \citep{Kalas2004}. Infrared observations confirmed the size of the disk, which has an inner radius of about 9 AU \citep{MacGregor2013, Matthews2015} and a peak brightness of about $~40$ AU \citep{MacGregor2013}. Its outer halo extends to 210 AU \citep{Kalas2004}. The disk exhibits complex structures from large radial cavities to clumps and warps. It also shows asymmetries in the vertical distribution of the dust \citep{Schneider2014, Wisniewski2019}.

The two known transiting planets in the system, AU Mic b and c, are both Neptune-sized bodies, with masses of 11.7$\pm$5.0~$\mathrm{M}_\oplus$ and 22.2$\pm$6.7~$\mathrm{M}_\oplus$, respectively \citep{Zicher2022}. The radii of planets b and c were determined to be 3.55$\pm$0.13~$\mathrm{R}_\oplus$ and 2.56$\pm$0.12~$\mathrm{R}_\oplus$ \citep{Szabó2022}, respectively. However, because of the strong activity and the spot-covered surface of the host star AU Mic, the precision of the estimates on the planetary radii may be lower. The inner planet, AU~Mic~b, has a mean orbital period of about 8.463 days \citep{Gilbert2022} and is in a 7:4 spin-orbit resonance with the star \citep{Szabó2021}. The outer planet, AU~Mic~c, has a mean orbital period of 18.859 days \citep{Gilbert2022}. Both planets exhibit significant transit-timing variations (TTVs) of the order of tens of minutes \citep{Szabó2022}. These large variations are likely due to the gravitational effects of additional planets in the system. \citet{Wittrock2023} found that the observed TTVs could be explained with an Earth-sized planet that orbits between AU Mic b and c. A more precise determination of the superperiod of the TTVs may help us to confirm the presence of AU Mic d.  \\

\begin{figure*}[h]
 \begin{tabular}{cc}
 AU Mic b, 2022 & AU Mic c, 2022 \\
    \includegraphics[bb=5 50 450 400,width=0.9\columnwidth]{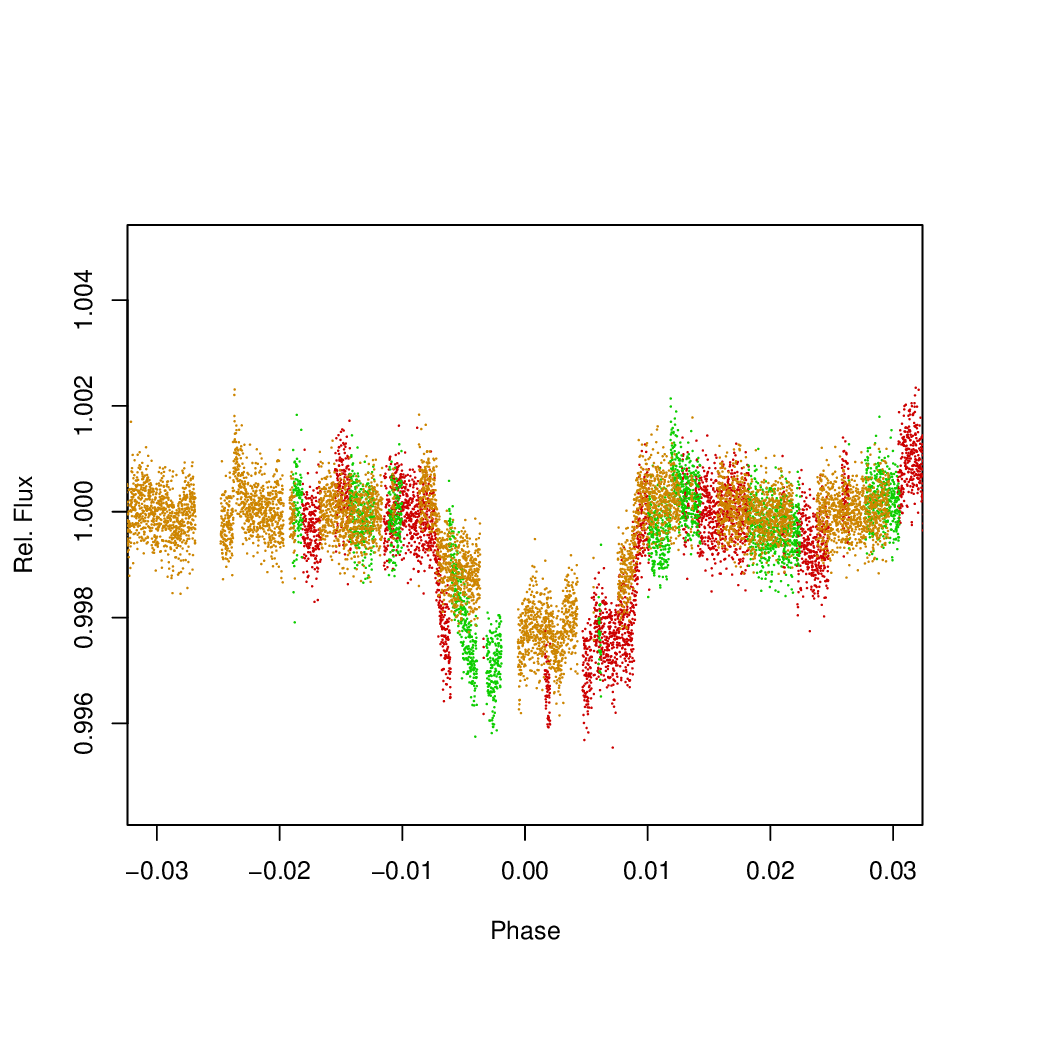} &
    \includegraphics[bb=5 50 450 400,width=0.9\columnwidth]{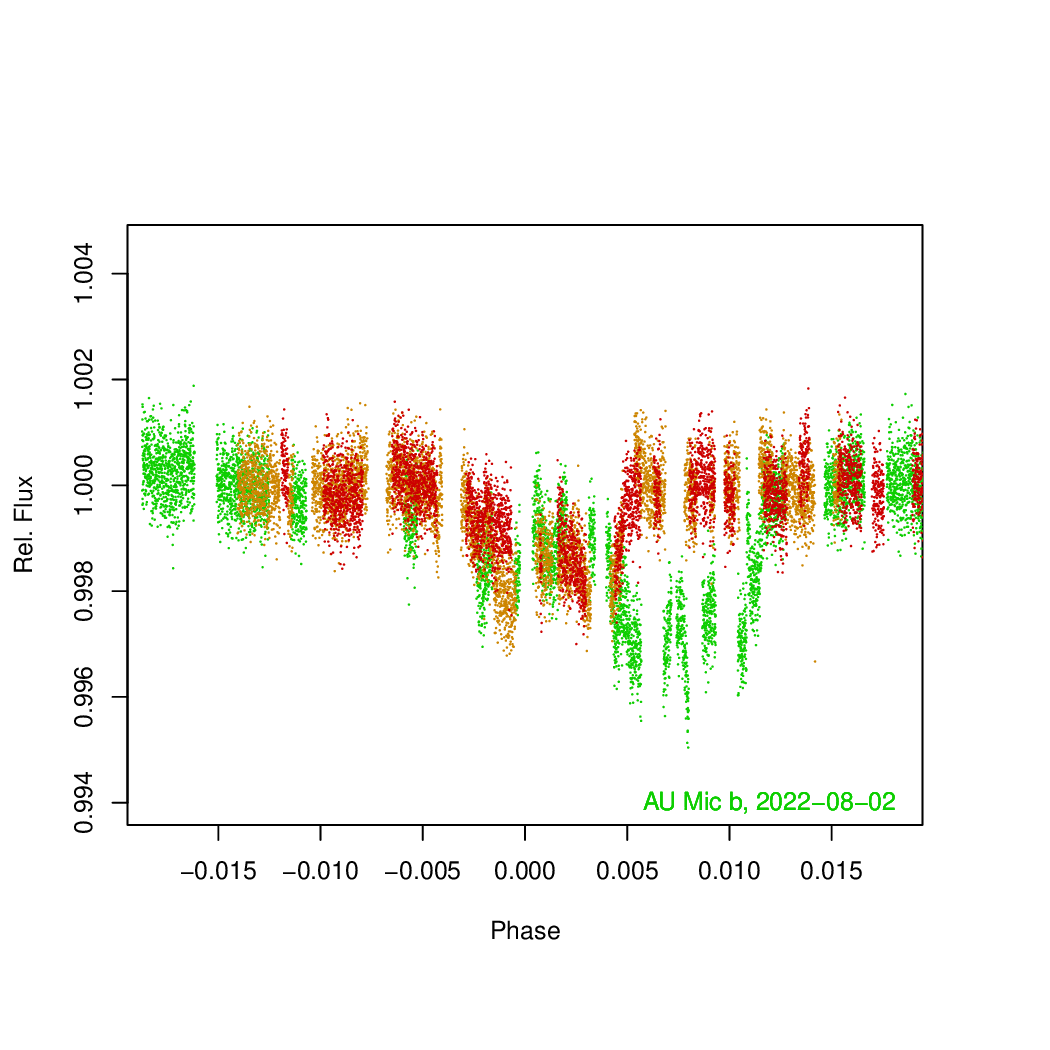} \\
     AU Mic b, 2023 & AU Mic c, 2023 \\
    \includegraphics[bb=5 50 450 400,width=0.9\columnwidth]{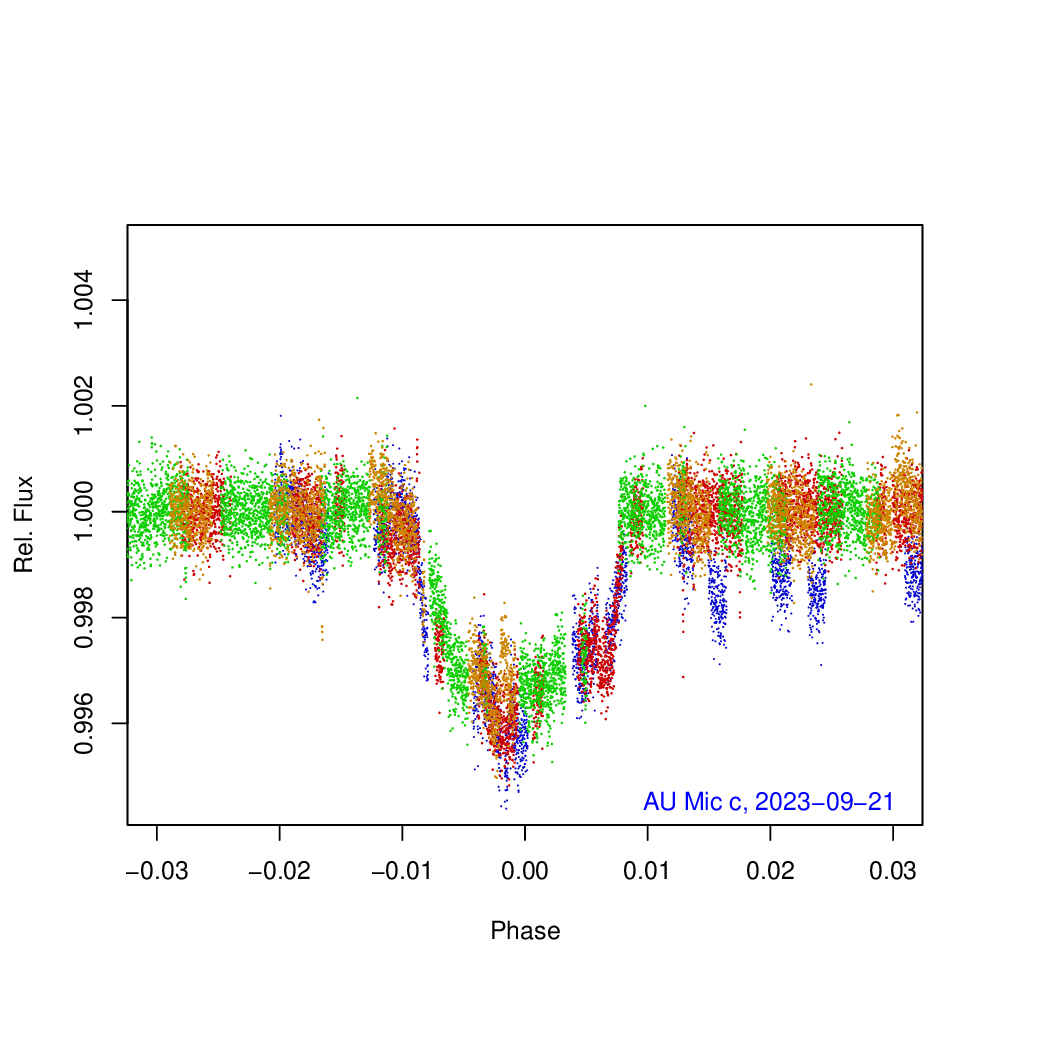} &
    \includegraphics[bb=5 50 450 400,width=0.9\columnwidth]{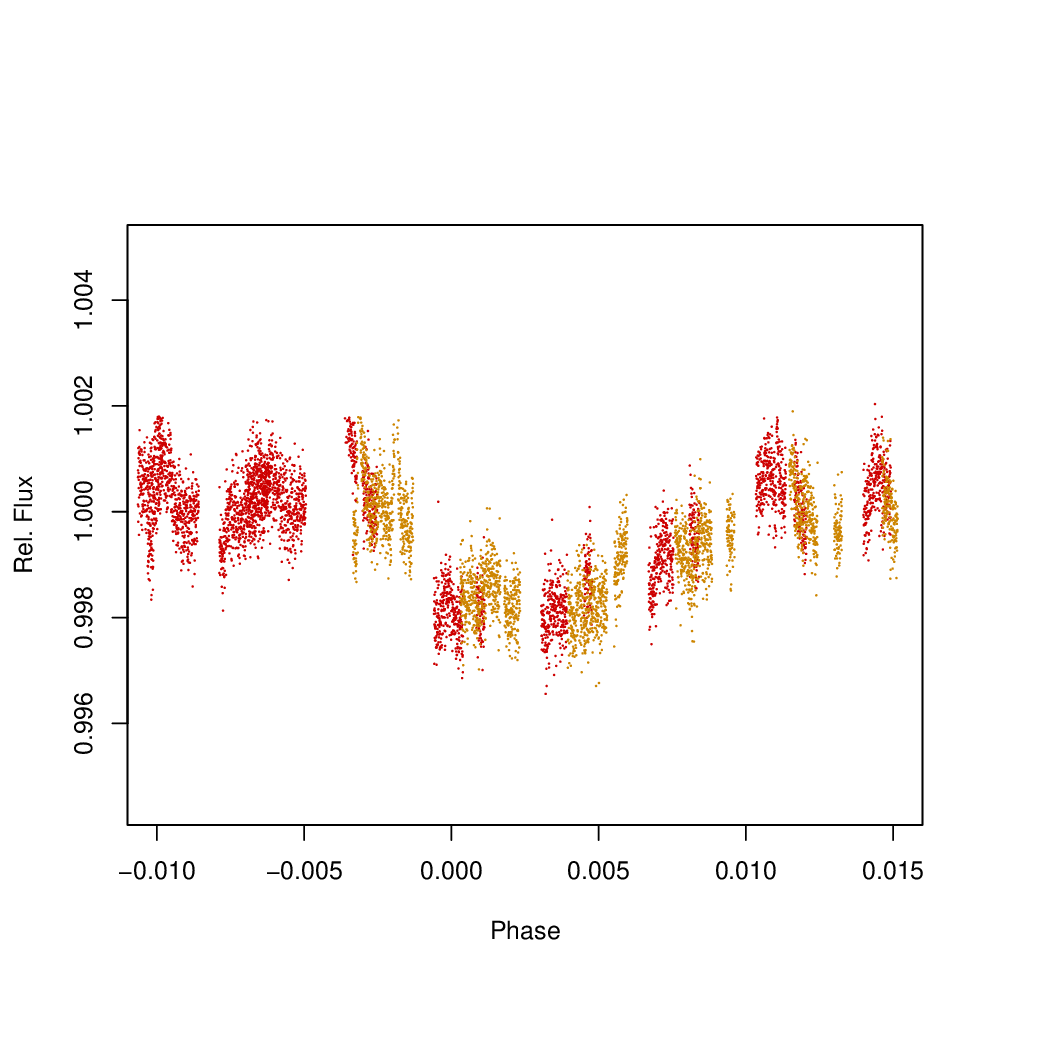} \\
 \end{tabular}
     \caption{Phased transit light curves of the 2022 and 2023 measurements of the AU~Mic~b (left panels) and c transits (right panels) after flare removal. The ephemeris and period used here are $T_{0,b} = 2\,458\,330.38416$\,d and $P_{b} = 8.4631427$\,d, and $T_{0,c} = 2459454.8973$~BJD, $P_{c} = 18.85882$~d. The TTV is very prominent for both planets compared to the earlier ephemeris. AU Mic b transited after the AU Mic c transit on 22 August 2022 (top right panel), and similarly, planet c transited after planet b on 21 September 2023 (bottom left panel).}
\label{figure:Cheops_raw_transits}
\end{figure*}

The CHaracterizing ExOPlanet Satellite (CHEOPS) is a small-class ESA mission with the primary objective of conducting follow-up studies of known exoplanetary systems using ultra-high precision photometry \citep{Benz2021}. With its observational strategy, stability, the effective aperture of 30~cm, and the broad bandpass of 330-1100 nm \citep{Deline2020}, CHEOPS is suitable for detecting planetary transits over an extended portion of the sky around a large variety of host stars \citep{Fortier2024}. In recent years, CHEOPS has proven to be a valuable tool for constraining planetary radii \citep{Lendl2020, Delrez2021, Bonfanti2021} and determining TTV in multiplanet systems \citep{Szabó2022, Delrez2023}. 

We present high-precision photometric observations of the transits of AU Mic b and c that were carried out by CHEOPS. In Sect. 2 we describe our observations and data reduction methods. In Sect. 3 we analyse the transit light curves and describe the dynamical analysis of the system based on the TTV measurements. We present and discuss our results in Sect. 4. 

\section{Observations and data reduction}

\begin{table*}[t]
\centering
\caption{Overview of the \texttt{Allesfitter} priors and best-fitting parameters of AU\,Mic\,b and c obtained based on the data from 2022 and 2023.}
\label{cheops-parameters-tab}
\begin{tabular}{llll}
\hline
Parameter [unit] & Prior & Value (2022) & Value (2023) \\
\hline
\multicolumn{4}{c}{Orbital and planetary parameters: AU\,Mic\,b}\\
$T_\mathrm{c}$ [$\mathrm{BJD}_\mathrm{TDB}$]             & $\mathcal{U}$(2459769.0500, 2459769.2000)$^1$ & $2459769.12638\pm0.00035$               & $2460149.95497_{-0.00086}^{+0.0012}$\\
                                                     & $\mathcal{U}$(2460149.9000, 2460150.0500)$^2$ & \\
$P_\mathrm{orb}$ [d]                                                         & $\mathcal{N}$(8.4631427, 0.0000005)$^3$           & $8.46314271\pm0.00000050$                & $8.46314280_{-0.00000051}^{+0.00000046}$\\
$R_\mathrm{p}/R_\mathrm{s}$                                              & $\mathcal{U}$(0.01, 0.1)                                  & $0.04700_{-0.00073}^{+0.00077}$          & $0.0517\pm0.0011$\\
$(R_\mathrm{p} + R_\mathrm{s})/a$                                        & $\mathcal{U}$(0.05, 0.06)                                 & $0.0556_{-0.0013}^{+0.0016}$             & $0.0578_{-0.0016}^{+0.0014}$\\
$\cos i$                                                                                     & $\mathcal{U}$(0.0, 0.0349)                                    & $0.0154_{-0.0054}^{+0.0047}$             & $0.0213_{-0.0045}^{+0.0035}$\\
\hline
\multicolumn{4}{c}{Orbital and planetary parameters: AU\,Mic\,c}\\
$T_\mathrm{c}$ [$\mathrm{BJD}_\mathrm{TDB}$]             & $\mathcal{U}$(2459756.6000, 2459756.7500)$^1$ & $2459756.6677_{-0.0052}^{+0.0040}$      & $2460133.8718_{-0.0074}^{+0.0096}$\\
                                                     & $\mathcal{U}$(2460133.8000, 2460133.9500)$^2$ & \\
$P_\mathrm{orb}$ [d]                                                         & $\mathcal{N}$(18.85882, 0.00005)$^3$                      & $18.858819\pm0.000049$                   & $18.858827_{-0.000052}^{+0.000049}$\\
$R_\mathrm{p}/R_\mathrm{s}$                                              & $\mathcal{U}$(0.01, 0.1)                                  & $0.0354\pm0.0016$                        & $0.0309_{-0.0033}^{+0.0028}$\\
$(R_\mathrm{p} + R_\mathrm{s})/a$                                        & $\mathcal{U}$(0.030, 0.036)                                   & $0.0338\pm0.0011$                        & $0.0336_{-0.0010}^{+0.0011}$\\
$\cos i$                                                                                     & $\mathcal{U}$(0.0, 0.0349)                                & $0.0200_{-0.0037}^{+0.0027}$             & $0.0214_{-0.0074}^{+0.0033}$\\
\hline
\multicolumn{4}{c}{LD and GP parameters}\\
$q_1$                                                                                    & $\mathcal{N}$(0.5100, 0.1)                                    & $0.485\pm0.089$                          & $0.619\pm0.085$\\
$q_2$                                                                                    & $\mathcal{N}$(0.2324, 0.1)                                    & $0.279\pm0.091$                          & $0.380\pm0.092$\\
$\log \sigma$ [$\log$ relative flux]                         & $\mathcal{U}$(-10.0, -6.0)                                   & $-7.7085_{-0.0037}^{+0.0039}$            & $-7.7548\pm0.0044$  \\ 
$\log S_\mathrm{0}$                                                                  & $\mathcal{U}$(-25.0, -15.0)                                       & $-21.05\pm0.12$                          & $-20.66_{-0.11}^{+0.11}$\\ 
$\log \omega_\mathrm{0}$ [$\log$ d$^{-1}$]                       & $\mathcal{U}$(4.0, 8.0)                                            & $5.599\pm0.051$                          & $5.898\pm0.047$\\
\hline
\end{tabular}
\tablefoot{The additional priors were set as follows: $T_\mathrm{eff} = 3665 \pm 31~\mathrm{K}$, $R_\mathrm{s} = 0.82 \pm 0.02~\mathrm{R}_\odot$, and $M_\mathrm{s} = 0.60 \pm 0.04~\mathrm{M}_\odot$ \citep{Donati2023}. $^1$The prior for the mid-transit time in 2022. $^2$The prior for the mid-transit time in 2023. $^3$Based on \citet{Szabó2022}.}
\end{table*}

\subsection{CHEOPS photometry}
We observed six and three transits of AU~Mic~b and c during the observation period 2022, respectively. In 2023, seven additional transits were observed by CHEOPS for AU~Mic~b, and four more transit for AU~Mic~c. On two occasions, a double transit of both planets b and c was observed, once in each year. Since AU Mic is a relatively bright star with a brightness of G=7\fm843 in the Gaia G band \citep{Gaia2023}, it was possible to perform short-cadence photometry (with exposures of 3~seconds). However, due to the strong flaring activity of the star, only eight transits of planet b and six for planet c were appropriate for further analysis following visual inspection. {The efficiency of the observations varied between 50.2--79.1$\%$ in 2022 and 50.0--79.3$\%$ in 2023.} The observation log for the entire campaign of 2022-2023 is available \href{https://doi.org/10.5281/zenodo.14637451}{here}. 

In this analysis, photometry was extracted using the CHEOPS imagettes alone. The imagettes are small images centred around the target star, with a radius of 30 pixels, and unlike sub-array images, they do not have to be co-added on board of the telescope. They can instead be downloaded individually. Photometric extraction of imagettes was carried out using PSF imagette photometric extraction (PIPE), a tool that was specifically developed for this purpose using point-spread function (PSF) photometry \citep{Brandeker2024}. In this case, the PIPE photometry has a shorter cadence than the CHEOPS data reduction pipeline (DRP) on the aperture photometry of the sub-arrays, but the signal-to-noise ratio (S/N) is similar. The shorter cadence of the imagettes can be of crucial importance in the case of active stars, such as AU Mic, to better identify and adequately analyse flares. We provide a detailed analysis of the flares of AU Mic in an accompanying publication (Kriskovics et al., in prep).

Since CHEOPS continuously rolls during its 98.77-minute orbit in order to keep the radiators from facing towards Earth, the field of view of the telescope also rotates around the target. This effect, combined with the irregularly shaped CHEOPS PSF, causes systematic variations in the background flux in phase with the roll angle, which has to be appropriately handled before the data are analysed. 

AU Mic belongs to the most active stars, which are characterised by very frequent energetic flares and usually complex flares. All transit light curves were contaminated by more or less separated short flares, which could have been successfully removed before the fitting process with {\tt{Allesfitter}}, described in Sect. 3.1. However, several transit light curves of both AU~Mic~b and AU~Mic~c were compromised by long-lasting flare complexes that occurred very close to the transit or appeared during the transit. Their contamination deteriorated the transit shape so much they could not be reliably separated if the precision of the parameters from clearer transits was to be achieved. Therefore, we decided to remove these severely contaminated transits from the transit analysis. The CHEOPS light curves of the transits we selected for parameter fitting are shown in Fig.~\ref{figure:Cheops_raw_transits}, and the transit light curves that we omitted because of severe flare contamination are plotted in Fig.~\ref{fig:AU_Mic_planetary_transits_bad}.

\subsection{Pre-processing of the light curves}


The CHEOPS light curves of AU~Mic were corrected for roll-angle systematics using a parametric estimation derived from the data, with the photometric points phased by the roll angle. After we mased out the transits and the evident flares, the data were phased in the roll-angle domain. A further $2.5\sigma$ clipping was applied to exclude outliers such as potential flares and light-curve transients. The roll systematics model was a sixth-order Fourier polynomial, whose coefficients were determined by a least-squares fit to the phased out-of-transit points, and we then subtracted the derived parametric model from the entire dataset. The corrected data set was then used as input for the light-curve modeller.

This correction method was simpler than the non-parametric models we applied in our earlier work (e.g. \citealt{Szabó2021}) because the new-generation fitting algorithms can properly handle the flares and the high-order residuals of the rolling systematics that survived the subtraction of the Fourier polynomial. We achieved a more reliable estimate of the planet parameter uncertainties and a more robust handling of the data in general by letting the fitting procedure separate the various systematics in the light curve. 

The flares of AU Mic are known to appear frequently, and their amplitude can well exceed the transit depth of the planets. Quite often, several flares merge into a flare complex with a timescale commensurable to the duration of the transits \citep{Szabó2021,Szabó2022}. The flares can therefore introduce biases in the estimation of the baseline flux (which propagates to the transit parameters, most importantly, the transit depth), and if they appear during the transit, they can compress the transit depth and distort the shape of the light curve.

To filter out flares, manual removal was followed in our earlier publications \citep{Szabó2021,Szabó2022}. In this current work, we changed this to a semi-automated process. First, light-curve segments that certainly had significant flares were marked by an artificial-intelligence application, which is the same process as for collecting the database of AU Mic flares for a dedicated analysis (see Kriskovics et al., in prep. for more details). The marked segments of the light curve were then removed. In the next step, we repeated the manual flare-removal process that we followed in our earlier publications, and we removed the low-amplitude flares that were not identified by the AI algorithms, with the aim of filtering out all suspicious light curve parts and possible biases due to flares from the analysis. Figure \ref{figure:Cheops_raw_transits} shows the phased transit light curves after the complete flare-filtering process, and this data set was the input for the light-curve analysis.

\section{Data analysis}

\subsection{Refining the transit parameters}
\label{allesfit}

In order to derive orbital and planetary parameters of AU\,Mic\,b and c based on the CHEOPS photometry, we employed the {\tt{Allesfitter}}\footnote{See \url{https://www.allesfitter.com}} software package \citep{allesfitter-code, allesfitter-paper}. This is a public software for modelling photometric and radial-velocity (RV) data. It can accommodate multiple exoplanets, multi-star systems, star spots, stellar flares, transit-time variations, and various noise models. It automatically runs a nested-sampling or Markov chain Monte Carlo fit. We opted for the nested-sampling fit with default settings. Several fundamental parameters were optimised during the fitting procedure, including the reference mid-transit time $T_\mathrm{c}$, the orbital period $P_\mathrm{orb}$, the planet-to-star radius ratio $R_\mathrm{p}/R_\mathrm{s}$, the scaled sum of the fractional radii $(R_\mathrm{p} + R_\mathrm{s})/a$, and the cosine of the orbit inclination angle ($\cos i$). The quadratic limb-darkening (LD) law was applied during the fitting procedure. The $u_1$ and $u_2$ LD coefficients were first linearly interpolated based on the stellar parameters of $T_\mathrm{eff} = 3665 \pm 31\,\mathrm{K}$ and $\log\,g = 4.52 \pm 0.05\,\mathrm{[cgs]}$ found by \citet{Donati2023}. We used the tables of coefficients calculated for the CHEOPS passband using the {\tt{PHOENIX-COND}} models by \citet{Claret2}. We then converted these LD coefficients into $q_1$ and $q_2$ \citep{Kipping3}. The $q_1$ and $q_2$ parameters were fitted during the procedure. In order to model the flux baseline, we applied a Gaussian process (GP) regression method using the {\tt{SHOTerm}} (simple harmonic oscillator – SHO) plus {\tt{JitterTerm}} kernel, implemented in the {\tt{Celerite}}\footnote{See \url{https://celerite.readthedocs.io/en/stable}} \citep{Kallinger1, Foreman2, Barros1} package, with a fixed quality factor of $Q_0 = 1/\sqrt{2}$, as is common for quasi-periodic stellar variability. The regression was made by using $\log \omega_0$ and $\log S_0$ with bounds on the values of these parameters to be input by the user. The parameter $\log \omega_0$ reflects the frequency (period), and $\log S_0$ reflects the scaled power (amplitude) of the signal. The instrumental noise in the CHEOPS data was sampled using the $\log \sigma$ parameter. We assumed a circular orbit of AU\,Mic\,b and c.

\begin{table}[h!]
\centering
\caption{Overview of the \texttt{Allesfitter}-derived parameters of AU\,Mic\,b and c obtained based on the data from 2022 and 2023.}
\label{cheops-parameters-tab2}
\begin{tabular}{lll}
\hline
Parameter [unit]                                                & Value (2022)                                            & Value (2023)       \\         
\hline
\multicolumn{3}{c}{Orbital and planetary parameters: AU\,Mic\,b}\\
$R_\mathrm{s}/a$                                                & $0.0531_{-0.0013}^{+0.0016}$                    & $0.0550_{-0.0015}^{+0.0013}$         \\
$a/R_\mathrm{s}$                                                & $18.83_{-0.54}^{+0.47}$                             & $18.18_{-0.42}^{+0.51}$             \\ 
$R_\mathrm{p}/a$                                                & $0.002497_{-0.000083}^{+0.000094}$      & $0.002843_{-0.00011}^{+0.000098}$\\ 
$R_\mathrm{p}$ [$\mathrm{R_{\oplus}}$]  & $4.20\pm0.12$                                               & $4.62\pm0.15$\\ 
$R_\mathrm{p}$ [$\mathrm{R_{Jup}}$]     & $0.375\pm0.011$                                                 & $0.412\pm0.013$\\ 
$a$ [$\mathrm{R_{\odot}}$]                              & $15.42\pm0.56$                                              & $14.92_{-0.49}^{+0.55}$\\ 
$a$ [au]                                                                & $0.0717\pm0.0026$                                     & $0.0694_{-0.0023}^{+0.0025}$\\ 
$i$ [deg]                                                               & $89.12_{-0.27}^{+0.31}$                           & $88.78_{-0.20}^{+0.26}$\\ 
$b$                                                                     & $0.290_{-0.098}^{+0.077}$                                 & $0.386_{-0.074}^{+0.057}$\\ 
$T_\mathrm{14}^{\star}$ [h]                     & $3.457_{-0.019}^{+0.021}$                               & $3.480_{-0.062}^{+0.048}$\\ 
$T_\mathrm{23}^{\star\star}$ [h]                & $3.118_{-0.022}^{+0.021}$                       & $3.084_{-0.074}^{+0.056}$\\ 
$T_\mathrm{eq}^{\diamond}$ [K]                  & $546.4_{-8.1}^{+8.9}$                                   & $555.7_{-8.6}^{+7.9}$\\ 
$T_\mathrm{d}^\ddagger$ [relative flux] & $0.002633\pm0.000088$                       & $0.00330\pm0.00011$\\ 
\hline
\multicolumn{3}{c}{Orbital and planetary parameters: AU\,Mic\,c}\\
$R_\mathrm{s}/a$                                                & $0.0326_{-0.0010}^{+0.0011}$            & $0.03260_{-0.00098}^{+0.0010}$\\
$a/R_\mathrm{s}$                                                & $30.7\pm1.0$                                                & $30.68\pm0.96$\\ 
$R_\mathrm{p}/a$                                                & $0.001155\pm0.000069$                   & $0.00100_{-0.00011}^{+0.00010}$\\ 
$R_\mathrm{p}$ [$\mathrm{R_{\oplus}}$]  & $3.17\pm0.16$                                               & $2.76_{-0.29}^{+0.26}$\\ 
$R_\mathrm{p}$ [$\mathrm{R_{Jup}}$]     & $0.283\pm0.014$                                                 & $0.246_{-0.026}^{+0.023}$\\ 
$a$ [$\mathrm{R_{\odot}}$]                              & $25.14\pm1.0$                                           & $25.14\pm1.0$\\ 
$a$ [au]                                                                & $0.1169\pm0.0047$                                     & $0.1169\pm0.0047$\\ 
$i$ [deg]                                                               & $88.85_{-0.16}^{+0.21}$                           & $88.78_{-0.19}^{+0.43}$\\ 
$b$                                                                     & $0.615_{-0.11}^{+0.074}$                                  & $0.658_{-0.23}^{+0.093}$\\ 
$T_\mathrm{14}^{\star}$ [h]                     & $3.90_{-0.25}^{+0.31}$                      & $3.73_{-0.40}^{+0.64}$\\ 
$T_\mathrm{23}^{\star\star}$ [h]                & $3.48_{-0.30}^{+0.36}$                                  & $3.34_{-0.48}^{+0.74}$\\ 
$T_\mathrm{eq}^{\diamond}$ [K]                  & $428.0\pm7.8$                                                   & $428.0\pm7.8$\\ 
$T_\mathrm{d}^\ddagger$ [relative flux] & $0.001376\pm0.000087$                   & $0.00104_{-0.00014}^{+0.00012}$\\ 
\hline
\multicolumn{3}{c}{Stellar parameters}\\
$u_\mathrm{1}$                                              & $0.39\pm0.13$                                           & $0.60\pm0.15$\\ 
$u_\mathrm{2}$                                              & $0.30_{-0.13}^{+0.14}$                          & $0.19\pm0.14$\\ 
$\rho_\mathrm{s}$ [g~cm$^{-3}$]                 & $1.65\pm0.19$                                       & $1.56_{-0.12}^{+0.14}$\\ 
\hline
\end{tabular}
\tablefoot{$^{\star}$Total transit duration between first and fourth contact. $^{\star\star}$Full transit duration between second and third contact. $^{\diamond}$Assuming an albedo of 0.3 and uniform heat redistribution. $^{\ddagger}$Transit depth.}
\end{table}

\begin{figure*}
\centering
\centerline{
\includegraphics[width=80mm]{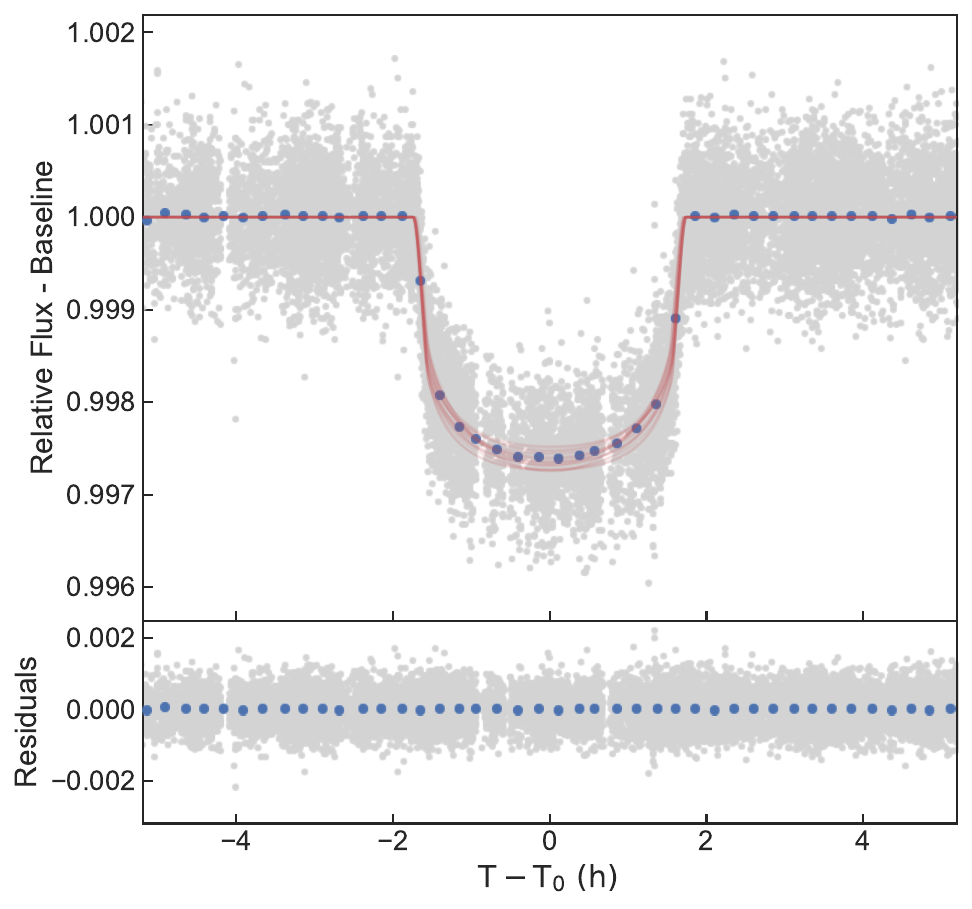}
\includegraphics[width=80mm]{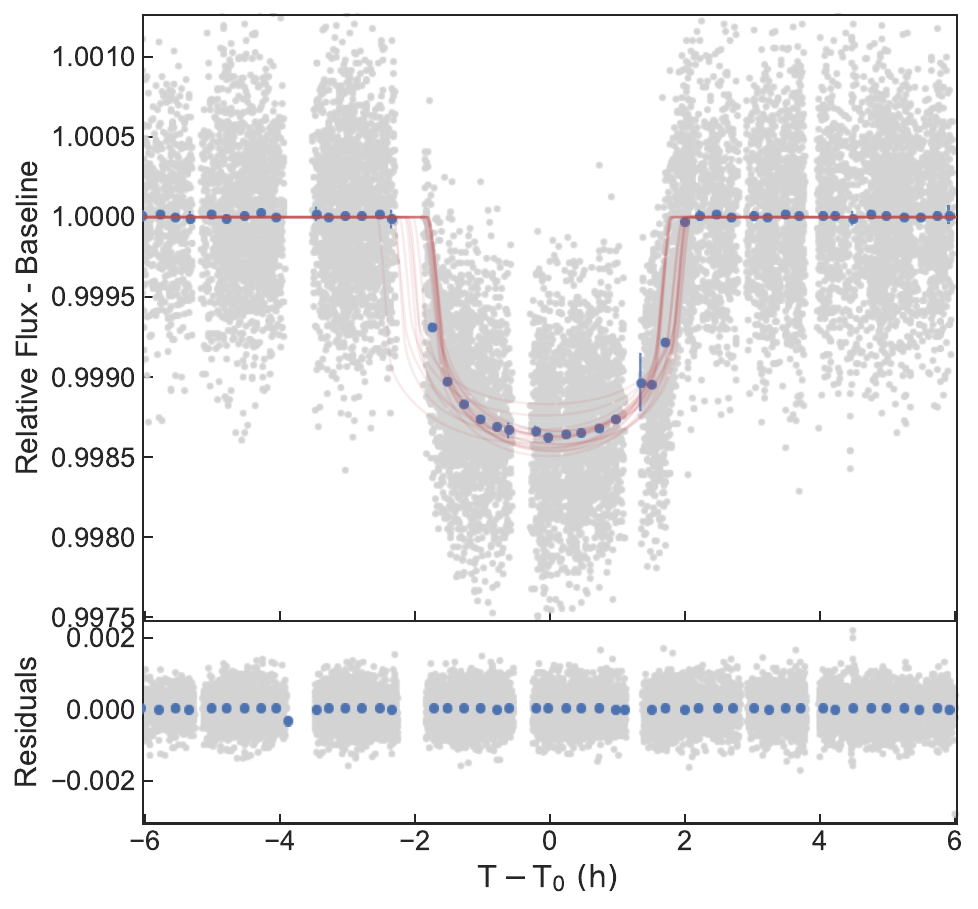}}
\centerline{
\includegraphics[width=80mm]{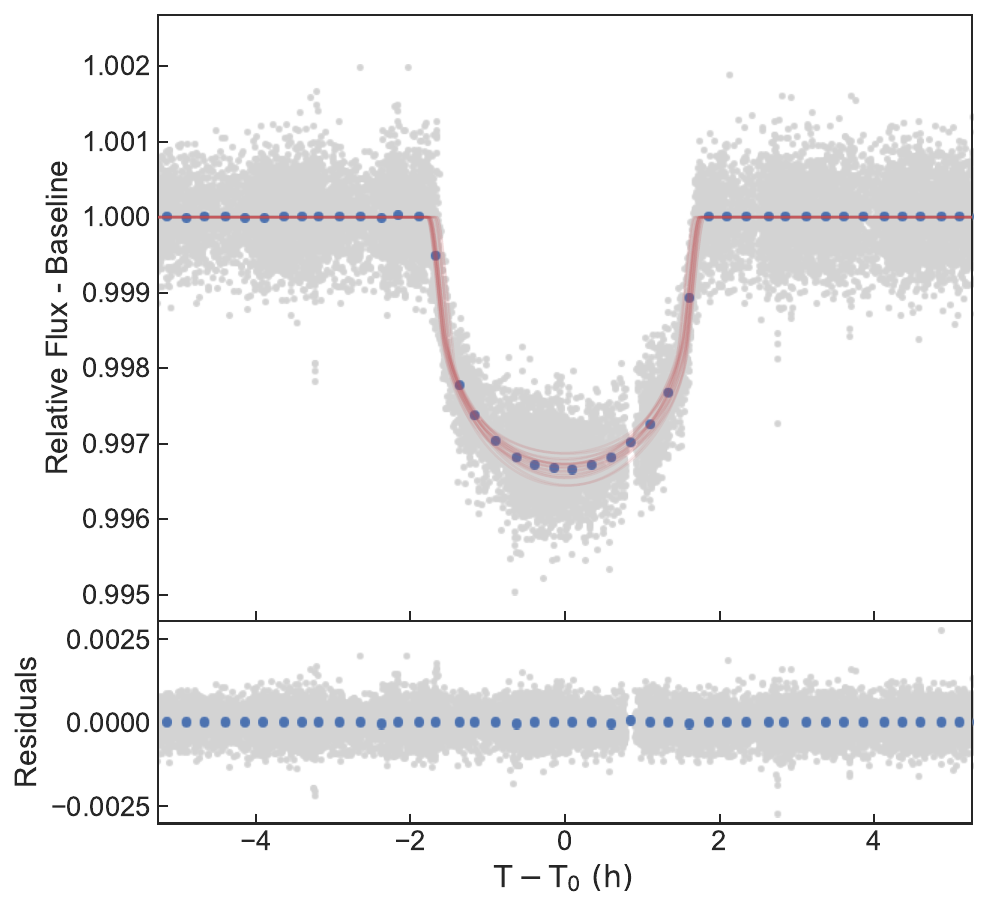}
\includegraphics[width=80mm]{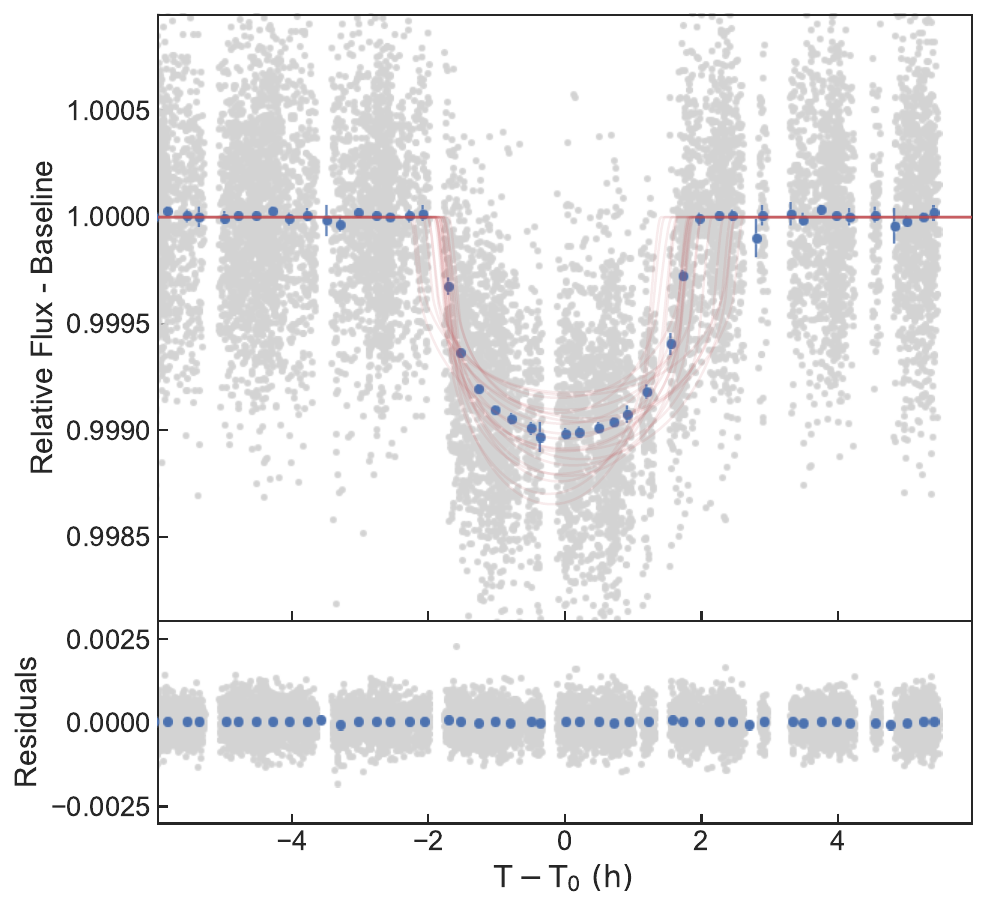}}

\caption{Stacked and binned CHEOPS transit light curves of AU\,Mic\,b (left panels) and AU\,Mic\,c (right panels) from 2022 (top panels) and from 2023 (bottom panels), overplotted with the best-fitting {\tt{Allesfitter}} models (20 curves from random posterior samples). The residuals are also shown.}
\label{fig:AU_Mic_planetary_transits_joint}
\end{figure*}

During the analysis with {\tt{Allesfitter}}, we followed the strategy described below. We simultaneously fitted all AU\,Mic\,b and c transits per observing year, that is, the transits from 2022 and 2023 were fitted separately. The main reason for splitting the data into two groups was the possibility of comparing the evolution of the transit parameters of the planets. In most cases, we applied broad uniform priors on the parameters. The applied priors and the fitted parameters are listed in Table \ref{cheops-parameters-tab} and the derived parameters are listed in Table \ref{cheops-parameters-tab2}. The corresponding transit light curves, overplotted with the best-fitting {\tt{Allesfitter}} models, are shown in Fig. \ref{fig:AU_Mic_planetary_transits_joint}.     

\subsection{TTV analysis}

\begin{figure}
\includegraphics[viewport=25 73 440 390, width=\columnwidth]{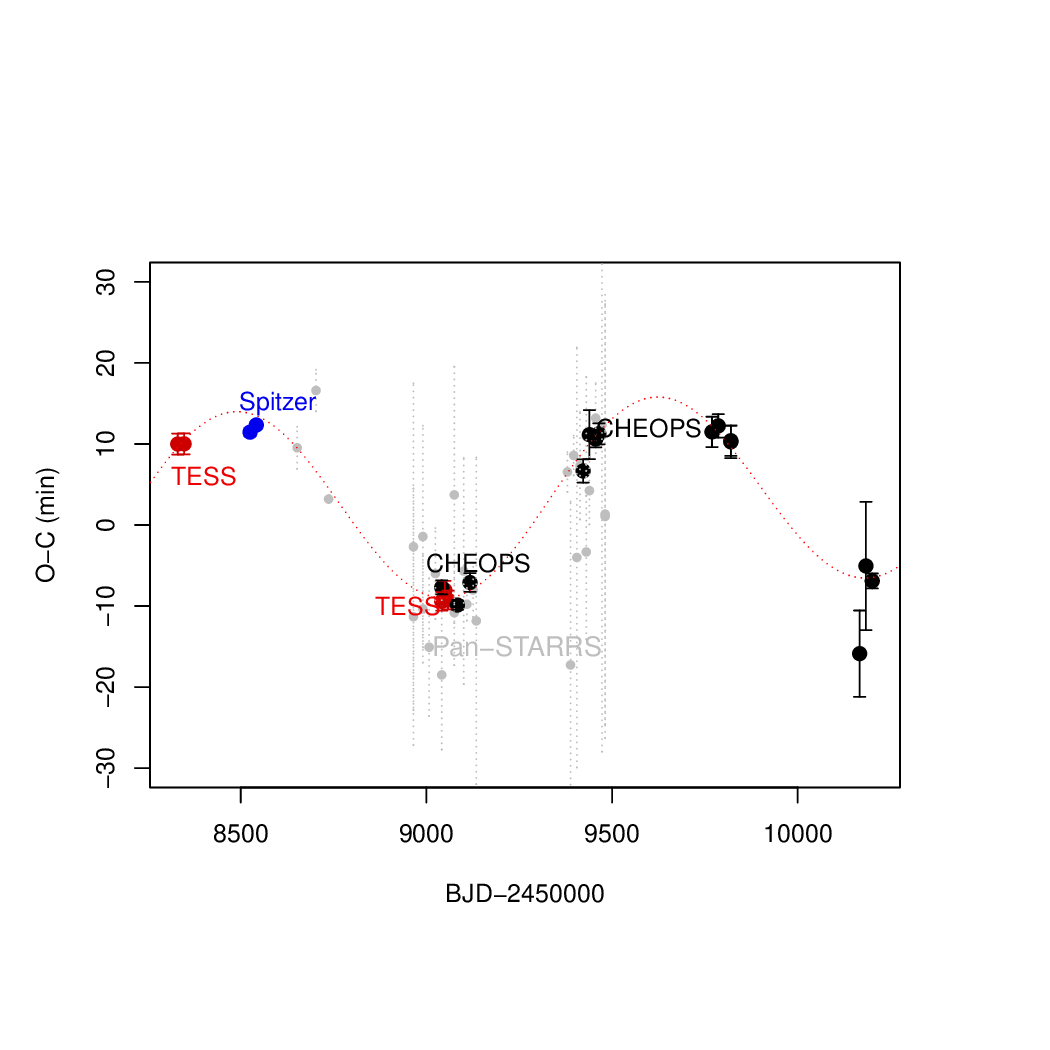}
\includegraphics[viewport=25 73 440 390, width=\columnwidth]{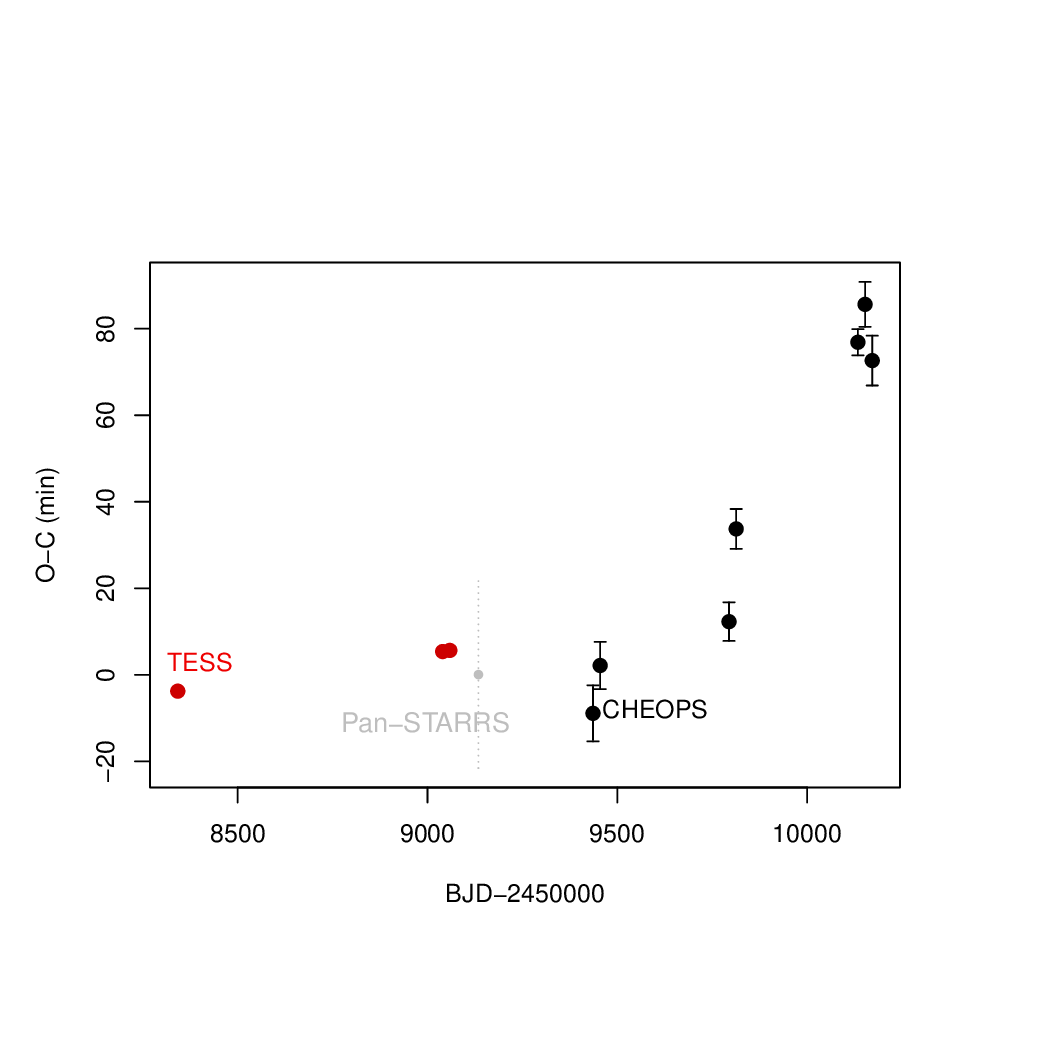}
\caption{TTV diagrams of AU\,Mic b (upper panel) calculated with $T_c=2\,458\,330.38416$ and $P_{\rm mean}=8.4631427$~d, and AU\,Mic c (lower panel) calculated with $T_c=2\,459\,454.8973$ and $P_{\rm mean}=18.85882$~d. We included Transiting Exoplanet Survey Satellite (TESS, red symbols), Spitzer (blue symbols), Earth-based Pan-STARRS (light grey symbols) and CHEOPS (black symbols) measurements. The harmonic fit to AU\,Mic\,b data (dotted red line) illustrates the most probable shape of a periodic TTV fitted to the data. This is shown to illustrate the trend of the distribution without any dynamical interpretation.}
\label{fig:ttvs}
\end{figure}

The $O-C$ diagrams of the transit mid-time for AU Mic b and c are shown in Fig.~\ref{fig:ttvs}. Observations from previous years are complemented with CHEOPS data from the 2022 and 2023 observation periods. Both planets showed large-amplitude TTVs, and planet c deviated significantly from those of previous years. The difference between the mid-transit times reached $\sim$90~minutes. Tables~\ref{table:ttvsb} and \ref{table:ttvsc} show the exact mid-transit times and $O-C$ values for planets b and c, respectively. Based on $T_c = 2\,458\,330.38416$~d and $P_{\rm mean} = 8.4631427$~d for planet b and $T_c=2\,459\,454.8973$ and $P_{\rm mean}=18.85882$~d for planet~c. The superperiod of the TTVs exhibited by planet b (according to the fit in Fig. \ref{fig:ttvs}) is 1150 days, which is slightly shorter than reported in \citet{Szabó2022}. The reason is the longer dataset that is now available (almost 1.5 superperiods) than was the case of the analysis in \citet{Szabó2022} (less than one superperiod), and the pattern is better defined. The peak-to-peak TTV amplitude of the best-fit periodic TTV model to AU~Mic~b is 24.5 minutes, which agrees well with the predictions in \citet{Szabó2022}. The presence of a third planet, AU Mic d, has been suggested by \citet{Wittrock2023} based on TTV signals. In order to test this possibility, we conducted dynamical simulations assuming a third planet in the system.
\\

We used \trades{}\footnote{\url{https://github.com/lucaborsato/trades}.}
\citep{Borsato2014A&A...571A..38B,Borsato2019MNRAS.484.3233B,Borsato2021MNRAS.506.3810B},
a \texttt{Fortran90-Python} software, to model four configurations of the AU\,Mic system.
In the initial configuration (1), we focused exclusively on the system comprising the two transiting planets,
b and c, and only modelled the transit times during the orbital integration.
In the subsequent two configurations (2) and (3), 
we additionally incorporated planet d \citep{Wittrock2023},
and as for (1), we only modelled the transit times of b and c.
Only in the fourth configuration (4)
did we also include the SERVAL RVs from \citet{Zicher2022}.
Another difference between the four configurations is the parameters that were fitted in the analysis.
Generally, we set the stellar mass from \citet{Plavchan2020}, 
and fitted \citep[as in ][]{Nascimbeni2023A&A...673A..42N} for 
the planet-to-star mass ratios ($M_\mathrm{p}/M_\star$),
the periods ($P$),
the eccentricities ($e$) and argument of pericenters ($\omega$) in the form $\sqrt{e}\cos\omega$, $\sqrt{e}\sin\omega$,
and the mean longitude ($\lambda$\footnote{The mean longitude is defined as $\lambda+\mathcal{M}+\omega+\Omega$,
where $\mathcal{M}$ is the mean anomaly and $\Omega$ is the longitude of the ascending node.})
for all the planets.
We fitted the longitude of the ascending node ($\Omega$) of planet c in all four configurations,
and of d in configurations (2), (3), and (4).
The inclination ($i$) of planet c in configurations (3), (4) and of d in (2), (3), and (4) was fitted as well.
In configuration (4) we also fitted for an RV offset ($\gamma$) and a jitter term (in $\log_2$).
The RV data are characterised by strong activity; but by default, \trades{} does not implement an activity model. Because the problem is so complex,
we decided not to test it.
The total numbers of fitted parameters were 11, 18, 19, and 21
for configurations (1), (2), (3), and (4), respectively.
All parameters were defined at the reference time $\mathrm{BJD_{TDB}}-2450000 = 8329$.
All priors on the parameters are listed in Table~\ref{tab:trades_parameters}
The parameters of b and c were set to the values in Table~\ref{cheops-parameters-tab2}, 
and $\Omega_\mathrm{b}$ was set to $180^{\circ}$ 
(assuming the same reference system as in \citealt{Winn2010} and adopted in \citealt{Borsato2014A&A...571A..38B}).\par

For each configuration,
we first performed a run of \trades{} with \pyde{} \citep{Parviainen2016}
allowing for 200\,000 steps
and used the best-fit configuration set as a starting point for \emcee{} \citep{Foreman2013}.
Then we ran \emcee{} for 
1\,000\,000 steps
(with a conservative thinning factor of 100).
For each configuration, we used as walkers for \emcee{}
the same number of parameter sets as used in \pyde{},
100 for configurations (1), (2), and (3),
and 200\footnote{We doubled the number of parameter sets/walkers due to the increased number of
fitting parameters and the availability of a much more performing server.} for (4).
Within \emcee{} we used a $80\% \div 20\%$ combination of
the differential evolution \citep{DEMOVE2014ApJS..210...11N}
and snooker \citep{terBraak2008} proposal samplers. 
After we confirmed that the chains converged through
visual inspection, 
auto-correlation function (ACF), 
Gelman-Rubin statistics \citep{GelmanRubin1992},
and the Geweke criterion \citep{geweke1991},
we discarded the first 
350\,000, 500\,000, 600\,000, and 800\,000 steps
as burn-in for configurations (1), (2), (3), and (4), respectively.\par
We computed the high-density interval (HDI) at 
68.27\% 
as the $1\sigma$
equivalent for each fitted and physical parameter of the posterior distribution.
We defined the maximum-a-posteriori (MAP) configuration set as the best-fit parameter,
that is, the set of parameters that maximised the log-probability\footnote{In case of 
all uniform-uninformative priors, the log-probability $\ln{\mathcal{P}}$ is equal
to the log-likelihood $\ln{\mathcal{L}}$.} $\ln{\mathcal{P}}$,
whose parameters were completely within all fitted $1\sigma$ HDI.
The best-fit solution with 100 random samples drawn from the $1\sigma$ posterior distribution is shown in 
Fig.~\ref{fig:trades_oc_1}, \ref{fig:trades_oc_2},
\ref{fig:trades_oc_3}, and \ref{fig:trades_oc_4} with the $O-C$ plots for planets b and c
for all four configurations, 
and the RV model in Fig.~\ref{fig:trades_rv_4} for configuration (4).

\begin{table}[h]
    \centering
\caption{Observed mid-transit times and $O-C$ values of AU\,Mic\,b based on TESS, Spitzer, and
CHEOPS observations. }
    \begin{tabular}{llrr}
\hline
Designation & Transit Time & $O-C$ & Err \\
            &  [BJD$-$2\,450\,000]  & [min] & [min] \\
\hline
TESS S1\#{}1$^a$& 8330.3911$\pm$0.0009 & 10.00& 1.33 \\
TESS S1\#{}2$^a$& 8347.3174$\pm$0.0009 & 10.02& 1.33\\
Spitzer\#{}1$^b$ & 8525.04509$\pm$0.0010 &12.45 & 1.43\\
TESS S27\#{}1$^a$& 9041.2816$\pm$0.0008 &-9.42&1.17\\
TESS S27\#{}2$^a$& 9049.7457$\pm$0.0008 &-8.05&1.17\\
TESS S27\#{}3$^a$& 9058.2080$\pm$0.0008 &-9.40&1.17 \\
CHEOPS 20-07-10$^a$& 9041.2828$\pm$0.0006 &-7.70 & 0.87 \\
CHEOPS 20-08-21$^a$& 9083.5970$\pm$0.0004 &-9.88 & 0.58\\
CHEOPS 20-09-24$^a$& 9117.4515$\pm$0.0008 &-7.08 & 1.17\\
CHEOPS 21-07-26$^c$& 9422.1342$\pm$0.0010 & 8.40&1.43\\
CHEOPS 21-08-12$^c$& 9439.0636$\pm$0.0021 & 9.55&3.15\\
CHEOPS 21-08-29$^c$& 9455.9895$\pm$0.0007 & 10.65&1.0\\
CHEOPS 21-09-06$^c$& 9464.4531$\pm$0.0009 & 11.25&1.27\\
CHEOPS 22-08-08$^d$& 9769.1264$\pm$0.0013 &  11.49 & 1.9\\
CHEOPS 22-08-25$^d$& 9786.0532$\pm$0.0013 &  12.23 & 1.9\\
CHEOPS 22-08-28$^d$& 9819.9045$\pm$0.0013 &  10.26 & 1.9\\
CHEOPS 23-08-10$^d$& 10166.8751$\pm$0.0037 & -15.88 & 5.2\\
CHEOPS 23-08-27$^d$& 10183.8089$\pm$0.0055 &  -5.06 &7.9\\
CHEOPS 23-09-13$^d$& 10200.7339$\pm$0.0064 &  -6.91 &9.2\\
\hline
    \end{tabular}
        \tablefoot{$T_c = 2\,458\,330.38416$\,d and $P_{\rm mean} = 8.4631427$\,d were used to determine $O-C$ values.  The references for the transit times are $a$: \cite{Szabó2021}, 
$b$: \cite{Plavchan2020}, $c$: \cite{Szabó2022}
$d$: This research. }
    
    \label{table:ttvsb}
\end{table}

\begin{table}[h]
    \centering
\caption{Observed mid-transit times and $O-C$ values of AU\,Mic\,c based on TESS and CHEOPS observations.}
    \begin{tabular}{llrr}
\hline
Designation & Transit Time & O-C & Err \\
            &  [BJD$-$2\,450\,000]  & [min] & [min] \\
\hline
TESS S1\#{}1$^a$& 8342.22432$\pm$0.0004 & -3.75 & 0.71 \\
TESS S27\#{}1$^a$& 9040.00697$\pm$0.0005 & 5.34&0.88\\
TESS S27\#{}2$^a$& 9058.86596$\pm$0.0006 & 5.58&0.98\\
CHEOPS 21-08-09$^b$& 9436.0323$\pm$0.0045 & -8.90&6.48\\
CHEOPS 21-08-28$^b$& 9454.8988$\pm$0.0050 & 2.16&5.47\\
CHEOPS 22-08-02$^c$& 9794.3646$\pm$0.0032 & 12.30 &4.7 \\
CHEOPS 22-08-21$^c$& 9813.2383$\pm$0.0031 &  33.72 &4.6 \\
CHEOPS 23-07-08$^c$& 10133.8682$\pm$0.0028 &  76.87 & 4.0\\
CHEOPS 23-07-27$^c$& 10152.7331$\pm$0.0036 & 85.62 & 5.2\\
\hline
    \end{tabular}
    \tablefoot{$T_c = 2\,459\,454.8973$\,d and $P_{\rm mean} = 18.85882$\,d were used to determine $O-C$ values.  The references for the transit times are $a$: \cite{Gilbert2022}, $b$: \cite{Szabó2022}, $c$: This research. }
    \label{table:ttvsc}
\end{table}

\section{Results and discussion}

\begin{table*}
    \centering
    \caption{Best-fitting parameters of AU\,Mic\,b compared to results from previous works.}
    \label{tab:solutionsb}
    \resizebox{\textwidth}{!}{
    \begin{tabular}{llllllll}
    \hline
    \noalign{\smallskip}
      & 
      2023$^{\star}$ &
      2022$^{\star}$ &
  2021$^{\star\star}$ &
  2020$^{\diamond}$  &   
  2018-2020$^{\ddagger}$ &
  2018-2020$^{\ast}$  &  
  2018$^{\bullet}$  \\
      \noalign{\smallskip}
\hline
      \noalign{\smallskip}
      \hline
      \noalign{\smallskip}
$R_p/R_\star$  & 
  0.0517$\pm$ 0.00011&
  0.04700$^{+0.00077}_{-0.00073}$&
  0.0433$\pm$ 0.0017  &
  0.0531 $\pm$ 0.0023  &     
  0.0512 $\pm$  0.0020 & 
  0.0526$^{+0.0003}_{-0.0002}$ &
  0.0514$\pm$0.0013 
  \\
$a/R_\star$ &  
  18.18$^{+0.51}_{-0.42}$ &
  18.83$^{+0.47}_{-0.54}$ &
  18.95 $\pm$ 0.35  &
  19.24 $\pm $   0.37  & 
  19.07 &
  19.1$^{+0.2}_{-0.4}$ &
  19.1$_{-1.6}^{+1.8}$ 
 \\
$R_p$ [R$_{\oplus}$] &
  4.62 $\pm$0.15 &
  4.20 $\pm$0.12 &
  3.55 $\pm$0.13 &
  4.36 $\pm $ 0.18 & 
  4.27 $\pm $ 0.17 &
  4.07 $\pm$ 0.17 &
4.29$\pm $0.20
  \\
$a$ [AU] &
  0.0694$^{+0.0025}_{-0.0023}$ &
  0.0717 $\pm$0.0026 &
  0.0654 $\pm$0.0012 &
  0.0678 $\pm $ 0.0013 & 
  0.0644 $_{-0.0054}^{+0.0056}$ &
  0.0645 $\pm$ 0.0013 &
  0.066$_{-0.006}^{+0.007}$  \\
$b$  &  
  $0.386_{-0.074}^{+0.057}$ &
  $0.290_{-0.098}^{+0.077}$  &
  0.17 $\pm$ 0.11 &
  0.09 $\pm $  0.05 & 
  0.26 $_{-0.17}^{+0.13}$ &
  0.18 $\pm$ 0.11&
  0.16$_{-0.11}^{+0.14}$  \\ %
\noalign{\smallskip}
\hline
    \end{tabular}
    }
    \tablefoot{{The parameters are compared to the results of \citet[P2020]{Plavchan2020}, \citet[M2021]{Martioli2021}, \citet[Sz2021]{Szabó2022}, \citet[G2022]{Gilbert2022}, and \citet[Sz2022]{Szabó2022}.} $^{\star}$This work, $^{\star\star}$Sz2022, $^{\diamond}$Sz2021, $^{\ddagger}$G2022, $^{\ast}$M2021, $^{\bullet}$P2020}
\end{table*}

\begin{table*}
    \centering
    \caption{Best-fitting parameters of AU\,Mic\,c compared to the results of M2021, G2022, and Sz2022.}
    \label{tab:solutionsc}
    \begin{tabular}{llllll}
    \hline
    \noalign{\smallskip}
      & 
      2023$^{\star}$ &
      2022$^{\star}$ &
  2021$^{\star\star}$ &  
  2018--2020$^{\ddagger}$ &
  2018--2020$^{\ast}$
  \\
      \noalign{\smallskip}
\hline
      \noalign{\smallskip}
      \hline
      \noalign{\smallskip}
$R_{\rm p}/R_\star$  &
  0.0309$_{-0.0033}^{+0.0028}$&
  0.0354$\pm$0.0016&
  0.0313$\pm$0.0016&  %
  0.0340$^{+0.0033}_{-0.0034}$ &
  0.0395$\pm$0.0011 \\
$a/R_\star$ &  
  30.68 $\pm$ 0.96 &
  30.7 $\pm$ 1.0 &
  28.8 $\pm$ 2.4 & %
  31.7 $^{+2.7}_{2.6}$ &
  29$\pm$ 3.0
 \\
$R_{\rm p}$ [R$_{\oplus}$] &
  2.76 $^{+0.26}_{-0.29}$ &
  3.17 $\pm$ 0.16 &
  2.56 $\pm$ 0.12 & %
  2.79 $_{-0.30}^{+0.31}$ &
  3.24 $\pm$ 0.16
  \\
$a$ [AU] &
  0.1169 $\pm$0.0047 &
  0.1169 $\pm$0.0047 &
  0.0993 $\pm$0.0085 & %
  0.110 $_{-0.010}^{+0.010}$ &
  0.1101 $\pm$ 0.0022 \\
$b$  & 
  $0.658_{-0.23}^{+0.093}$ &
  $0.615_{-0.11}^{+0.074}$ &
  0.58 $\pm$ 0.13 &
  0.30 $_{-0.20}^{+0.21}$ &
  0.51 $\pm$ 0.21 \\ %
\noalign{\smallskip}
\hline
    \end{tabular}
        \tablefoot{$^{\star}$This work, $^{\star\star}$Sz2022, $^{\ddagger}$G2022, $^{\ast}$M2021}
\end{table*}

\subsection{Fundamental planetary parameters}

In Tables \ref{tab:solutionsb} and \ref{tab:solutionsc}, we compare the fundamental transit parameters observed between 2018 and 2023, taken from this paper and from five earlier analyses in the literature. Here, \cite{Plavchan2020} contains the analysis of 2018 TESS data, \cite{Gilbert2022} and \cite{Martioli2021} are based on the combined TESS 2018--2020 data, and \cite{Szabó2021} and \cite{Szabó2022} are based on the CHEOPS observations from 2020 and 2021, respectively. We sorted the table columns according to the time of observation; and therefore, moving from left to right, the columns show the measurements in a decreasing order of the time of the observations.

Most parameters are consistent within the uncertainties, with the important exceptions of $R_p/R_s$ and $b$ for planet b. In the case of AU Mic~b, the $R_p/R_s$ parameter was found to be smaller in 2022 than in 2023 (0.0470 versus 0.0517). The transit depth in 2022 was slightly larger than what we observed in 2021 \citep[0.0433$\pm$0.001,][]{Szabó2022}, while it was determined to be in the interval of 0.0512--0.0531 in 2018 and 2020 \citep[][]{Szabó2022}, which is compatible to the transit depth parameter that we observed in 2023. The transit depth parameters of AU Mic~c are comparable to each other, with a slight decrease in the 2023 value.

The impact parameter $b$ has an average value of $0.22$ in the case of AU Mic b, but since the 2023 CHEOPS data and 2020 CHEOPS data are two outliers (with values of 0.386 and 0.09, respectively), the dataset is barely consistent. The values for the impact parameter of AU Mic c are consistent within the uncertainties, but appear to be slightly higher in the case of the more recent CHEOPS data.  

We followed the working hypothesis that the main reason for the varying transit parameters of the two planets can be the actual realisation of the spot distribution on the surface and their biases on the transit shape \citep{Szabó2021}. This is also corroborated by the observation that CHEOPS measurements seem to be more strongly biased (more year-to-year variations) than the TESS measurements, and CHEOPS is bluer and more sensitive to inhomogeneities in the surface temperature. Therefore, assuming a common reason behind the biases and expecting similarities (correlation, anti-correlation, at least in part) in the actual values of the parameters, we tested a possible correlation between the transit depth and the impact parameters of AU Mic b and c, and also possible correlations between the same parameter ($R_\mathrm{p}/R_\mathrm{s}$ and impact parameter of the two planets in comparison). We calculated Pearson's correlation coefficient (testing linear dependence) and Spearman's rank correlation coefficient (testing any monotonic dependence). We found no significant correlation (a coefficient value of zero was always within the error interval, and we also observed $p>0.1$ in all cases). However, the lack of a significant correlations does not rule out activity as a reason for the observed biases of the parameters, but is instead a result of the low number statistics, which is still not enough for a conclusive result. This will be better determined in the future when AU Mic has been observed for more years.

The observed instability of the transit parameters on a timescale of years confirms our observation that the transit parameter depends significantly on the actual spot pattern on the stellar disk, which is reproducible (we see similar biases for transits that repeat at similar stellar rotational longitudes; \cite{Szabó2021}). This proves that in case of AU Mic, the transit depth parameter is not exclusively characteristic of the planet itself, but the actual spot distribution causes bias the observed value. (Hence, calling it the ``transit depth parameter'' is more accurate because it is an $R_p/R_s$ value, but biased by the actual spot distribution).


\subsection{Possible solutions for AU Mic d and concluding remarks}

The parameters of the planets in the AU Mic system based on the dynamical analysis are shown in Table~\ref{dynal_results}. Configuration (1), in which only two planets, b and c, were assumed, is unable to explain the observed TTVs, as shown in Fig.~\ref{fig:trades_oc_1} and in the first column of Table~\ref{dynal_results}. We found that in accordance with \citet{Wittrock2023}, the TTVs exhibited by planets b and c can be explained by the presence of an additional third planetary body in the system on an orbit that falls between the orbits of the two transiting planets. It is worth noting, however, that the shapes and amplitudes of the fitted TTV signals of planets b and c differed significantly in the remaining three configurations, where a third perturbing body was always assumed. This resulted in differences between the derived parameters from configurations (2), (3), and (4). Since the BIC value was lowest for configuration (2), it was treated as the main solution. Therefore, our discussion continues with the parameters of configuration (2), which are shown in the second column of Table~\ref{dynal_results}. We found the orbital period of planet d to be around 12.6~days, which is compatible with the 12.6--12.7~-day AU Mic d solutions of \cite{Wittrock2023}, and most importantly, narrows the possible AU Mic d solutions to the class of middle AU Mic d models, that is, planet d is between the orbits of planets b and c. The mass of this third perturbing body is expected to be far lower than that of planets c or b, around $0.1 M_\oplus$, which suggests that it falls into the category of rocky exoplanets. The sub-Earth mass agrees with the result of \cite{Wittrock2023} (with values of about 0.5 and 1.0 $M_\oplus$ for the 12.6~d and 12.7~d solutions, respectively), but our mass parameter is still below these determinations and is about two Mars masses. 

The derived planetary masses for planets b and c in the different configurations differed significantly. While the mass of planet b in configuration (2) is higher than other published values, the mass of planet c is significantly lower than the masses provided in previous works. Although the latter is compatible with \citet{Martioli2021}, the estimated 4.4~M$_{\odot}$ for AU Mic c disagrees with the values provided by \citet{Cale2021, Zicher2022}, or \citet{Donati2023}. On the other hand, configuration (3), whose BIC value was only slightly lower, resulted in completely different planetary masses, to the point that in this case, AU Mic c had the higher mass. This configuration agrees with \citet{Zicher2022, Donati2023}, and most recently, \citet{Mallorquin2024}. This type of discrepancy between the masses of b and c is not new; the derived planetary masses in previous studies also contradicted each other (e.g. \citealt{Cale2021} and \citealt{Zicher2022}). However, an activity model was not incorporated in our dynamical analysis, which may explain the difference from the published values. Moreover, the orbital configurations may further contribute to the discrepancies between the estimated values of the planetary masses.

In order to gain further insight into the geometry of the system, 
we calculated the angle between the rotational axis of the star and the normal of the transit chord of the planet, 
$\lambda_{\mathrm{RM}}$, such that $\lambda_{\mathrm{RM}}=0^\circ$ if the orbital motion is prograde and 
the orbit of the planet is aligned with the stellar equator, 
and $\lambda_{\mathrm{RM}}=90^\circ$ and 270$^\circ$ for a prograde and retrograde polar orbit, respectively. 
The $\lambda_{\mathrm{RM}}$ for each transit was calculated based on the parameters from the dynamical analysis, 
such that a planet with $\Omega=180^\circ$ and $i=90^\circ$ corresponded to an aligned orbit ($\lambda_{\mathrm{RM}}=0^\circ$).
The evolution of the $\lambda_{\mathrm{RM}}$ in time due to the interactions between the planets is shown in Fig.~\ref{fig:lambdaRM}.
The resulting orbit for planet b shows an aligned system with a mean value of $\lambda_{\mathrm{RM}}=0.02^\circ$. 
Planet c, on the other hand, exhibited a 
misaligned orbital geometry with a mean value of $\lambda_{\mathrm{RM}}=171.91^\circ$, suggesting that planet c is significantly misaligned and presumably has a retrograde orbit. 
In agreement with our research, recent results based on Rossiter-McLaughlin measurements also suggest a misaligned orbit of AU Mic c 
\citep{Yu2025}.

The value for $\Omega$ of AU Mic c and planet d differs by 225$^\circ$. This suggests that the orbital geometry of AU Mic d does not allow observable transits. Moreover, the inferred low mass (in \citealt{Wittrock2023} and this work) suggests a small radius and that the individual AU Mic d transits would not be detectable by the currently working exoplanet telescopes (TESS and CHEOPS) because the transit depth is shallow.

The above dynamical analysis is based on a limited number of epochs (AU Mic years), distributed to 5--5 self-consistent groups of transits, and the fit has 12 derived parameters, which exceeds the number of transits currently available for AU Mic c. The current best-fit results mostly show the preferred parameter space of possible AU Mic d solutions rather than their exact quantitative values. We conclude that more data are needed to constrain the planetary parameters of planet d better. Several future AU Mic visits are expected by CHEOPS (at least until 2026) and TESS (at least one sector in August 2025), with additional targeted JWST and Earth-based observations. The orbit solutions will be upgraded in the close future, and we expect to see a convergence for the possible AU Mic d solution in a few years. 

\section*{Data availability}
This research made use of observations of AU Mic carried out with the CHEOPS telescope in 2022 and 2023. The full list of CHEOPS observations from the 2022-2023 campaign can be reached \href{https://doi.org/10.5281/zenodo.14637451}{here} along with the start and end date of each observation.
   
\begin{acknowledgements}
CHEOPS is an ESA mission in partnership with Switzerland with important contributions to the payload and the ground segment from Austria, Belgium, France, Germany, Hungary, Italy, Portugal, Spain, Sweden, and the United Kingdom. The CHEOPS Consortium would like to gratefully acknowledge the support received by all the agencies, offices, universities, and industries involved. Their flexibility and willingness to explore new approaches were essential to the success of this mission. CHEOPS data analysed in this article will be made available in the CHEOPS mission archive (\url{https://cheops.unige.ch/archive_browser/}). 
GyMSz acknowledges the support of the Hungarian National Research, Development and Innovation Office (NKFIH) grant K-125015, a a PRODEX Experiment Agreement No. 4000137122, the Lendület LP2018-7/2021 grant of the Hungarian Academy of Science and the support of the city of Szombathely. 
LBo, GBr, VNa, IPa, GPi, RRa, GSc, VSi, and TZi acknowledge support from CHEOPS ASI-INAF agreement n. 2019-29-HH.0. 
DG gratefully acknowledges financial support from the CRT foundation under Grant No. 2018.2323 “Gaseousor rocky? Unveiling the nature of small worlds”. 
ML acknowledges support of the Swiss National Science Foundation under grant number PCEFP2\_194576. 
MNG is the ESA CHEOPS Project Scientist and Mission Representative, and as such also responsible for the Guest Observers (GO) Programme. MNG does not relay proprietary information between the GO and Guaranteed Time Observation (GTO) Programmes, and does not decide on the definition and target selection of the GTO Programme. 
TWi acknowledges support from the UKSA and the University of Warwick. 
ABr was supported by the SNSA. 
ZG was supported by the VEGA grant of the Slovak Academy of Sciences No. 2/0031/22 and by the Slovak Research and Development Agency - the contract No. APVV-20-0148. 
YAl acknowledges support from the Swiss National Science Foundation (SNSF) under grant 200020\_192038. 
We acknowledge financial support from the Agencia Estatal de Investigación of the Ministerio de Ciencia e Innovación MCIN/AEI/10.13039/501100011033 and the ERDF “A way of making Europe” through projects PID2019-107061GB-C61, PID2019-107061GB-C66, PID2021-125627OB-C31, and PID2021-125627OB-C32, from the Centre of Excellence “Severo Ochoa” award to the Instituto de Astrofísica de Canarias (CEX2019-000920-S), from the Centre of Excellence “María de Maeztu” award to the Institut de Ciències de l’Espai (CEX2020-001058-M), and from the Generalitat de Catalunya/CERCA programme. 
We acknowledge financial support from the Agencia Estatal de Investigación of the Ministerio de Ciencia e Innovación MCIN/AEI/10.13039/501100011033 and the ERDF “A way of making Europe” through projects PID2019-107061GB-C61, PID2019-107061GB-C66, PID2021-125627OB-C31, and PID2021-125627OB-C32, from the Centre of Excellence “Severo Ochoa'' award to the Instituto de Astrofísica de Canarias (CEX2019-000920-S), from the Centre of Excellence “María de Maeztu” award to the Institut de Ciències de l’Espai (CEX2020-001058-M), and from the Generalitat de Catalunya/CERCA programme. 
S.C.C.B. acknowledges support from FCT through FCT contracts nr. IF/01312/2014/CP1215/CT0004. 
C.B. acknowledges support from the Swiss Space Office through the ESA PRODEX program. 
This work has been carried out within the framework of the NCCR PlanetS supported by the Swiss National Science Foundation under grants 51NF40\_182901 and 51NF40\_205606. 
ACC acknowledges support from STFC consolidated grant number ST/V000861/1, and UKSA grant number ST/X002217/1. 
ACMC acknowledges support from the FCT, Portugal, through the CFisUC projects UIDB/04564/2020 and UIDP/04564/2020, with DOI identifiers 10.54499/UIDB/04564/2020 and 10.54499/UIDP/04564/2020, respectively. 
A.C., A.D., B.E., K.G., and J.K. acknowledge their role as ESA-appointed CHEOPS Science Team Members. 
P.E.C. is funded by the Austrian Science Fund (FWF) Erwin Schroedinger Fellowship, program J4595-N. 
This project was supported by the CNES. 
A.De. 
This work was supported by FCT - Funda\c{c}\~{a}o para a Ci\^{e}ncia e a Tecnologia through national funds and by FEDER through COMPETE2020 through the research grants UIDB/04434/2020, UIDP/04434/2020, 2022.06962.PTDC. 
O.D.S.D. is supported in the form of work contract (DL 57/2016/CP1364/CT0004) funded by national funds through FCT. 
B.-O. D. acknowledges support from the Swiss State Secretariat for Education, Research and Innovation (SERI) under contract number MB22.00046. 
A.C., A.D., B.E., K.G., and J.K. acknowledge their role as ESA-appointed CHEOPS Science Team Members. 
This project has received funding from the Swiss National Science Foundation for project 200021\_200726. It has also been carried out within the framework of the National Centre of Competence in Research PlanetS supported by the Swiss National Science Foundation under grant 51NF40\_205606. The authors acknowledge the financial support of the SNSF. 
MF and CMP gratefully acknowledge the support of the Swedish National Space Agency (DNR 65/19, 174/18). 
M.G. is an F.R.S.-FNRS Senior Research Associate. 
CHe acknowledges support from the European Union H2020-MSCA-ITN-2019 under Grant Agreement no. 860470 (CHAMELEON). 
KGI is the ESA CHEOPS Project Scientist and is responsible for the ESA CHEOPS Guest Observers Programme. She does not participate in, or contribute to, the definition of the Guaranteed Time Programme of the CHEOPS mission through which observations described in this paper have been taken, nor to any aspect of target selection for the programme. 
K.W.F.L. was supported by Deutsche Forschungsgemeinschaft grants RA714/14-1 within the DFG Schwerpunkt SPP 1992, Exploring the Diversity of Extrasolar Planets. 
This work was granted access to the HPC resources of MesoPSL financed by the Region Ile de France and the project Equip@Meso (reference ANR-10-EQPX-29-01) of the programme Investissements d'Avenir supervised by the Agence Nationale pour la Recherche. 
PM acknowledges support from STFC research grant number ST/R000638/1. 
This work was also partially supported by a grant from the Simons Foundation (PI Queloz, grant number 327127). 
NCSa acknowledges funding by the European Union (ERC, FIERCE, 101052347). Views and opinions expressed are however those of the author(s) only and do not necessarily reflect those of the European Union or the European Research Council. Neither the European Union nor the granting authority can be held responsible for them. 
A. S. acknowledges support from the Swiss Space Office through the ESA PRODEX program. 
S.G.S. acknowledge support from FCT through FCT contract nr. CEECIND/00826/2018 and POPH/FSE (EC). 
The Portuguese team thanks the Portuguese Space Agency for the provision of financial support in the framework of the PRODEX Programme of the European Space Agency (ESA) under contract number 4000142255. 
V.V.G. is an F.R.S-FNRS Research Associate. 
JV acknowledges support from the Swiss National Science Foundation (SNSF) under grant PZ00P2\_208945. 
EV acknowledges support from the ‘DISCOBOLO’ project funded by the    Spanish Ministerio de Ciencia, Innovación y Universidades undergrant PID2021-127289NB-I00. 
NAW acknowledges UKSA grant ST/R004838/1. 
TZi acknowledges NVIDIA Academic Hardware Grant Program for the use of the Titan V GPU card and the Italian MUR Departments of Excellence grant 2023-2027 “Quantum Frontiers”. 

\end{acknowledgements}

\bibliographystyle{aa} 
\bibliography{references}

\begin{thebibliography}{52}
\expandafter\ifx\csname natexlab\endcsname\relax\def\natexlab#1{#1}\fi

\bibitem[{{Barros} {et~al.}(2020){Barros}, {Demangeon}, {D{\'\i}az}, {Cabrera},
  {Santos}, {Faria}, \& {Pereira}}]{Barros1}
{Barros}, S.~C.~C., {Demangeon}, O., {D{\'\i}az}, R.~F., {et~al.} 2020, \aap,
  634, A75

\bibitem[{{Benz} {et~al.}(2021){Benz}, {Broeg}, {Fortier}, {Rando}, {Beck},
  {Beck}, {Queloz}, {Ehrenreich}, {Maxted}, {Isaak}, {Billot}, {Alibert},
  {Alonso}, {Ant{\'o}nio}, {Asquier}, {Bandy}, {B{\'a}rczy}, {Barrado},
  {Barros}, {Baumjohann}, {Bekkelien}, {Bergomi}, {Biondi}, {Bonfils},
  {Borsato}, {Brandeker}, {Busch}, {Cabrera}, {Cessa}, {Charnoz}, {Chazelas},
  {Collier Cameron}, {Corral Van Damme}, {Cortes}, {Davies}, {Deleuil},
  {Deline}, {Delrez}, {Demangeon}, {Demory}, {Erikson}, {Farinato}, {Fossati},
  {Fridlund}, {Futyan}, {Gandolfi}, {Garcia Munoz}, {Gillon}, {Guterman},
  {Gutierrez}, {Hasiba}, {Heng}, {Hernandez}, {Hoyer}, {Kiss}, {Kovacs},
  {Kuntzer}, {Laskar}, {Lecavelier des Etangs}, {Lendl}, {L{\'o}pez}, {Lora},
  {Lovis}, {L{\"u}ftinger}, {Magrin}, {Malvasio}, {Marafatto}, {Michaelis}, {de
  Miguel}, {Modrego}, {Munari}, {Nascimbeni}, {Olofsson}, {Ottacher},
  {Ottensamer}, {Pagano}, {Palacios}, {Pall{\'e}}, {Peter}, {Piazza}, {Piotto},
  {Pizarro}, {Pollaco}, {Ragazzoni}, {Ratti}, {Rauer}, {Ribas}, {Rieder},
  {Rohlfs}, {Safa}, {Salatti}, {Santos}, {Scandariato}, {S{\'e}gransan},
  {Simon}, {Smith}, {Sordet}, {Sousa}, {Steller}, {Szab{\'o}}, {Szoke},
  {Thomas}, {Tschentscher}, {Udry}, {Van Grootel}, {Viotto}, {Walter},
  {Walton}, {Wildi}, \& {Wolter}}]{Benz2021}
{Benz}, W., {Broeg}, C., {Fortier}, A., {et~al.} 2021, Experimental Astronomy,
  51, 109

\bibitem[{{Bonfanti} {et~al.}(2021){Bonfanti}, {Delrez}, {Hooton}, {Wilson},
  {Fossati}, {Alibert}, {Hoyer}, {Mustill}, {Osborn}, {Adibekyan}, {Gandolfi},
  {Salmon}, {Sousa}, {Tuson}, {Van Grootel}, {Cabrera}, {Nascimbeni}, {Maxted},
  {Barros}, {Billot}, {Bonfils}, {Borsato}, {Broeg}, {Davies}, {Deleuil},
  {Demangeon}, {Fridlund}, {Lacedelli}, {Lendl}, {Persson}, {Santos},
  {Scandariato}, {Szab{\'o}}, {Collier Cameron}, {Udry}, {Benz}, {Beck},
  {Ehrenreich}, {Fortier}, {Isaak}, {Queloz}, {Alonso}, {Asquier}, {Bandy},
  {B{\'a}rczy}, {Barrado}, {Barrag{\'a}n}, {Baumjohann}, {Beck}, {Bekkelien},
  {Bergomi}, {Brandeker}, {Busch}, {Cessa}, {Charnoz}, {Chazelas}, {Corral Van
  Damme}, {Demory}, {Erikson}, {Farinato}, {Futyan}, {Garcia Mu{\~n}oz},
  {Gillon}, {Guedel}, {Guterman}, {Hasiba}, {Heng}, {Hernandez}, {Kiss},
  {Kuntzer}, {Laskar}, {Lecavelier des Etangs}, {Lovis}, {Magrin}, {Malvasio},
  {Marafatto}, {Michaelis}, {Munari}, {Olofsson}, {Ottacher}, {Ottensamer},
  {Pagano}, {Pall{\'e}}, {Peter}, {Piazza}, {Piotto}, {Pollacco}, {Ragazzoni},
  {Rando}, {Ratti}, {Rauer}, {Ribas}, {Rieder}, {Rohlfs}, {Safa}, {Salatti},
  {S{\'e}gransan}, {Simon}, {Smith}, {Sordet}, {Steller}, {Thomas},
  {Tschentscher}, {Van Eylen}, {Viotto}, {Walter}, {Walton}, {Wildi}, \&
  {Wolter}}]{Bonfanti2021}
{Bonfanti}, A., {Delrez}, L., {Hooton}, M.~J., {et~al.} 2021, \aap, 646, A157

\bibitem[{{Borsato} {et~al.}(2019){Borsato}, {Malavolta}, {Piotto}, {Buchhave},
  {Mortier}, {Rice}, {Collier Cameron}, {Coffinet}, {Sozzetti}, {Charbonneau},
  {Cosentino}, {Dumusque}, {Figueira}, {Latham}, {Lopez-Morales}, {Mayor},
  {Micela}, {Molinari}, {Pepe}, {Phillips}, {Poretti}, {Udry}, \&
  {Watson}}]{Borsato2019MNRAS.484.3233B}
{Borsato}, L., {Malavolta}, L., {Piotto}, G., {et~al.} 2019, \mnras, 484, 3233

\bibitem[{{Borsato} {et~al.}(2014){Borsato}, {Marzari}, {Nascimbeni}, {Piotto},
  {Granata}, {Bedin}, \& {Malavolta}}]{Borsato2014A&A...571A..38B}
{Borsato}, L., {Marzari}, F., {Nascimbeni}, V., {et~al.} 2014, \aap, 571, A38

\bibitem[{{Borsato} {et~al.}(2021){Borsato}, {Piotto}, {Gandolfi},
  {Nascimbeni}, {Lacedelli}, {Marzari}, {Billot}, {Maxted}, {Sousa}, {Cameron},
  {Bonfanti}, {Wilson}, {Serrano}, {Garai}, {Alibert}, {Alonso}, {Asquier},
  {B{\'a}rczy}, {Bandy}, {Barrado}, {Barros}, {Baumjohann}, {Beck}, {Beck},
  {Benz}, {Bonfils}, {Brandeker}, {Broeg}, {Cabrera}, {Charnoz}, {Csizmadia},
  {Davies}, {Deleuil}, {Delrez}, {Demangeon}, {Demory}, {des Etangs},
  {Ehrenreich}, {Erikson}, {Escud{\'e}}, {Fortier}, {Fossati}, {Fridlund},
  {Gillon}, {Guedel}, {Hasiba}, {Heng}, {Hoyer}, {Isaak}, {Kiss}, {Kopp},
  {Laskar}, {Lendl}, {Lovis}, {Magrin}, {Munari}, {Olofsson}, {Ottensamer},
  {Pagano}, {Pall{\'e}}, {Peter}, {Pollacco}, {Queloz}, {Ragazzoni}, {Rando},
  {Rauer}, {Ribas}, {S{\'e}gransan}, {Santos}, {Scandariato}, {Simon}, {Smith},
  {Steller}, {Szab{\'o}}, {Thomas}, {Udry}, {Van Grootel}, \&
  {Walton}}]{Borsato2021MNRAS.506.3810B}
{Borsato}, L., {Piotto}, G., {Gandolfi}, D., {et~al.} 2021, \mnras, 506, 3810

\bibitem[{{Brandeker} {et~al.}(2024){Brandeker}, {Patel}, \&
  {Morris}}]{Brandeker2024}
{Brandeker}, A., {Patel}, J.~A., \& {Morris}, B.~M. 2024, {PIPE: Extracting PSF
  photometry from CHEOPS data}, Astrophysics Source Code Library, record
  ascl:2404.002

\bibitem[{{Butler} {et~al.}(1981){Butler}, {Byrne}, {Andrews}, \&
  {Doyle}}]{Butler1981}
{Butler}, C.~J., {Byrne}, P.~B., {Andrews}, A.~D., \& {Doyle}, J.~G. 1981,
  \mnras, 197, 815

\bibitem[{{Cale} {et~al.}(2021){Cale}, {Reefe}, {Plavchan}, {Tanner}, {Gaidos},
  {Gagn{\'e}}, {Gao}, {Kane}, {B{\'e}jar}, {Lodieu}, {Anglada-Escud{\'e}},
  {Ribas}, {Pall{\'e}}, {Quirrenbach}, {Amado}, {Reiners}, {Caballero}, {Rosa
  Zapatero Osorio}, {Dreizler}, {Howard}, {Fulton}, {Xuesong Wang}, {Collins},
  {El Mufti}, {Wittrock}, {Gilbert}, {Barclay}, {Klein}, {Martioli},
  {Wittenmyer}, {Wright}, {Addison}, {Hirano}, {Tamura}, {Kotani}, {Narita},
  {Vermilion}, {Lee}, {Geneser}, {Teske}, {Quinn}, {Latham}, {Esquerdo},
  {Calkins}, {Berlind}, {Zohrabi}, {Stibbards}, {Kotnana}, {Jenkins},
  {Twicken}, {Henze}, {Kidwell}, {Burke}, {Villase{\~n}or}, \&
  {Boyd}}]{Cale2021}
{Cale}, B.~L., {Reefe}, M., {Plavchan}, P., {et~al.} 2021, \aj, 162, 295

\bibitem[{{Claret}(2021)}]{Claret2}
{Claret}, A. 2021, Research Notes of the American Astronomical Society, 5, 13

\bibitem[{{Deline} {et~al.}(2020){Deline}, {Queloz}, {Chazelas}, {Sordet},
  {Wildi}, {Fortier}, {Broeg}, {Futyan}, \& {Benz}}]{Deline2020}
{Deline}, A., {Queloz}, D., {Chazelas}, B., {et~al.} 2020, \aap, 635, A22

\bibitem[{{Delrez} {et~al.}(2021){Delrez}, {Ehrenreich}, {Alibert}, {Bonfanti},
  {Borsato}, {Fossati}, {Hooton}, {Hoyer}, {Pozuelos}, {Salmon}, {Sulis},
  {Wilson}, {Adibekyan}, {Bourrier}, {Brandeker}, {Charnoz}, {Deline},
  {Guterman}, {Haldemann}, {Hara}, {Oshagh}, {Sousa}, {Van Grootel}, {Alonso},
  {Anglada-Escud{\'e}}, {B{\'a}rczy}, {Barrado}, {Barros}, {Baumjohann},
  {Beck}, {Bekkelien}, {Benz}, {Billot}, {Bonfils}, {Broeg}, {Cabrera},
  {Collier Cameron}, {Davies}, {Deleuil}, {Delisle}, {Demangeon}, {Demory},
  {Erikson}, {Fortier}, {Fridlund}, {Futyan}, {Gandolfi}, {Garcia Mu{\~n}oz},
  {Gillon}, {Guedel}, {Heng}, {Kiss}, {Laskar}, {Lecavelier des Etangs},
  {Lendl}, {Lovis}, {Maxted}, {Nascimbeni}, {Olofsson}, {Osborn}, {Pagano},
  {Pall{\'e}}, {Piotto}, {Pollacco}, {Queloz}, {Rauer}, {Ragazzoni}, {Ribas},
  {Santos}, {Scandariato}, {S{\'e}gransan}, {Simon}, {Smith}, {Steller},
  {Szab{\'o}}, {Thomas}, {Udry}, \& {Walton}}]{Delrez2021}
{Delrez}, L., {Ehrenreich}, D., {Alibert}, Y., {et~al.} 2021, Nature Astronomy,
  5, 775

\bibitem[{{Delrez} {et~al.}(2023){Delrez}, {Leleu}, {Brandeker}, {Gillon},
  {Hooton}, {Collier Cameron}, {Deline}, {Fortier}, {Queloz}, {Bonfanti}, {Van
  Grootel}, {Wilson}, {Egger}, {Alibert}, {Alonso}, {Anglada}, {Asquier},
  {B{\'a}rczy}, {Barrado y Navascues}, {Barros}, {Baumjohann}, {Beck}, {Beck},
  {Benz}, {Billot}, {Bonf{\i}ls}, {Borsato}, {Broeg}, {Buder}, {Cabrera},
  {Cessa}, {Charnoz}, {Csizmadia}, {Cubillos}, {Davies}, {Deleuil},
  {Demangeon}, {Demory}, {Ehrenreich}, {Erikson}, {Fossati}, {Fridlund},
  {Gandolfi}, {G{\"u}del}, {Hasiba}, {Hoyer}, {Isaak}, {Jenkins}, {Kiss},
  {Laskar}, {Latham}, {Lecavelier des Etangs}, {Lendl}, {Lovis}, {Luque},
  {Magrin}, {Maxted}, {Mordasini}, {Nascimbeni}, {Olofsson}, {Ottensamer},
  {Pagano}, {Pall{\'e}}, {Peter}, {Piotto}, {Pollacco}, {Ragazzoni}, {Rando},
  {Rauer}, {Ribas}, {Ricker}, {Santos}, {Scandariato}, {Seager},
  {S{\'e}gransan}, {Simon}, {Smith}, {Sousa}, {Steller}, {Szab{\'o}}, {Thomas},
  {Udry}, {Vanderspek}, {Venturini}, {Viotto}, {Walton}, \&
  {Winn}}]{Delrez2023}
{Delrez}, L., {Leleu}, A., {Brandeker}, A., {et~al.} 2023, \aap, 678, A200

\bibitem[{{Donati} {et~al.}(2023){Donati}, {Cristofari}, {Finociety}, {Klein},
  {Moutou}, {Gaidos}, {Cadieux}, {Artigau}, {Correia}, {Bou{\'e}}, {Cook},
  {Carmona}, {Lehmann}, {Bouvier}, {Martioli}, {Morin}, {Fouqu{\'e}},
  {Delfosse}, {Doyon}, {H{\'e}brard}, {Alencar}, {Laskar}, {Arnold}, {Petit},
  {K{\'o}sp{\'a}l}, {Vidotto}, {Folsom}, \& {collaboration}}]{Donati2023}
{Donati}, J.~F., {Cristofari}, P.~I., {Finociety}, B., {et~al.} 2023, \mnras,
  525, 455

\bibitem[{{Foreman-Mackey} {et~al.}(2017){Foreman-Mackey}, {Agol},
  {Ambikasaran}, \& {Angus}}]{Foreman2}
{Foreman-Mackey}, D., {Agol}, E., {Ambikasaran}, S., \& {Angus}, R. 2017, \aj,
  154, 220

\bibitem[{{Foreman-Mackey} {et~al.}(2013){Foreman-Mackey}, {Hogg}, {Lang}, \&
  {Goodman}}]{Foreman2013}
{Foreman-Mackey}, D., {Hogg}, D.~W., {Lang}, D., \& {Goodman}, J. 2013, \pasp,
  125, 306

\bibitem[{{Fortier} {et~al.}(2024){Fortier}, {Simon}, {Broeg}, {Olofsson},
  {Deline}, {Wilson}, {Maxted}, {Brandeker}, {Collier Cameron}, {Beck},
  {Bekkelien}, {Billot}, {Bonfanti}, {Bruno}, {Cabrera}, {Delrez}, {Demory},
  {Futyan}, {Flor{\'e}n}, {G{\"u}nther}, {Heitzmann}, {Hoyer}, {Isaak},
  {Sousa}, {Stalport}, {Turin}, {Verhoeve}, {Akinsanmi}, {Alibert}, {Alonso},
  {B{\'a}nhidi}, {B{\'a}rczy}, {Barrado}, {Barros}, {Baumjohann}, {Baycroft},
  {Beck}, {Benz}, {B{\'\i}r{\'o}}, {B{\'o}di}, {Bonfils}, {Borsato}, {Charnoz},
  {Cseh}, {Csizmadia}, {Cs{\'a}nyi}, {Cubillos}, {Davies}, {Davis}, {Deleuil},
  {Demangeon}, {Derekas}, {Dransfield}, {Ducrot}, {Ehrenreich}, {Erikson},
  {Fari{\~n}a}, {Fossati}, {Fridlund}, {Gandolfi}, {Garai}, {Garcia}, {Gillon},
  {G{\'o}mez Maqueo Chew}, {G{\'o}mez-Mu{\~n}oz}, {Granata}, {G{\"u}del},
  {Guterman}, {Heged{\"u}s}, {Helling}, {Jehin}, {Kalup}, {Kilkenny}, {Kiss},
  {Kriskovics}, {Lam}, {Laskar}, {Lecavelier des Etangs}, {Lendl}, {Lopez
  Pina}, {Luntzer}, {Magrin}, {Miller}, {Modrego Contreras}, {Mordasini},
  {Munari}, {Murray}, {Nascimbeni}, {Ottacher}, {Ottensamer}, {Pagano},
  {P{\'a}l}, {Pall{\'e}}, {Pasetti}, {Pedersen}, {Peter}, {Petrucci}, {Piotto},
  {Pizarro-Rubio}, {Pollacco}, {Pribulla}, {Queloz}, {Ragazzoni}, {Rando},
  {Rauer}, {Ribas}, {Sabin}, {Santos}, {Scandariato}, {Schanche},
  {Schroffenegger}, {Scutt}, {Sebastian}, {S{\'e}gransan}, {Seli}, {Smith},
  {Southworth}, {Standing}, {Szab{\'o}}, {Szak{\'a}ts}, {Thomas}, {Timmermans},
  {Triaud}, {Udry}, {Van Grootel}, {Venturini}, {Villaver}, {Vink{\'o}},
  {Walton}, {Wells}, \& {Wolter}}]{Fortier2024}
{Fortier}, A., {Simon}, A.~E., {Broeg}, C., {et~al.} 2024, \aap, 687, A302

\bibitem[{{Gaia Collaboration} {et~al.}(2023){Gaia Collaboration}, {Vallenari},
  {Brown}, {Prusti}, {de Bruijne}, {Arenou}, {Babusiaux}, {Biermann},
  {Creevey}, {Ducourant}, {Evans}, {Eyer}, {Guerra}, {Hutton}, {Jordi},
  {Klioner}, {Lammers}, {Lindegren}, {Luri}, {Mignard}, {Panem}, {Pourbaix},
  {Randich}, {Sartoretti}, {Soubiran}, {Tanga}, {Walton}, {Bailer-Jones},
  {Bastian}, {Drimmel}, {Jansen}, {Katz}, {Lattanzi}, {van Leeuwen}, {Bakker},
  {Cacciari}, {Casta{\~n}eda}, {De Angeli}, {Fabricius}, {Fouesneau},
  {Fr{\'e}mat}, {Galluccio}, {Guerrier}, {Heiter}, {Masana}, {Messineo},
  {Mowlavi}, {Nicolas}, {Nienartowicz}, {Pailler}, {Panuzzo}, {Riclet}, {Roux},
  {Seabroke}, {Sordo}, {Th{\'e}venin}, {Gracia-Abril}, {Portell}, {Teyssier},
  {Altmann}, {Andrae}, {Audard}, {Bellas-Velidis}, {Benson}, {Berthier},
  {Blomme}, {Burgess}, {Busonero}, {Busso}, {C{\'a}novas}, {Carry}, {Cellino},
  {Cheek}, {Clementini}, {Damerdji}, {Davidson}, {de Teodoro}, {Nu{\~n}ez
  Campos}, {Delchambre}, {Dell'Oro}, {Esquej}, {Fern{\'a}ndez-Hern{\'a}ndez},
  {Fraile}, {Garabato}, {Garc{\'\i}a-Lario}, {Gosset}, {Haigron}, {Halbwachs},
  {Hambly}, {Harrison}, {Hern{\'a}ndez}, {Hestroffer}, {Hodgkin}, {Holl},
  {Jan{\ss}en}, {Jevardat de Fombelle}, {Jordan}, {Krone-Martins}, {Lanzafame},
  {L{\"o}ffler}, {Marchal}, {Marrese}, {Moitinho}, {Muinonen}, {Osborne},
  {Pancino}, {Pauwels}, {Recio-Blanco}, {Reyl{\'e}}, {Riello}, {Rimoldini},
  {Roegiers}, {Rybizki}, {Sarro}, {Siopis}, {Smith}, {Sozzetti}, {Utrilla},
  {van Leeuwen}, {Abbas}, {{\'A}brah{\'a}m}, {Abreu Aramburu}, {Aerts},
  {Aguado}, {Ajaj}, {Aldea-Montero}, {Altavilla}, {{\'A}lvarez}, {Alves},
  {Anders}, {Anderson}, {Anglada Varela}, {Antoja}, {Baines}, {Baker},
  {Balaguer-N{\'u}{\~n}ez}, {Balbinot}, {Balog}, {Barache}, {Barbato},
  {Barros}, {Barstow}, {Bartolom{\'e}}, {Bassilana}, {Bauchet}, {Becciani},
  {Bellazzini}, {Berihuete}, {Bernet}, {Bertone}, {Bianchi}, {Binnenfeld},
  {Blanco-Cuaresma}, {Blazere}, {Boch}, {Bombrun}, {Bossini}, {Bouquillon},
  {Bragaglia}, {Bramante}, {Breedt}, {Bressan}, {Brouillet}, {Brugaletta},
  {Bucciarelli}, {Burlacu}, {Butkevich}, {Buzzi}, {Caffau}, {Cancelliere},
  {Cantat-Gaudin}, {Carballo}, {Carlucci}, {Carnerero}, {Carrasco},
  {Casamiquela}, {Castellani}, {Castro-Ginard}, {Chaoul}, {Charlot}, {Chemin},
  {Chiaramida}, {Chiavassa}, {Chornay}, {Comoretto}, {Contursi}, {Cooper},
  {Cornez}, {Cowell}, {Crifo}, {Cropper}, {Crosta}, {Crowley}, {Dafonte},
  {Dapergolas}, {David}, {David}, {de Laverny}, {De Luise}, {De March}, {De
  Ridder}, {de Souza}, {de Torres}, {del Peloso}, {del Pozo}, {Delbo},
  {Delgado}, {Delisle}, {Demouchy}, {Dharmawardena}, {Di Matteo}, {Diakite},
  {Diener}, {Distefano}, {Dolding}, {Edvardsson}, {Enke}, {Fabre}, {Fabrizio},
  {Faigler}, {Fedorets}, {Fernique}, {Fienga}, {Figueras}, {Fournier},
  {Fouron}, {Fragkoudi}, {Gai}, {Garcia-Gutierrez}, {Garcia-Reinaldos},
  {Garc{\'\i}a-Torres}, {Garofalo}, {Gavel}, {Gavras}, {Gerlach}, {Geyer},
  {Giacobbe}, {Gilmore}, {Girona}, {Giuffrida}, {Gomel}, {Gomez},
  {Gonz{\'a}lez-N{\'u}{\~n}ez}, {Gonz{\'a}lez-Santamar{\'\i}a},
  {Gonz{\'a}lez-Vidal}, {Granvik}, {Guillout}, {Guiraud},
  {Guti{\'e}rrez-S{\'a}nchez}, {Guy}, {Hatzidimitriou}, {Hauser}, {Haywood},
  {Helmer}, {Helmi}, {Sarmiento}, {Hidalgo}, {Hilger}, {H{\l}adczuk}, {Hobbs},
  {Holland}, {Huckle}, {Jardine}, {Jasniewicz}, {Jean-Antoine Piccolo},
  {Jim{\'e}nez-Arranz}, {Jorissen}, {Juaristi Campillo}, {Julbe}, {Karbevska},
  {Kervella}, {Khanna}, {Kontizas}, {Kordopatis}, {Korn}, {K{\'o}sp{\'a}l},
  {Kostrzewa-Rutkowska}, {Kruszy{\'n}ska}, {Kun}, {Laizeau}, {Lambert},
  {Lanza}, {Lasne}, {Le Campion}, {Lebreton}, {Lebzelter}, {Leccia}, {Leclerc},
  {Lecoeur-Taibi}, {Liao}, {Licata}, {Lindstr{\o}m}, {Lister}, {Livanou},
  {Lobel}, {Lorca}, {Loup}, {Madrero Pardo}, {Magdaleno Romeo}, {Managau},
  {Mann}, {Manteiga}, {Marchant}, {Marconi}, {Marcos}, {Marcos Santos},
  {Mar{\'\i}n Pina}, {Marinoni}, {Marocco}, {Marshall}, {Martin Polo},
  {Mart{\'\i}n-Fleitas}, {Marton}, {Mary}, {Masip}, {Massari},
  {Mastrobuono-Battisti}, {Mazeh}, {McMillan}, {Messina}, {Michalik}, {Millar},
  {Mints}, {Molina}, {Molinaro}, {Moln{\'a}r}, {Monari}, {Mongui{\'o}},
  {Montegriffo}, {Montero}, {Mor}, {Mora}, {Morbidelli}, {Morel}, {Morris},
  {Muraveva}, {Murphy}, {Musella}, {Nagy}, {Noval}, {Oca{\~n}a}, {Ogden},
  {Ordenovic}, {Osinde}, {Pagani}, {Pagano}, {Palaversa}, {Palicio},
  {Pallas-Quintela}, {Panahi}, {Payne-Wardenaar}, {Pe{\~n}alosa Esteller},
  {Penttil{\"a}}, {Pichon}, {Piersimoni}, {Pineau}, {Plachy}, {Plum}, {Poggio},
  {Pr{\v{s}}a}, {Pulone}, {Racero}, {Ragaini}, {Rainer}, {Raiteri}, {Rambaux},
  {Ramos}, {Ramos-Lerate}, {Re Fiorentin}, {Regibo}, {Richards}, {Rios Diaz},
  {Ripepi}, {Riva}, {Rix}, {Rixon}, {Robichon}, {Robin}, {Robin}, {Roelens},
  {Rogues}, {Rohrbasser}, {Romero-G{\'o}mez}, {Rowell}, {Royer}, {Ruz Mieres},
  {Rybicki}, {Sadowski}, {S{\'a}ez N{\'u}{\~n}ez}, {Sagrist{\`a} Sell{\'e}s},
  {Sahlmann}, {Salguero}, {Samaras}, {Sanchez Gimenez}, {Sanna},
  {Santove{\~n}a}, {Sarasso}, {Schultheis}, {Sciacca}, {Segol}, {Segovia},
  {S{\'e}gransan}, {Semeux}, {Shahaf}, {Siddiqui}, {Siebert}, {Siltala},
  {Silvelo}, {Slezak}, {Slezak}, {Smart}, {Snaith}, {Solano}, {Solitro},
  {Souami}, {Souchay}, {Spagna}, {Spina}, {Spoto}, {Steele},
  {Steidelm{\"u}ller}, {Stephenson}, {S{\"u}veges}, {Surdej}, {Szabados},
  {Szegedi-Elek}, {Taris}, {Taylor}, {Teixeira}, {Tolomei}, {Tonello}, {Torra},
  {Torra}, {Torralba Elipe}, {Trabucchi}, {Tsounis}, {Turon}, {Ulla}, {Unger},
  {Vaillant}, {van Dillen}, {van Reeven}, {Vanel}, {Vecchiato}, {Viala},
  {Vicente}, {Voutsinas}, {Weiler}, {Wevers}, {Wyrzykowski}, {Yoldas}, {Yvard},
  {Zhao}, {Zorec}, {Zucker}, \& {Zwitter}}]{Gaia2023}
{Gaia Collaboration}, {Vallenari}, A., {Brown}, A.~G.~A., {et~al.} 2023, \aap,
  674, A1

\bibitem[{Gelman \& Rubin(1992)}]{GelmanRubin1992}
Gelman, A. \& Rubin, D.~B. 1992, Statistical Science, 7, 457

\bibitem[{Geweke(1991)}]{geweke1991}
Geweke, J.~F. 1991, {Evaluating the accuracy of sampling-based approaches to
  the calculation of posterior moments}, Staff Report 148, Federal Reserve Bank
  of Minneapolis

\bibitem[{{Gilbert} {et~al.}(2022){Gilbert}, {Barclay}, {Quintana},
  {Walkowicz}, {Vega}, {Schlieder}, {Monsue}, {Cale}, {Collins}, {Gaidos}, {El
  Mufti}, {Reefe}, {Plavchan}, {Tanner}, {Wittenmyer}, {Wittrock}, {Jenkins},
  {Latham}, {Ricker}, {Rose}, {Seager}, {Vanderspek}, \& {Winn}}]{Gilbert2022}
{Gilbert}, E.~A., {Barclay}, T., {Quintana}, E.~V., {et~al.} 2022, \aj, 163,
  147

\bibitem[{{G{\"u}nther} \& {Daylan}(2019)}]{allesfitter-code}
{G{\"u}nther}, M.~N. \& {Daylan}, T. 2019, {Allesfitter: Flexible Star and
  Exoplanet Inference From Photometry and Radial Velocity}, Astrophysics Source
  Code Library

\bibitem[{{G{\"u}nther} \& {Daylan}(2021)}]{allesfitter-paper}
{G{\"u}nther}, M.~N. \& {Daylan}, T. 2021, \apjs, 254, 13

\bibitem[{{Hebb} {et~al.}(2007){Hebb}, {Petro}, {Ford}, {Ardila}, {Toledo},
  {Minniti}, {Golimowski}, \& {Clampin}}]{Hebb2007}
{Hebb}, L., {Petro}, L., {Ford}, H.~C., {et~al.} 2007, \mnras, 379, 63

\bibitem[{{Ilin} \& {Poppenhaeger}(2022)}]{IlinPoppenhaeger2022}
{Ilin}, E. \& {Poppenhaeger}, K. 2022, \mnras, 513, 4579

\bibitem[{{Kalas} {et~al.}(2004){Kalas}, {Liu}, \& {Matthews}}]{Kalas2004}
{Kalas}, P., {Liu}, M.~C., \& {Matthews}, B.~C. 2004, Science, 303, 1990

\bibitem[{{Kallinger} {et~al.}(2014){Kallinger}, {De Ridder}, {Hekker},
  {Mathur}, {Mosser}, {Gruberbauer}, {Garc{\'\i}a}, {Karoff}, \&
  {Ballot}}]{Kallinger1}
{Kallinger}, T., {De Ridder}, J., {Hekker}, S., {et~al.} 2014, \aap, 570, A41

\bibitem[{{Kavanagh} {et~al.}(2021){Kavanagh}, {Vidotto}, {Klein}, {Jardine},
  {Donati}, \& {{\'O} Fionnag{\'a}in}}]{Kavanagh2021}
{Kavanagh}, R.~D., {Vidotto}, A.~A., {Klein}, B., {et~al.} 2021, \mnras, 504,
  1511

\bibitem[{{Kipping}(2013)}]{Kipping3}
{Kipping}, D.~M. 2013, \mnras, 435, 2152

\bibitem[{{Kochukhov} \& {Reiners}(2020)}]{Kochukhov2020}
{Kochukhov}, O. \& {Reiners}, A. 2020, \apj, 902, 43

\bibitem[{{Lendl} {et~al.}(2020){Lendl}, {Csizmadia}, {Deline}, {Fossati},
  {Kitzmann}, {Heng}, {Hoyer}, {Salmon}, {Benz}, {Broeg}, {Ehrenreich},
  {Fortier}, {Queloz}, {Bonfanti}, {Brandeker}, {Collier Cameron}, {Delrez},
  {Garcia Mu{\~n}oz}, {Hooton}, {Maxted}, {Morris}, {Van Grootel}, {Wilson},
  {Alibert}, {Alonso}, {Asquier}, {Bandy}, {B{\'a}rczy}, {Barrado}, {Barros},
  {Baumjohann}, {Beck}, {Beck}, {Bekkelien}, {Bergomi}, {Billot}, {Biondi},
  {Bonfils}, {Bourrier}, {Busch}, {Cabrera}, {Cessa}, {Charnoz}, {Chazelas},
  {Corral Van Damme}, {Davies}, {Deleuil}, {Demangeon}, {Demory}, {Erikson},
  {Farinato}, {Fridlund}, {Futyan}, {Gandolfi}, {Gillon}, {Guterman}, {Hasiba},
  {Hernandez}, {Isaak}, {Kiss}, {Kuntzer}, {Lecavelier des Etangs},
  {L{\"u}ftinger}, {Laskar}, {Lovis}, {Magrin}, {Malvasio}, {Marafatto},
  {Michaelis}, {Munari}, {Nascimbeni}, {Olofsson}, {Ottacher}, {Ottensamer},
  {Pagano}, {Pall{\'e}}, {Peter}, {Piazza}, {Piotto}, {Pollacco}, {Ratti},
  {Rauer}, {Ragazzoni}, {Rando}, {Ribas}, {Rieder}, {Rohlfs}, {Safa}, {Santos},
  {Scandariato}, {S{\'e}gransan}, {Simon}, {Singh}, {Smith}, {Sordet}, {Sousa},
  {Steller}, {Szab{\'o}}, {Thomas}, {Tschentscher}, {Udry}, {Viotto}, {Walter},
  {Walton}, {Wildi}, \& {Wolter}}]{Lendl2020}
{Lendl}, M., {Csizmadia}, S., {Deline}, A., {et~al.} 2020, \aap, 643, A94

\bibitem[{{MacGregor} {et~al.}(2013){MacGregor}, {Wilner}, {Rosenfeld},
  {Andrews}, {Matthews}, {Hughes}, {Booth}, {Chiang}, {Graham}, {Kalas},
  {Kennedy}, \& {Sibthorpe}}]{MacGregor2013}
{MacGregor}, M.~A., {Wilner}, D.~J., {Rosenfeld}, K.~A., {et~al.} 2013, \apjl,
  762, L21

\bibitem[{{Mallorqu{\'\i}n} {et~al.}(2024){Mallorqu{\'\i}n}, {B{\'e}jar},
  {Lodieu}, {Zapatero Osorio}, {Yu}, {Su{\'a}rez Mascare{\~n}o}, {Damasso},
  {Sanz-Forcada}, {Ribas}, {Reiners}, {Quirrenbach}, {Amado}, {Caballero},
  {Aigrain}, {Barrag{\'a}n}, {Dreizler}, {Fern{\'a}ndez-Mart{\'\i}n}, {Goffo},
  {Henning}, {Kaminski}, {Klein}, {Luque}, {Montes}, {Morales}, {Nagel},
  {Pall{\'e}}, {Reffert}, {Schlecker}, \& {Schweitzer}}]{Mallorquin2024}
{Mallorqu{\'\i}n}, M., {B{\'e}jar}, V.~J.~S., {Lodieu}, N., {et~al.} 2024,
  \aap, 689, A132

\bibitem[{{Mamajek} \& {Bell}(2014)}]{MamajekBell2014}
{Mamajek}, E.~E. \& {Bell}, C. P.~M. 2014, \mnras, 445, 2169

\bibitem[{{Martioli} {et~al.}(2021){Martioli}, {H{\'e}brard}, {Correia},
  {Laskar}, \& {Lecavelier des Etangs}}]{Martioli2021}
{Martioli}, E., {H{\'e}brard}, G., {Correia}, A.~C.~M., {Laskar}, J., \&
  {Lecavelier des Etangs}, A. 2021, \aap, 649, A177

\bibitem[{{Mathioudakis} \& {Doyle}(1991)}]{MathioudakisDoyle1991}
{Mathioudakis}, M. \& {Doyle}, J.~G. 1991, \aap, 244, 433

\bibitem[{{Matthews} {et~al.}(2015){Matthews}, {Kennedy}, {Sibthorpe},
  {Holland}, {Booth}, {Kalas}, {MacGregor}, {Wilner}, {Vandenbussche},
  {Olofsson}, {Blommaert}, {Brandeker}, {Dent}, {de Vries}, {Di Francesco},
  {Fridlund}, {Graham}, {Greaves}, {Heras}, {Hogerheijde}, {Ivison}, {Pantin},
  \& {Pilbratt}}]{Matthews2015}
{Matthews}, B.~C., {Kennedy}, G., {Sibthorpe}, B., {et~al.} 2015, \apj, 811,
  100

\bibitem[{{Nascimbeni} {et~al.}(2023){Nascimbeni}, {Borsato}, {Zingales},
  {Piotto}, {Pagano}, {Beck}, {Broeg}, {Ehrenreich}, {Hoyer}, {Majidi},
  {Granata}, {Sousa}, {Wilson}, {Van Grootel}, {Bonfanti}, {Salmon}, {Mustill},
  {Delrez}, {Alibert}, {Alonso}, {Anglada}, {B{\'a}rczy}, {Barrado}, {Barros},
  {Baumjohann}, {Beck}, {Benz}, {Bergomi}, {Billot}, {Bonfils}, {Brandeker},
  {Cabrera}, {Charnoz}, {Collier Cameron}, {Csizmadia}, {Cubillos}, {Davies},
  {Deleuil}, {Deline}, {Demangeon}, {Demory}, {Erikson}, {Fortier}, {Fossati},
  {Fridlund}, {Gandolfi}, {Gillon}, {G{\"u}del}, {Isaak}, {Kiss}, {Laskar},
  {Lecavelier des Etangs}, {Lendl}, {Lovis}, {Luque}, {Magrin}, {Maxted},
  {Mordasini}, {Olofsson}, {Ottensamer}, {Pall{\'e}}, {Peter}, {Piazza},
  {Pollacco}, {Queloz}, {Ragazzoni}, {Rando}, {Ratti}, {Rauer}, {Ribas},
  {Santos}, {Scandariato}, {S{\'e}gransan}, {Simon}, {Smith}, {Steinberger},
  {Steller}, {Szab{\'o}}, {Thomas}, {Udry}, {Venturini}, {Walton}, \&
  {Wolter}}]{Nascimbeni2023A&A...673A..42N}
{Nascimbeni}, V., {Borsato}, L., {Zingales}, T., {et~al.} 2023, \aap, 673, A42

\bibitem[{{Nelson} {et~al.}(2014){Nelson}, {Ford}, \&
  {Payne}}]{DEMOVE2014ApJS..210...11N}
{Nelson}, B., {Ford}, E.~B., \& {Payne}, M.~J. 2014, \apjs, 210, 11

\bibitem[{{Oshagh} {et~al.}(2013){Oshagh}, {Santos}, {Boisse}, {Bou{\'e}},
  {Montalto}, {Dumusque}, \& {Haghighipour}}]{Oshagh2013}
{Oshagh}, M., {Santos}, N.~C., {Boisse}, I., {et~al.} 2013, \aap, 556, A19

\bibitem[{{Parviainen} {et~al.}(2016){Parviainen}, {Pall{\'e}}, {Nortmann},
  {Nowak}, {Iro}, {Murgas}, \& {Aigrain}}]{Parviainen2016}
{Parviainen}, H., {Pall{\'e}}, E., {Nortmann}, L., {et~al.} 2016, \aap, 585,
  A114

\bibitem[{{Plavchan} {et~al.}(2020){Plavchan}, {Barclay}, {Gagn{\'e}}, {Gao},
  {Cale}, {Matzko}, {Dragomir}, {Quinn}, {Feliz}, {Stassun}, {Crossfield},
  {Berardo}, {Latham}, {Tieu}, {Anglada-Escud{\'e}}, {Ricker}, {Vanderspek},
  {Seager}, {Winn}, {Jenkins}, {Rinehart}, {Krishnamurthy}, {Dynes}, {Doty},
  {Adams}, {Afanasev}, {Beichman}, {Bottom}, {Bowler}, {Brinkworth}, {Brown},
  {Cancino}, {Ciardi}, {Clampin}, {Clark}, {Collins}, {Davison},
  {Foreman-Mackey}, {Furlan}, {Gaidos}, {Geneser}, {Giddens}, {Gilbert},
  {Hall}, {Hellier}, {Henry}, {Horner}, {Howard}, {Huang}, {Huber}, {Kane},
  {Kenworthy}, {Kielkopf}, {Kipping}, {Klenke}, {Kruse}, {Latouf}, {Lowrance},
  {Mennesson}, {Mengel}, {Mills}, {Morton}, {Narita}, {Newton}, {Nishimoto},
  {Okumura}, {Palle}, {Pepper}, {Quintana}, {Roberge}, {Roccatagliata},
  {Schlieder}, {Tanner}, {Teske}, {Tinney}, {Vanderburg}, {von Braun}, {Walp},
  {Wang}, {Wang}, {Weigand}, {White}, {Wittenmyer}, {Wright}, {Youngblood},
  {Zhang}, \& {Zilberman}}]{Plavchan2020}
{Plavchan}, P., {Barclay}, T., {Gagn{\'e}}, J., {et~al.} 2020, \nat, 582, 497

\bibitem[{{Schneider} {et~al.}(2014){Schneider}, {Grady}, {Hines}, {Stark},
  {Debes}, {Carson}, {Kuchner}, {Perrin}, {Weinberger}, {Wisniewski},
  {Silverstone}, {Jang-Condell}, {Henning}, {Woodgate}, {Serabyn},
  {Moro-Martin}, {Tamura}, {Hinz}, \& {Rodigas}}]{Schneider2014}
{Schneider}, G., {Grady}, C.~A., {Hines}, D.~C., {et~al.} 2014, \aj, 148, 59

\bibitem[{{Szab{\'o}} {et~al.}(2021){Szab{\'o}}, {Gandolfi}, {Brandeker},
  {Csizmadia}, {Garai}, {Billot}, {Broeg}, {Ehrenreich}, {Fortier}, {Fossati},
  {Hoyer}, {Kiss}, {Lecavelier des Etangs}, {Maxted}, {Ribas}, {Alibert},
  {Alonso}, {Anglada Escud{\'e}}, {B{\'a}rczy}, {Barros}, {Barrado},
  {Baumjohann}, {Beck}, {Beck}, {Bekkelien}, {Bonfils}, {Benz}, {Borsato},
  {Busch}, {Cabrera}, {Charnoz}, {Collier Cameron}, {Van Damme}, {Davies},
  {Delrez}, {Deleuil}, {Demangeon}, {Demory}, {Erikson}, {Fridlund}, {Futyan},
  {Garc{\'\i}a Mu{\~n}oz}, {Gillon}, {Guedel}, {Guterman}, {Heng}, {Isaak},
  {Lacedelli}, {Laskar}, {Lendl}, {Lovis}, {Luntzer}, {Magrin}, {Nascimbeni},
  {Olofsson}, {Osborn}, {Ottensamer}, {Pagano}, {Pall{\'e}}, {Peter}, {Piazza},
  {Piotto}, {Pollacco}, {Queloz}, {Ragazzoni}, {Rando}, {Rauer}, {Santos},
  {Scandariato}, {S{\'e}gransan}, {Serrano}, {Sicilia}, {Simon}, {Smith},
  {Sousa}, {Steller}, {Thomas}, {Udry}, {Van Grootel}, {Walton}, \&
  {Wilson}}]{Szabó2021}
{Szab{\'o}}, G.~M., {Gandolfi}, D., {Brandeker}, A., {et~al.} 2021, \aap, 654,
  A159

\bibitem[{{Szab{\'o}} {et~al.}(2022){Szab{\'o}}, {Garai}, {Brandeker},
  {Gandolfi}, {Wilson}, {Deline}, {Olofsson}, {Fortier}, {Queloz}, {Borsato},
  {Kiefer}, {Lecavelier des Etangs}, {Lendl}, {Serrano}, {Sulis}, {Ulmer Moll},
  {Van Grootel}, {Alibert}, {Alonso}, {Anglada}, {B{\'a}rczy}, {Barrado y
  Navascues}, {Barros}, {Baumjohann}, {Beck}, {Beck}, {Benz}, {Billot},
  {Bonfanti}, {Bonfils}, {Broeg}, {Cabrera}, {Charnoz}, {Collier Cameron},
  {Csizmadia}, {Davies}, {Deleuil}, {Delrez}, {Demangeon}, {Demory},
  {Ehrenreich}, {Erikson}, {Fossati}, {Fridlund}, {Gillon}, {G{\"u}del},
  {Heng}, {Hoyer}, {Isaak}, {Kiss}, {Laskar}, {Lovis}, {Magrin}, {Maxted},
  {Mecina}, {Nascimbeni}, {Ottensamer}, {Pagano}, {Pall{\'e}}, {Peter},
  {Piotto}, {Pollacco}, {Ragazzoni}, {Rando}, {Rauer}, {Ribas}, {Santos},
  {Sarajlic}, {Scandariato}, {S{\'e}gransan}, {Simon}, {Smith}, {Sousa},
  {Steller}, {Thomas}, {Udry}, {Verrecchia}, {Walton}, \&
  {Wolter}}]{Szabó2022}
{Szab{\'o}}, G.~M., {Garai}, Z., {Brandeker}, A., {et~al.} 2022, \aap, 659, L7

\bibitem[{ter Braak \& Vrugt(2008)}]{terBraak2008}
ter Braak, C. J.~F. \& Vrugt, J.~A. 2008, Statistics and Computing, 18, 435

\bibitem[{{Tsikoudi} \& {Kellett}(2000)}]{TsikoudiKellett2000}
{Tsikoudi}, V. \& {Kellett}, B.~J. 2000, \mnras, 319, 1147

\bibitem[{{Winn}(2010)}]{Winn2010}
{Winn}, J.~N. 2010, in Exoplanets, ed. S.~{Seager}, 55--77

\bibitem[{{Wisniewski} {et~al.}(2019){Wisniewski}, {Kowalski}, {Davenport},
  {Schneider}, {Grady}, {Hebb}, {Lawson}, {Augereau}, {Boccaletti}, {Brown},
  {Debes}, {Gaspar}, {Henning}, {Hines}, {Kuchner}, {Lagrange}, {Milli},
  {Sezestre}, {Stark}, \& {Thalmann}}]{Wisniewski2019}
{Wisniewski}, J.~P., {Kowalski}, A.~F., {Davenport}, J. R.~A., {et~al.} 2019,
  \apjl, 883, L8

\bibitem[{{Wittrock} {et~al.}(2023){Wittrock}, {Plavchan}, {Cale}, {Barclay},
  {Ludwig}, {Schwarz}, {M{\'e}karnia}, {Triaud}, {Abe}, {Suarez}, {Guillot},
  {Conti}, {Collins}, {Waite}, {Kielkopf}, {Collins}, {Dreizler}, {El Mufti},
  {Feliz}, {Gaidos}, {Geneser}, {Horne}, {Kane}, {Lowrance}, {Martioli},
  {Radford}, {Reefe}, {Roccatagliata}, {Shporer}, {Stassun}, {Stockdale},
  {Tan}, {Tanner}, \& {Vega}}]{Wittrock2023}
{Wittrock}, J.~M., {Plavchan}, P.~P., {Cale}, B.~L., {et~al.} 2023, \aj, 166,
  232

\bibitem[{{Yu} {et~al.}(2025){Yu}, {Garai}, {Cretignier}, {Szab{\'o}},
  {Aigrain}, {Gandolfi}, {Bryant}, {Correia}, {Klein}, {Brandeker}, {Owen},
  {G{\"u}nther}, {Winn}, {Heitzmann}, {Cegla}, {Wilson}, {Gill}, {Kriskovics},
  {Barrag{\'a}n}, {Boldog}, {Nielsen}, {Billot}, {Lafarga}, {Meech}, {Alibert},
  {Alonso}, {B{\'a}rczy}, {Barrado}, {Barros}, {Baumjohann}, {Bayliss}, {Benz},
  {Bergomi}, {Borsato}, {Broeg}, {Cameron}, {Csizmadia}, {Cubillos}, {Davies},
  {Deleuil}, {Deline}, {Demangeon}, {Demory}, {Derekas}, {Doyle}, {Edwards},
  {Egger}, {Ehrenreich}, {Erikson}, {Fortier}, {Fossati}, {Fridlund}, {Gazeas},
  {Gillon}, {G{\"u}del}, {Helling}, {Isaak}, {Kiss}, {Korth}, {Lam}, {Laskar},
  {Lecavelier des Etangs}, {Lendl}, {Magrin}, {Maxted}, {McCormac},
  {Mer{\'\i}n}, {Mordasini}, {Nascimbeni}, {O'Brien}, {Olofsson}, {Ottensamer},
  {Pagano}, {Pall{\'e}}, {Peter}, {Piazza}, {Piotto}, {Pollacco}, {Queloz},
  {Ragazzoni}, {Rando}, {Rauer}, {Ribas}, {Santos}, {Scandariato},
  {S{\'e}gransan}, {Simon}, {Smith}, {Sousa}, {Southworth}, {Stalport},
  {Steinberger}, {Sulis}, {Udry}, {Ulmer}, {Ulmer-Moll}, {Van Grootel},
  {Venturini}, {Villaver}, {Walton}, \& {Wheatley}}]{Yu2025}
{Yu}, H., {Garai}, Z., {Cretignier}, M., {et~al.} 2025, \mnras, 536, 2046

\bibitem[{{Zicher} {et~al.}(2022){Zicher}, {Barrag{\'a}n}, {Klein}, {Aigrain},
  {Owen}, {Gandolfi}, {Lagrange}, {Serrano}, {Kaye}, {Nielsen}, {Rajpaul},
  {Grandjean}, {Goffo}, \& {Nicholson}}]{Zicher2022}
{Zicher}, N., {Barrag{\'a}n}, O., {Klein}, B., {et~al.} 2022, \mnras, 512, 3060

\end{thebibliography}

\begin{appendix}
\onecolumn
\section{Observations of AU Mic with CHEOPS}
\label{allesfit-ind}

\begin{figure}[h!]
    \centering
    \resizebox{13cm}{10cm}{
    \includegraphics[bb=5 50 450 400,width=\columnwidth]{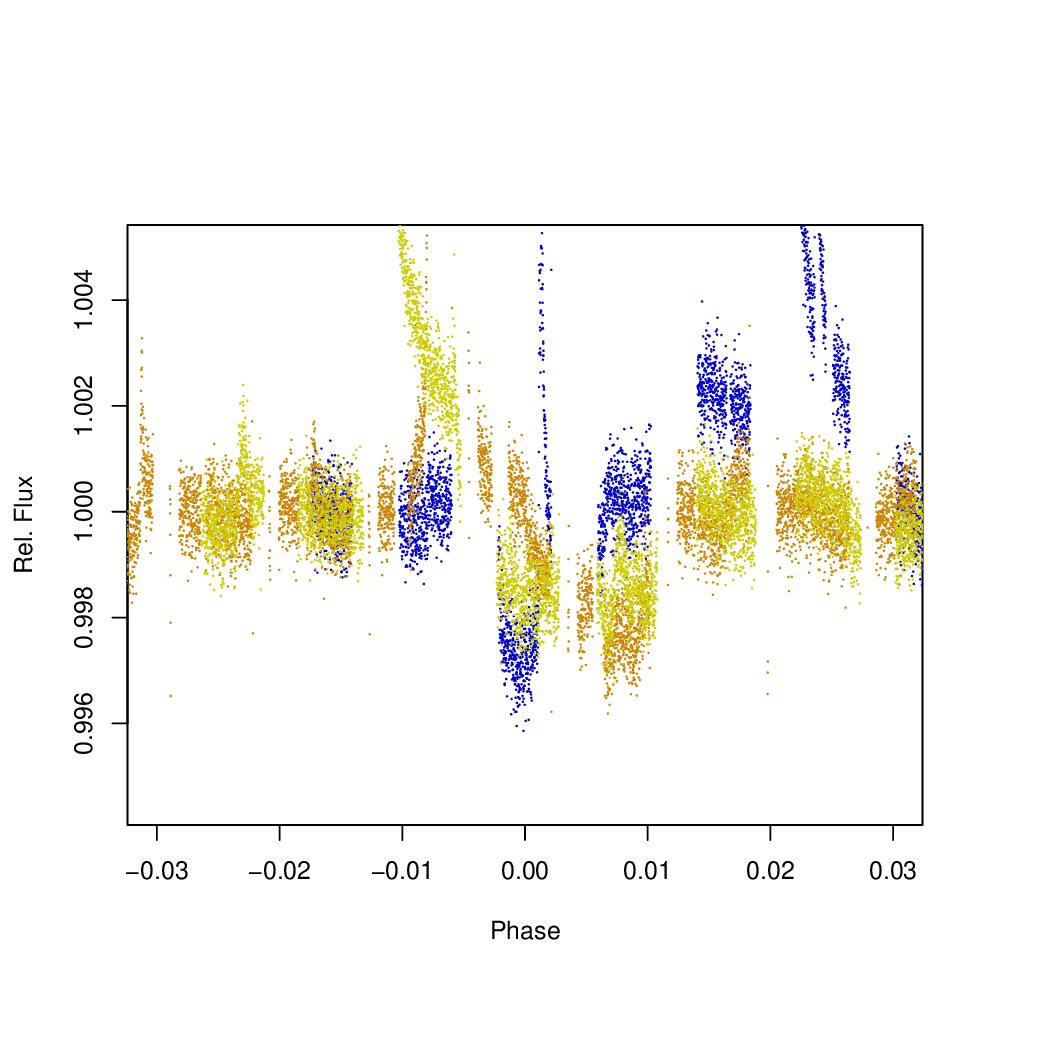}}
    \resizebox{13cm}{10cm}{
    \includegraphics[bb=5 50 450 400,width=\columnwidth]{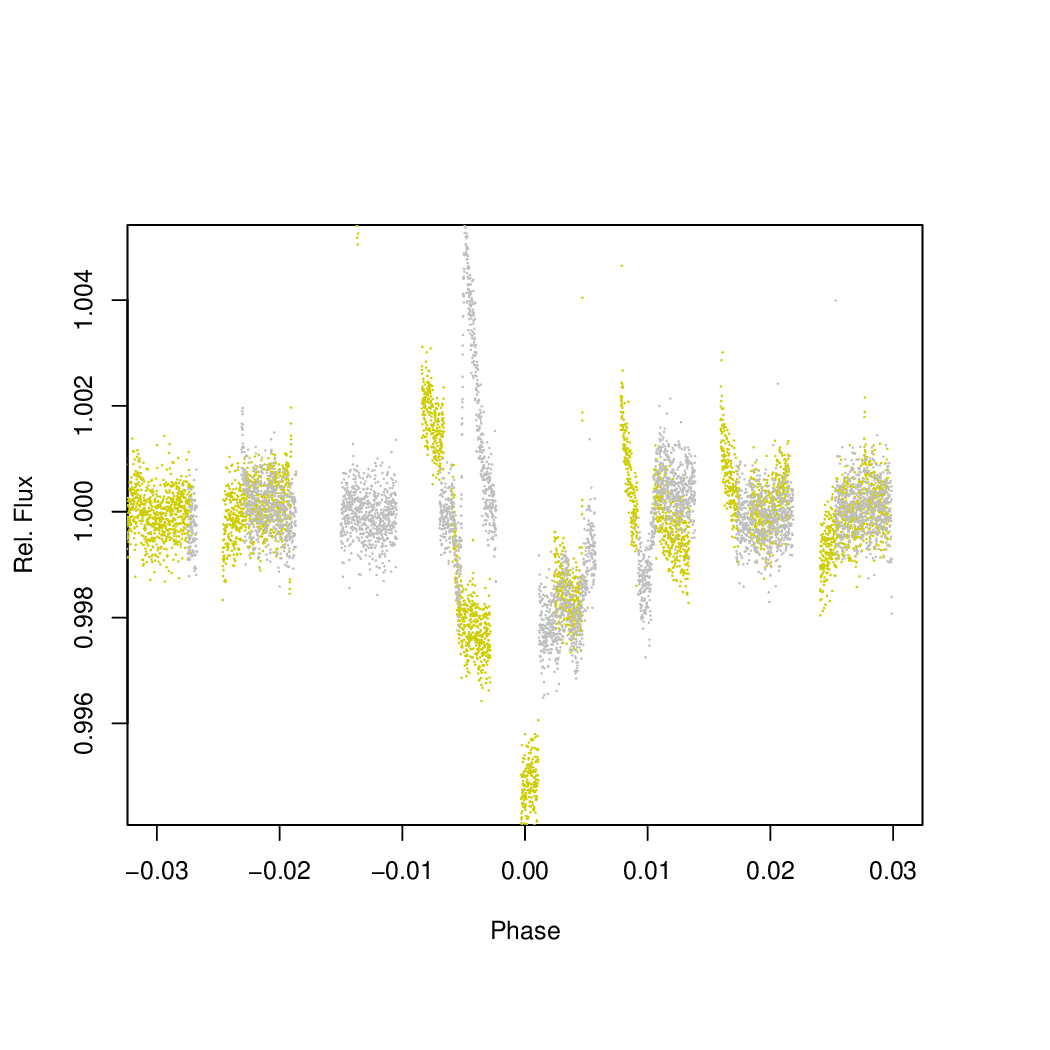}}
    \caption{Light curves of AU Mic b in 2022 (upper panel) and 2023 (lower panel) that were omitted from parameter fitting due to flare complexes during the transits. }
    \label{fig:AU_Mic_planetary_transits_bad}
\end{figure}

Transits shown in Fig.~\ref{fig:AU_Mic_planetary_transits_bad} were omitted from the analysis, due to the significant flaring activity of the host star during the planetary transits.

\section{Dynamical analyses}

\begin{figure*}[h!]
    \centering
    \resizebox{9cm}{8cm}{
    \includegraphics[width=\columnwidth]{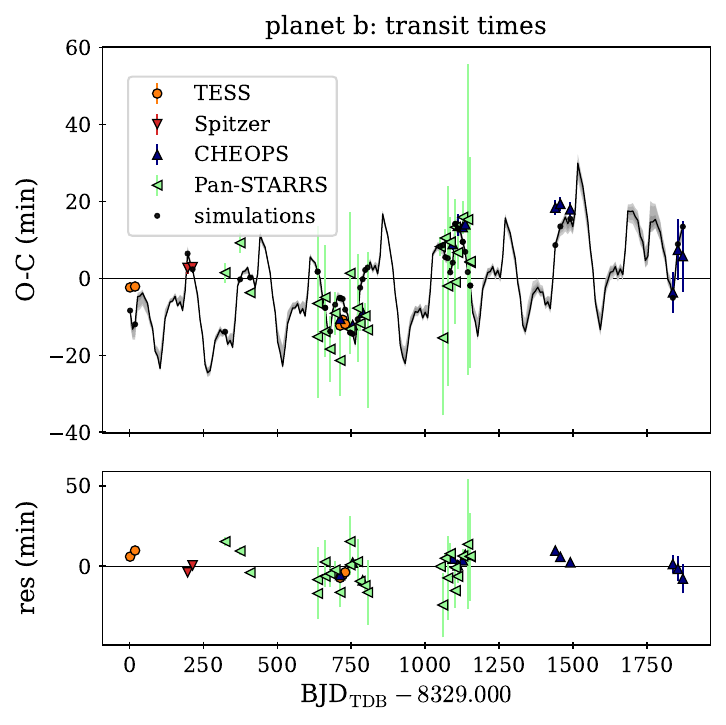}}
    \resizebox{9cm}{8cm}{\includegraphics[width=\columnwidth]{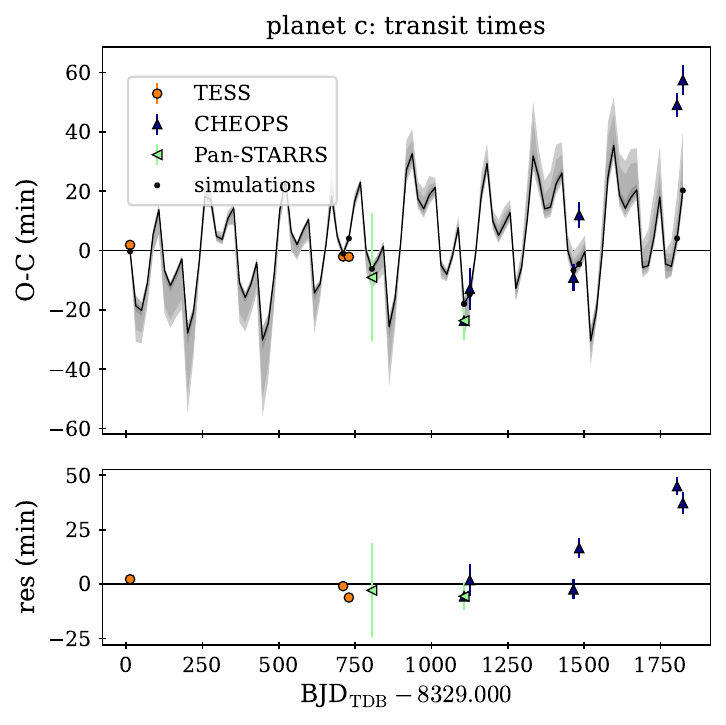}}
    \caption{\trades{} modelling of AU~Mic~b (left panels) and AU~Mic~c (right panels) of configuration (1). 
    Top panel: $(O-C)$ diagram, observed $T_0$s are plotted with coloured markers with black stroke (different colour and marker for each telescope), 
    while the simulated $O-C$ values computed by the best-fit \trades{} dynamical model are plotted as black line with black circles (corresponding to the observations). 
    Grey-shaded areas correspond to 1, 2, and $3\sigma$ of random samples drawn from the posterior distribution of \trades{}. 
    Bottom panel: Residuals computed as the difference between observed and simulated $T_0$s.}
    \label{fig:trades_oc_1}
\end{figure*}

\begin{figure*}[h!]
    \centering
    \resizebox{9cm}{8cm}{
    \includegraphics[width=\columnwidth]{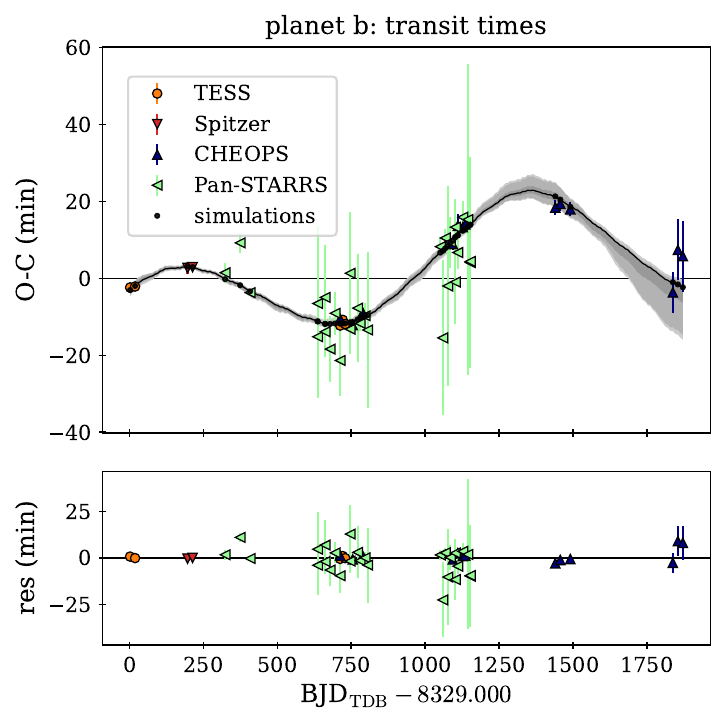}}
    \resizebox{9cm}{8cm}{\includegraphics[width=\columnwidth]{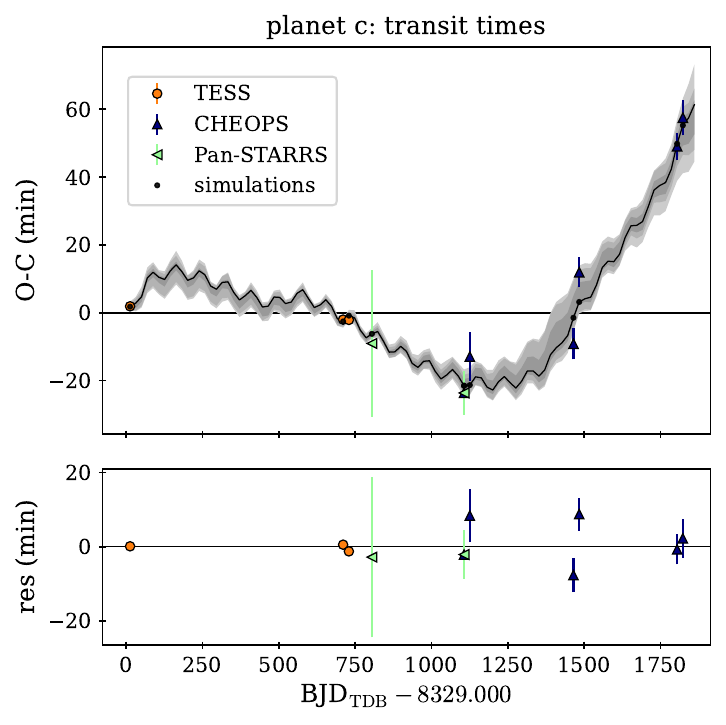}}
    \caption{\trades{} modelling of AU~Mic~b (left panels) and AU~Mic~c (right panels) of configuration (2).
    See Fig.~\ref{fig:trades_oc_1} for a description.}
    \label{fig:trades_oc_2}
\end{figure*}

\begin{figure*}
    \centering
    \includegraphics[width=\columnwidth]{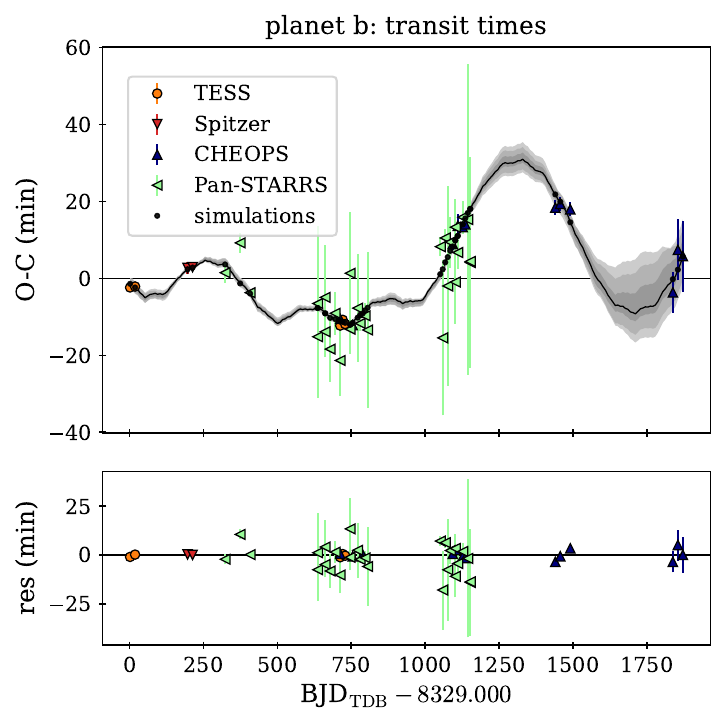} \includegraphics[width=\columnwidth]{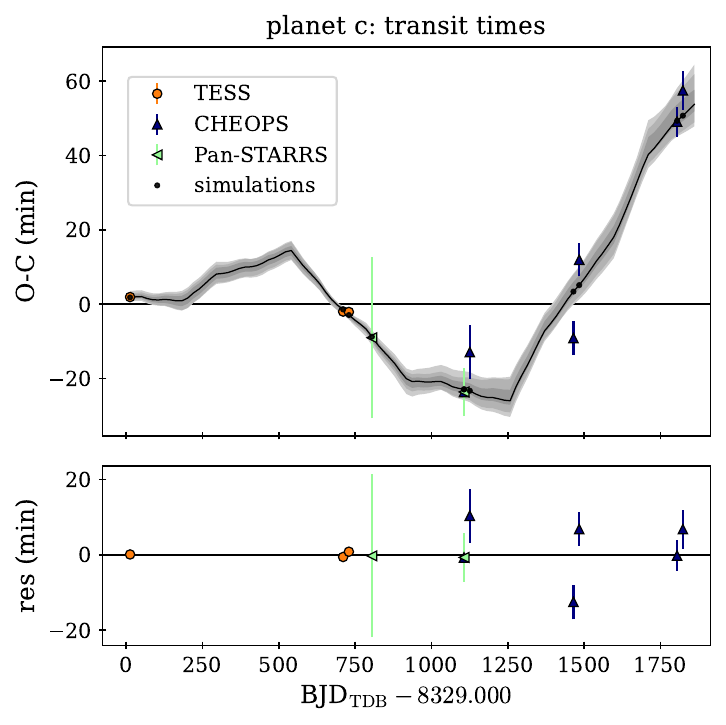}
    \caption{\trades{} modelling of AU~Mic~b (left panels) and AU~Mic~c (right panels) of configuration (3).
    See Fig.~\ref{fig:trades_oc_1} for a description.}
    \label{fig:trades_oc_3}
\end{figure*}

\begin{figure*}
    \sidecaption
    \centering
    \includegraphics[width=\columnwidth]{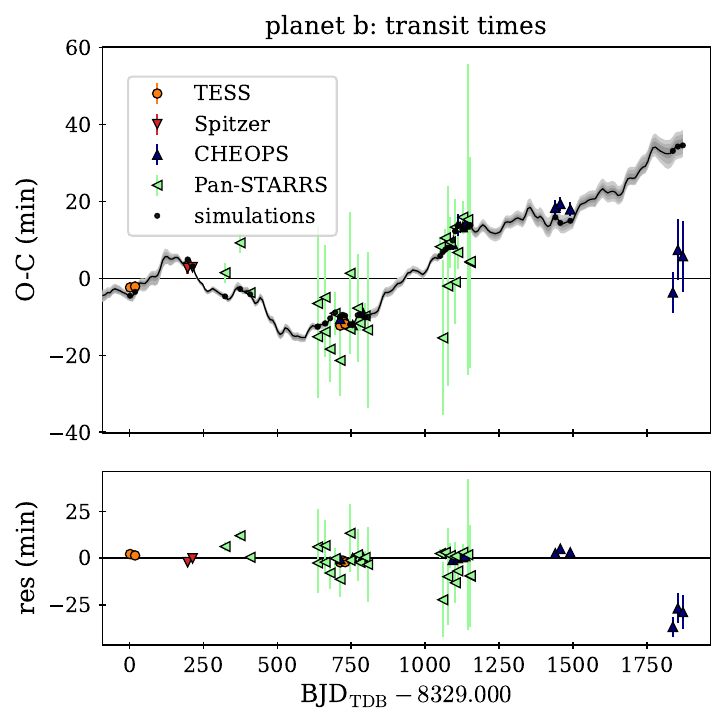} \includegraphics[width=\columnwidth]{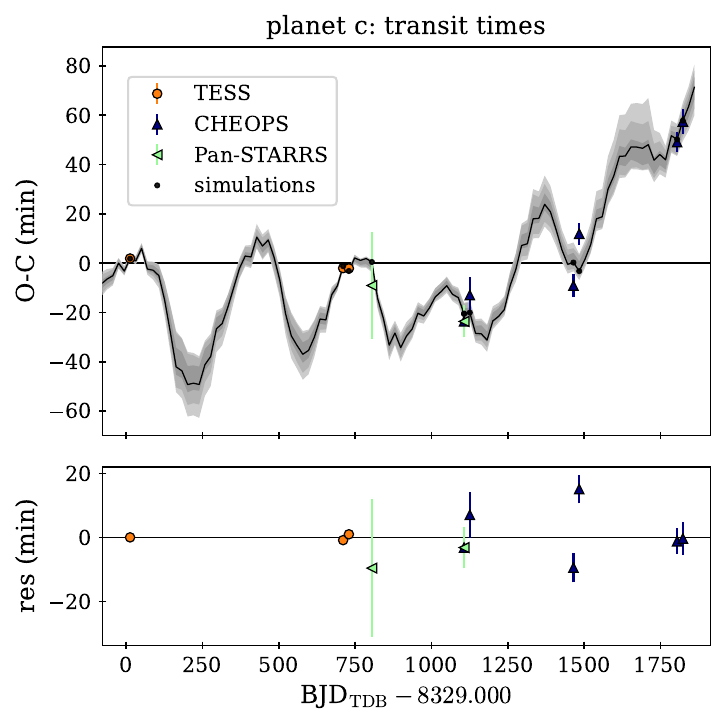}
    \caption{\trades{} modelling of AU~Mic~b (left panels) and AU~Mic~c (right panels) of the configuration (4).
    See Fig.~\ref{fig:trades_oc_1} for a description.}
    \label{fig:trades_oc_4}
\end{figure*}

\begin{figure*}
    \sidecaption
    \centering
    \includegraphics[width=\columnwidth]{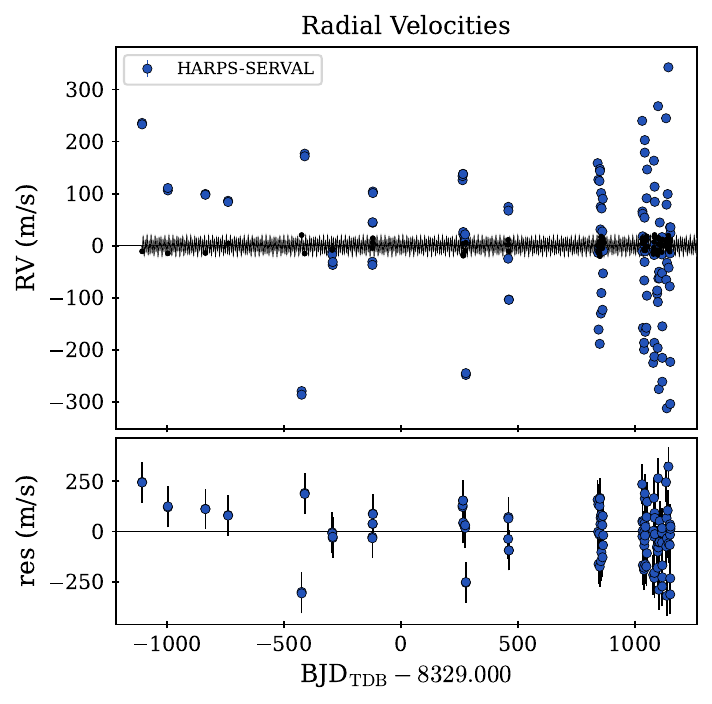}
    \caption{\trades{} RV modelling of AU~Mic system of the configuration (4).
    Black error bars in the lower panel (residuals) take into account the RV jitter term summed in quadrature to the observed uncertainty.}
    \label{fig:trades_rv_4}
\end{figure*}

\begin{figure*}
    \centering
    \includegraphics[width=\columnwidth]{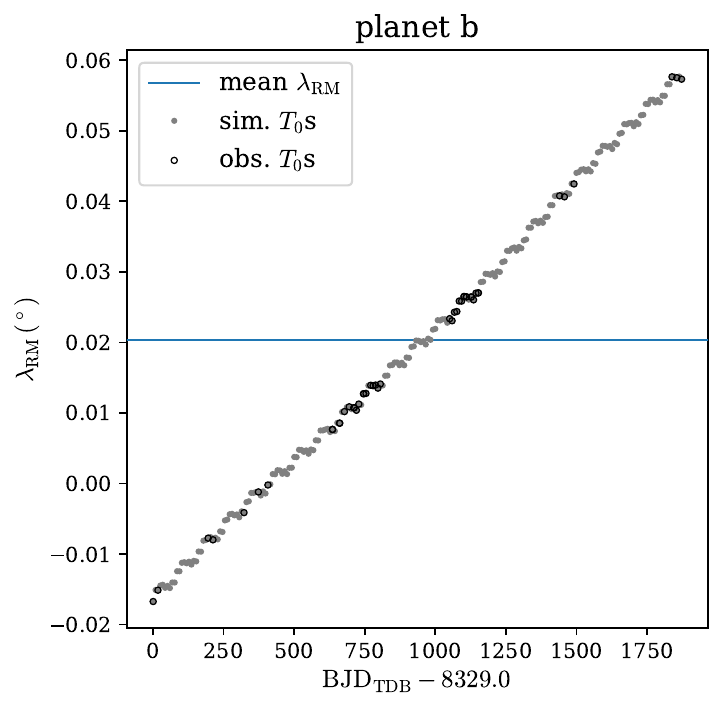} \includegraphics[width=\columnwidth]{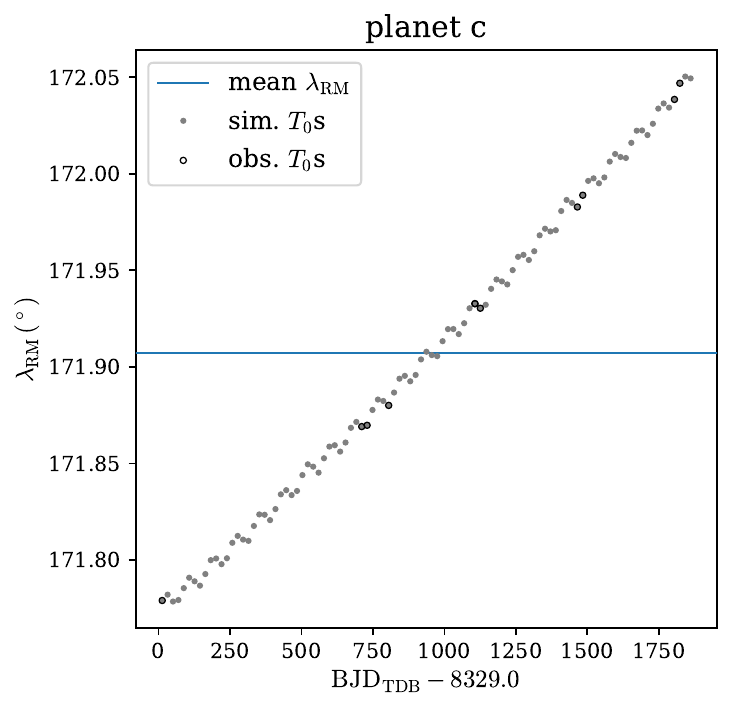}
    \caption{Projected spin-orbit misalignment ($\lambda_\mathrm{Rm}$) evolution computed at simulated transit during \trades{} modelling of AU~Mic~b (left panels) and AU~Mic~c (right panels) of configuration (2).
    Gray circles are all the synthetic transit times, highlighted as black-open circle the corresponding observed transit times.
    The horizontal-blue line is the mean value of $\lambda_\mathrm{Rm}$.
    }
    \label{fig:lambdaRM}
\end{figure*}

Summary of the results of the four dynamical analyses with \trades{}.
Priors and MAP best-fit solution for each configuration are listed in Table~\ref{tab:trades_parameters}.
Observed-Calculated diagrams of the MAP dynamical solution are shown in Fig.~\ref{fig:trades_oc_1}, \ref{fig:trades_oc_2},
\ref{fig:trades_oc_3}, and \ref{fig:trades_oc_4} for configurations (1), (2), (3) and (4),
respectively.
In Fig.~\ref{fig:trades_rv_4} the RV model from the MAP for the \trades{} configuration (4) is shown.
Figure~\ref{fig:lambdaRM} shows the evolution of the projected spin-orbit misalignment ($\lambda_\mathrm{RM}$) 
of the configuration (2)
at synthetic and observed transit times of planet b and c, respectively.

\twocolumn

\begin{sidewaystable*}
\centering\tiny\renewcommand{\arraystretch}{1.3}
\caption{\label{tab:trades_parameters}
Posterior and derived parameters of the AU~Mic system from the four dynamical analyses with TRADES.
}
\label{dynal_results}
\begin{tabular}{lcccccccc}
\hline\hline
configuration & \multicolumn{2}{c}{(1)} & \multicolumn{2}{c}{(2)} & \multicolumn{2}{c}{(3)} & \multicolumn{2}{c}{(4)} \\
planets       & \multicolumn{2}{c}{b, c} & \multicolumn{2}{c}{b, c, d} & \multicolumn{2}{c}{b, c, d} & \multicolumn{2}{c}{b, c, d} \\
              & \multicolumn{2}{c}{$i_\plc$ fixed, no RV} & \multicolumn{2}{c}{$i_\plc$ fixed, no RV} & \multicolumn{2}{c}{$i_\plc$ fitted, no RV} & \multicolumn{2}{c}{$i_\plc$ fitted, with RV} \\
\hline
Parameter (unit) & Prior & MAP (HDI) & Prior & MAP (HDI) & Prior & MAP (HDI) & Prior & MAP (HDI) \\
\hline
\textit{fitted parameters} & & & & & & & & \\
$M_\plb/M_\star\, (M_\odot/M_\star)\times10^{-6}$  
    & \unif{0.057}{319.520} & $110_{-4}^{+6}$ 
    & \unif{0.057}{319.520} & $159_{-10}^{+9}$ 
    & \unif{0.057}{319.520} & $28_{-2}^{+3}$
    & \unif{0.057}{319.520} & $187_{-15}^{+16}$\\
$P_\plb$~(days)                                    
    & \unif{8.3}{8.6} & $8.461705_{-0.000099}^{+0.000016}$ 
    & \unif{8.3}{8.6} & $8.463364_{-0.000031}^{+0.000017}$ 
    & \unif{8.3}{8.6} & $8.462697_{-0.000035}^{+0.000037}$
    & \unif{8.3}{8.6} & $8.463067_{-0.000084}^{+0.000053}$\\
$\sqrt{e}\cos{\omega}_\plb$ 
    & \unif{-0.71}{0.71} & $0.5429_{-0.0206}^{+0.0098}$ 
    & \unif{-0.71}{0.71} & $0.120_{-0.028}^{+0.027}$
    & \unif{-0.71}{0.71} & $0.05568_{-0.05032}^{+0.00096}$
    & \unif{-0.71}{0.71} & $-0.027_{-0.020}^{+0.011}$\\
$\sqrt{e}\sin{\omega}_\plb$
    & \unif{-0.71}{0.71} & $0.452_{-0.023}^{+0.014}$ 
    & \unif{-0.71}{0.71} & $0.5369_{-0.0028}^{+0.0259}$
    & \unif{-0.71}{0.71} & $-0.30225_{-0.01469}^{+0.00069}$
    & \unif{-0.71}{0.71} & $0.060_{-0.025}^{+0.032}$\\
$\lambda_\plb\, (^{\circ})$ 
    & \unif{0}{360} & $176.66_{-0.70}^{+1.75}$
    & \unif{0}{360} & $205_{-2}^{+1}$
    & \unif{0}{360} & $208.69_{-0.17}^{+1.81}$
    & \unif{0}{360} & $211.094_{-0.084}^{+0.178}$\\
$M_\plc/M_\star\, (M_\odot/M_\star)\times10^{-6}$  
    & \unif{0.057}{319.520} & $127_{-6}^{+2}$ 
    & \unif{0.057}{319.520} & $26_{-1}^{+7}$
    & \unif{0.057}{319.520} & $86_{-5}^{+8}$
    & \unif{0.057}{319.520} & $86_{-2}^{+8}$\\
$P_\plc$~(days)
    & \unif{18.6}{19.0} & $18.859052_{-0.000013}^{+0.000432}$
    & \unif{18.6}{19.0} & $18.86404_{-0.00014}^{+0.00135}$
    & \unif{18.6}{19.0} & $18.859167_{-0.000077}^{+0.000101}$
    & \unif{18.6}{19.0} & $18.85263_{-0.00057}^{+0.00063}$\\
$\sqrt{e}\cos{\omega}_\plc$ 
    & \unif{-0.71}{0.71} & $0.1608_{-0.0074}^{+0.0043}$
    & \unif{-0.71}{0.71} & $0.012_{-0.016}^{+0.012}$
    & \unif{-0.71}{0.71} & $0.0230_{-0.0028}^{+0.0250}$
    & \unif{-0.71}{0.71} & $-0.049_{-0.018}^{+0.016}$\\
$\sqrt{e}\sin{\omega}_\plc$
    & \unif{-0.71}{0.71} & $-0.4473_{-0.0129}^{+0.0039}$
    & \unif{-0.71}{0.71} & $0.0510_{-0.0587}^{+0.0035}$
    & \unif{-0.71}{0.71} & $-0.0548_{-0.0246}^{+0.0076}$
    & \unif{-0.71}{0.71} & $-0.142_{-0.017}^{+0.023}$\\
$\lambda_\plc\, (^{\circ})$
    & \unif{0}{360} & $202.46_{-0.31}^{+1.17}$
    & \unif{0}{360} & $189_{-2}^{+4}$
    & \unif{0}{360} & $155.51_{-2.29}^{+0.57}$
    & \unif{0}{360} & $72_{-1}^{+1}$\\
$i_\plc\, (^{\circ})$
    & fixed & - 
    & fixed & - 
    & \unif{80}{100} & $90.89_{-0.46}^{+0.98}$
    & \unif{80}{100} & $90.55_{-1.05}^{+0.37}$\\
$\Omega_\plc\, (^{\circ})$ 
    & \unif{0}{360} & $15.09_{-0.33}^{+1.36}$
    & \unif{0}{360} & $352_{-2}^{+4}$
    & \unif{0}{360} & $318.09_{-2.26}^{+0.59}$
    & \unif{0}{360} & $233_{-1}^{+2}$\\
$M_\pld/M_\star\, (M_\odot/M_\star)\times10^{-6}$  
    & - & - 
    & \unif{0.057}{319.520} & $1.22_{-0.12}^{+0.10}$
    & \unif{0.057}{319.520} & $1.4562_{-0.0061}^{+0.1593}$
    & \unif{0.057}{319.520} & $5.91_{-0.40}^{+0.50}$\\
$P_\pld$~(days) 
    & - & - 
    & \unif{11}{15} & $12.6173_{-0.0114}^{+0.0050}$
    & \unif{11}{15} & $13.6140_{-0.0016}^{+0.0026}$
    & \unif{11}{15} & $13.9276_{-0.0043}^{+0.0052}$\\
$\sqrt{e}\cos{\omega}_\pld$
    & - & - 
    & \unif{-0.71}{0.71} & $-0.2041_{-0.0423}^{+0.0084}$
    & \unif{-0.71}{0.71} & $0.059_{-0.012}^{+0.038}$
    & \unif{-0.71}{0.71} & $0.3067_{-0.0049}^{+0.0133}$\\
$\sqrt{e}\sin{\omega}_\pld$
    & - & - 
    & \unif{-0.71}{0.71} & $0.4993_{-0.0189}^{+0.0035}$
    & \unif{-0.71}{0.71} & $0.50045_{-0.01099}^{+0.00069}$
    & \unif{-0.71}{0.71} & $-0.2443_{-0.0078}^{+0.0219}$\\
$\lambda_\pld\, (^{\circ})$ 
    & - & - 
    & \unif{0}{360} & $18_{-2}^{+8}$
    & \unif{0}{360} & $348.65_{-2.99}^{+0.60}$
    & \unif{0}{360} & $7_{-3}^{+1}$\\
$i_\pld\, (^{\circ})$
    & - & - 
    & \unif{50}{130} & $127.08_{-0.52}^{+2.44}$
    & \unif{50}{130} & $108.86_{-0.89}^{+1.54}$
    & \unif{50}{130} & $101.07_{-2.73}^{+0.40}$\\
$\Omega_\pld\, (^{\circ})$
    & - & - 
    & \unif{0}{360} & $120_{-1}^{+5}$
    & \unif{0}{360} & $184.39_{-2.17}^{+0.22}$
    & \unif{0}{360} & $211.15_{-1.33}^{+0.86}$\\
$\gamma_\mathrm{RV}\, (\mps)$
    & - & - 
    & - & - 
    & - & - 
    & \unif{-5412}{5243} & $-100_{-10}^{+3}$\\
$\log_{2}\sigma_\mathrm{jitter, RV}$
    & - & - 
    & - & -
    & - & - 
    & \unif{-49.8}{6.6} & $6.6346_{-0.0041}^{+0.0091}$\\
\textit{derived parameters} & & & & & & & & \\
$M_\plb\, (M_\oplus)$
    & \bounds{0.01}{50} & $18_{-1}^{+2}$
    & \bounds{0.01}{50} & $26_{-2}^{+2}$
    & \bounds{0.01}{50} & $4.70_{-0.40}^{+0.55}$
    & \bounds{0.01}{50} & $31_{-3}^{+3}$\\
$e_\plb$ 
    & \bounds{0.0}{0.5} & $0.4988_{-0.0148}^{+0.0012}$
    & \bounds{0.0}{0.5} & $0.303_{-0.013}^{+0.040}$ $(2\sigma)$
    & \bounds{0.0}{0.5} & $0.0945_{-0.0029}^{+0.0063}$
    & \bounds{0.0}{0.5} & $0.0044_{-0.0023}^{+0.0033}$\\
$\omega_\plb\, (^{\circ})$
    & \unif{0}{360} & $40_{-2}^{+2}$
    & \unif{0}{360} & $77_{-3}^{+3}$
    & \unif{0}{360} & $-80_{-15}^{+3}$ $(2\sigma)$
    & \unif{0}{360} & $114_{-22}^{+17}$\\
$\mathcal{M}_\plb\, (^{\circ})$
    & \unif{0}{360} & $316.89_{-0.40}^{+1.35}$
    & \unif{0}{360} & $307.30_{-2.04}^{+0.99}$
    & \unif{0}{360} & $108.249_{-0.030}^{+11.401}$
    & \unif{0}{360} & $-83_{-17}^{+22}$\\
$M_\plc\, (M_\oplus)$
    & \bounds{0.01}{50} & $21_{-2}^{+1}$
    & \bounds{0.01}{50} & $4.40_{-0.23}^{+1.28}$
    & \bounds{0.01}{50} & $14.24_{-1.00}^{+1.71}$
    & \bounds{0.01}{50} & $14.31_{-0.73}^{+1.76}$\\
$e_\plc$
    & \bounds{0.0}{0.5} & $0.2260_{-0.0037}^{+0.0105}$ 
    & \bounds{0.0}{0.5} & $0.0027_{-0.0027}^{+0.0057}$ $(2\sigma)$
    & \bounds{0.0}{0.5} & $0.00353_{-0.00023}^{+0.00287}$
    & \bounds{0.0}{0.5} & $0.0226_{-0.0042}^{+0.0045}$\\
$\omega_\plc\, (^{\circ})$
    & \unif{0}{360} & $289.77_{-1.17}^{+0.52}$
    & \unif{0}{360} & $77_{-58}^{+40}$
    & \unif{0}{360} & $-67_{-9}^{+19}$
    & \unif{0}{360} & $-109_{-7}^{+8}$\\
$\mathcal{M}_\plc\, (^{\circ})$
    & \unif{0}{360} & $257.60_{-0.80}^{+1.45}$ 
    & \unif{0}{360} & $121_{-40}^{+58}$
    & \unif{0}{360} & $265_{-19}^{+9}$
    & \unif{0}{360} & $307_{-8}^{+7}$\\
$M_\pld\, (M_\oplus)$
    & - & - 
    & \bounds{0.01}{50} & $0.203_{-0.024}^{+0.022}$
    & \bounds{0.01}{50} & $0.2424_{-0.0084}^{+0.0333}$
    & \bounds{0.01}{50} & $0.983_{-0.087}^{+0.106}$\\
$e_\pld$
    & - & - 
    & \bounds{0.0}{0.5} & $0.2910_{-0.0057}^{+0.0028}$
    & \bounds{0.0}{0.5} & $0.2539_{-0.0086}^{+0.0017}$ $(2\sigma)$
    & \bounds{0.0}{0.5} & $0.1537_{-0.0059}^{+0.0033}$\\
$\omega_\pld\, (^{\circ})$
    & - & - 
    & \unif{0}{360} & $112.23_{-0.87}^{+5.02}$
    & \unif{0}{360} & $83_{-5}^{+1}$
    & \unif{0}{360} & $321_{-2}^{+3}$\\
$\mathcal{M}_\pld\, (^{\circ})$
    & - & - 
    & \unif{0}{360} & $145.02_{-1.89}^{+0.87}$
    & \unif{0}{360} & $80.97_{-0.82}^{+3.67}$
    & \unif{0}{360} & $194.56_{-4.14}^{+0.96}$\\
$\sigma_\mathrm{jitter, RV}\, (\mps)$
    & - & - 
    & - & - 
    & - & - 
    & \bounds{10^{-15}}{100} & $99.36_{-0.28}^{+0.63}$\\
$\mathrm{BIC} = -2\ln\mathcal{P} + n_\mathrm{fit}\ln{n_\mathrm{data}}$ 
    & \multicolumn{2}{c}{$208.1$} 
    & \multicolumn{2}{c}{$-536.6$} 
    & \multicolumn{2}{c}{$-520.3$}
    & \multicolumn{2}{c}{$1115.2$} \\
\hline
\end{tabular}
\tablefoot{
All the parameters have been defined at the reference time $\mathrm{BJD_{TDB}}-2450000 = 8329$.
Priors indicated 
as $\norm{\mu}{\sigma}$ means Gaussian-Normal distribution,
as $\unif{\mathrm{lower}}{\mathrm{upper}}$ means Uniform distribution,
and as $\bounds{\mathrm{lower}}{\mathrm{upper}}$ means the values have been used to force the boundaries,
e.g we sampled uniformly in $\sqrt{e}\cos(\omega)$ and $\sqrt{e}\sin(\omega)$, 
but we checked that $e$ was within the boundaries.
$\lambda$ is the mean longitude, 
that is $\lambda=\omega + \Omega + \mathcal{M}$, where
$\omega$ is the argument of the pericentre (or periastron),
$\Omega$ is the longitude of the ascending node,
and $\mathcal{M}$ is mean anomaly.
In some cases, the physical parameter converted from the MAP fitted parameters was not within the HDI at $68.27\%$,
so we reported also the value of the $2\sigma$ uncertainty from the HDI at $95.5\%$.
}
\end{sidewaystable*}

\end{appendix}

\end{document}